\documentclass[%
 reprint,
nofootinbib,
nobibnotes,
 amsmath,amssymb,
 aps,
floatfix,
]{revtex4-2}

\usepackage{enumitem} 
\usepackage{graphicx}
\usepackage{dcolumn}
\usepackage{bm}
\usepackage{amsmath,amssymb,amsfonts}%
\usepackage{amsthm}%
\usepackage{mathrsfs}%
\usepackage{xcolor}%
\usepackage{hyperref}
\usepackage{cleveref}
\usepackage{comment}
\usepackage[T1]{fontenc}
\usepackage{multirow}
\usepackage{bibunits}

\newcommand{\ECD}{\mathrm{ECD}}

\makeatletter
\def\maketitle{
	\@author@finish
	\title@column\titleblock@produce
	\suppressfloats[t]}
\makeatother 

\makeatletter
\patchcmd{\@ssect@ltx}
{\addcontentsline{toc}{#1}{\protect\numberline{}#8}}
{}
{}
{}
\makeatother

\usepackage{dsfont}

\begin{document}


\title{Quantum Error Correction of Qudits Beyond Break-even}

\author{Benjamin L. Brock}
\email{benjamin.brock@yale.edu}

\author{Shraddha Singh}

\author{Alec Eickbusch} \thanks{Present address: Google Quantum AI, Santa Barbara, CA}

\author{Volodymyr V. Sivak} \thanks{Present address: Google Quantum AI, Santa Barbara, CA}

\author{Andy Z. Ding}

\author{Luigi Frunzio}

\author{Steven M. Girvin}

\author{Michel H. Devoret}\email{michel.devoret@yale.edu}

\affiliation{\medskip Departments of Applied Physics and Physics, Yale University, New Haven, CT, USA \\ Yale Quantum Institute, Yale University, New Haven, CT, USA}

\date{\today}

\begin{abstract}

Hilbert space dimension is a key resource for quantum information processing. A large Hilbert space is not only an essential requirement for quantum error correction, but it can also be advantageous for realizing gates and algorithms more efficiently. There has thus been considerable experimental effort in recent years to develop quantum computing platforms using qudits (d-dimensional quantum systems with d \textgreater 2) as the fundamental unit of quantum information. Just as with qubits, quantum error correction of these qudits will be necessary in the long run, but to date error correction of logical qudits has not been demonstrated experimentally. Here we report the experimental realization of an error-corrected logical qutrit (d=3) and ququart (d=4) by employing the Gottesman-Kitaev-Preskill (GKP) bosonic code. Using a reinforcement learning agent, we optimize the GKP qutrit (ququart) as a ternary (quaternary) quantum memory and achieve beyond break-even error correction with a gain of 1.82 ± 0.03 (1.87 ± 0.03). This work represents a new way of leveraging the large Hilbert space of a harmonic oscillator for hardware-efficient quantum error correction.

\end{abstract}

\maketitle
\begin{bibunit}[apsrev4-2]

The number of quantum states available to a quantum computer, quantified by its Hilbert space dimension, is a fundamental and precious resource \cite{blume-kohout_climbing_2002, greentree_maximizing_2004}. Crucially, the goal of achieving quantum advantage at scale relies on the ability to manipulate an exponentially large Hilbert space with sub-exponentially many operations. This large Hilbert space is typically realized using $N$ qubits (two-level quantum systems), giving rise to a $2^{N}$-dimensional Hilbert space. However, most physical realizations of qubits have many more than two available states for storing quantum information. These valuable additional quantum states often go untapped, since the methods for working with qudits ($d$-level quantum systems with $d>2$) as the fundamental unit of quantum information are more complicated and less well developed than those for working with qubits \cite{wang_qudits_2020}. 

On the other hand, embracing these qudits can enable more efficient distillation of magic states \cite{campbell_enhanced_2014}, synthesis of gates \cite{ralph_efficient_2007, fedorov_implementation_2012}, compilation of algorithms \cite{bocharov_factoring_2017, kiktenko_scalable_2020, gokhale_asymptotic_2019, Chu2023}, and simulation of high-dimensional quantum systems \cite{meth_simulating_2023, sawaya_resource-efficient_2020}. For these reasons, there has been considerable experimental effort in recent years to develop qudit-based platforms for quantum computing, for example using donor spins in silicon \cite{asaad_coherent_2020, fernandez_de_fuentes_navigating_2024, yu_creation_2024}, ultracold atoms and molecules \cite{vilas_optical_2024, lindon_complete_2023, smith_quantum_2013, chaudhury_quantum_2007}, optical photons \cite{chi_programmable_2022, giovannini_characterization_2013, bent_experimental_2015, yoshikawa_heralded_2018, reimer_high-dimensional_2019, babazadeh_high-dimensional_2017, wang_multidimensional_2018, kues_-chip_2017, zhang_spinorbit_2022}, superconducting circuits \cite{champion_multi-frequency_2024, goss_high-fidelity_2022, morvan_qutrit_2021, neeley_emulation_2009, nguyen_empowering_2024, roy_two-qutrit_2023, roy_synthetic_2024, svetitsky_hidden_2014, wang_systematic_2024}, trapped ions \cite{hrmo_native_2023, leupold_sustained_2018, low_control_2023, ringbauer_universal_2022, senko_realization_2015}, and vacancy centers \cite{adambukulam_hyperfine_2024, singh_multi-photon_2022, soltamov_excitation_2019}. If qudits are to be useful in the long run, however, quantum error correction (QEC) will be necessary.

In this work we experimentally demonstrate QEC of logical qudits with $d>2$, using the Gottesman-Kitaev-Preskill (GKP) bosonic code \cite{gottesman_encoding_2001} to realize a logical qutrit ($d=3$) and ququart ($d=4$) encoded in grid states of an oscillator. Our optimized GKP qutrit (ququart) lives longer, on average, than the best physical qutrit (ququart) available in our system by a factor of $1.82\pm 0.03$ ($1.87\pm 0.03$), making this one of only a handful of experiments to beat the break-even point of QEC for quantum memories \cite{ofek_extending_2016, sivak_real-time_2023, ni_beating_2023, acharya_quantum_2024}. This experiment represents a new way of leveraging the large Hilbert space of an oscillator, building on previous realizations of GKP qubits \cite{fluhmann_encoding_2019, de_neeve_error_2022, campagne-ibarcq_quantum_2020, sivak_real-time_2023, lachance-quirion_autonomous_2024, konno_logical_2024, matsos_universal_2024} and bosonic codes \cite{cai_bosonic_2021, gertler_protecting_2021, ofek_extending_2016, ni_beating_2023}. Access to a higher-dimensional error-corrected manifold of quantum states may enable more hardware-efficient architectures for quantum information processing.

\section*{Error Correction of GKP Qudits}

\begin{figure}
\includegraphics[width=\columnwidth]{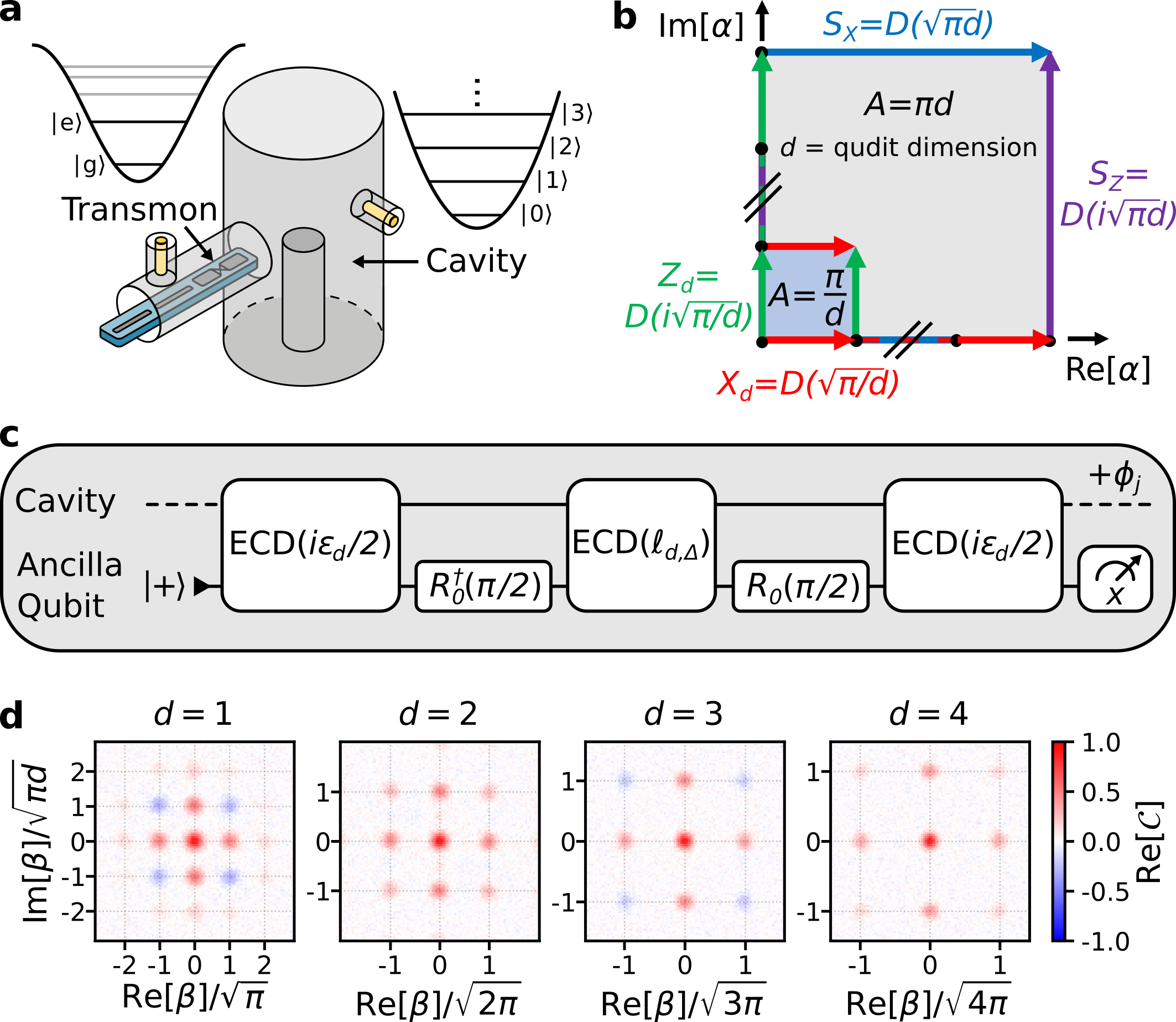}
\caption{\textbf{a}, Schematic of the experimental device. \textbf{b}, Geometric structure of the displacement operators that define the single-mode square GKP code. \textbf{c}, Circuit for one round of finite-energy GKP qudit stabilization, generalizing the small-Big-small (sBs) protocol \cite{royer_stabilization_2020}. The big echoed conditional displacement (ECD) gate \cite{eickbusch_fast_2022} of amplitude $\ell_{d,\Delta} = \sqrt{\pi d}\cosh(\Delta^{2})$ is approximately the stabilizer length, while the small ECD gates of amplitude $\varepsilon_{d}/2 = \sqrt{\pi d}\sinh(\Delta^{2})/2$ account for the finite-energy envelope size $\Delta$. At the end of sBs round $j$, the cavity phase is updated by $\phi_{j}$ (see Methods). \textbf{d}, Measured real part of the characteristic function of the maximally-mixed GKP qudit state for $d=1$ through $d=4$ with $\Delta = 0.3$, prepared by performing $300$ sBs rounds starting from the cavity in its vacuum state $|0\rangle$.
}\label{fig1}
\end{figure}

Our experimental device is the same as in Ref. \cite{sivak_real-time_2023}, and consists of a tantalum transmon \cite{place_new_2021, wang_towards_2022, ganjam_surpassing_2024} dispersively coupled to a 3D superconducting microwave cavity \cite{reagor_quantum_2016}, as shown in Fig.  \ref{fig1}a. The cavity hosts an oscillator mode (described by Fock states $\{|n\rangle : n\in\mathbb{Z}_{\geq 0}\}$ and mode operator $a$), which is used for storing our logical GKP states. The transmon hosts a qubit (described by ground and excited states $\{|g\rangle,|e\rangle \}$ and Pauli operators $\sigma_{x,y,z}$), which is used as an ancilla for controlling the oscillator and performing error correction. The cavity has an energy relaxation lifetime of $T_{1,c} = 631$ $\mathrm{\mu}$s and Ramsey coherence time $T_{2R,c} = 1030$ $\mathrm{\mu}$s, while the transmon has $T_{1,q} = 295$ $\mathrm{\mu}$s and Hahn-echo lifetime $T_{2E,q} = 286$ $\mathrm{\mu}$s \cite{noauthor_see_nodate}.

We employ the single-mode square GKP code \cite{gottesman_encoding_2001}, which is designed to be translationally symmetric in phase space. The structure of the code comes from the geometric phase associated with displacement operators $D(\alpha) = \exp(\alpha a^{\dagger} - \alpha^{*}a)$ in phase space, as depicted in Fig. \ref{fig1}b. Two displacements commute up to a phase given by twice the area $A$ they enclose, such that $D(\alpha_{1})D(\alpha_{2})=\exp(2iA)D(\alpha_{2})D(\alpha_{1})$ with $A=\mathrm{Im}[\alpha_{1}\alpha_{2}^{*}]$. The ideal code has stabilizer generators $S_{X} = D(\ell_{d})$ and $S_{Z} = D(i\ell_{d})$, where $\ell_{d}$ is the stabilizer length. In order for the two stabilizers of the code to have a common $+1$ eigenspace (i.e. the codespace) they must commute, which means they must enclose an area $\pi d$ in phase space for positive integer $d$, such that $\ell_{d} = \sqrt{\pi d}$, where $d$ is the dimension of the code space \cite{gottesman_encoding_2001, noauthor_see_nodate}. The codewords of this idealized logical qudit are grids of position eigenstates
\begin{equation}
|Z_{n}\rangle_{d} \propto \sum_{k=-\infty}^{\infty}|n\sqrt{2\pi/d} + k\sqrt{2\pi d}\rangle_{q}
\end{equation}
where $n=0, 1 , ..., d-1$ and $q = (a + a^{\dagger})/\sqrt{2}$ is the position operator. Note that with our choice of phase-space units, translations in position and displacements along the real axis of phase space differ in amplitude by a factor of $\sqrt{2}$. The logical operators of the code are the displacement operators $X_{d} = D(\sqrt{\pi/d})$ and $Z_{d} = D(i\sqrt{\pi/d})$, which act on the codespace as
\begin{equation}
    \begin{split}
    Z_{d}|Z_{n}\rangle_{d} &= (\omega_{d})^{n}|Z_{n}\rangle_{d}, \\
    X_{d}|Z_{n}\rangle_{d} &= |Z_{(n+1)\:\mathrm{mod}\:d}\rangle_{d}, \\
    \end{split}
\end{equation}
where $\omega_{d} = \exp(2\pi i/d)$ is the primitive $d$th root of unity. These operators $Z_{d}$ and $X_{d}$ are the generalized Pauli operators \cite{weyl_theory_1950, schwinger_unitary_1960}, which are unitary but no longer Hermitian for $d>2$. These operators obey the generalized commutation relation $Z_{d}X_{d} = \omega_{d}X_{d}Z_{d}$, determined by the area these displacements enclose in phase space. Compared to GKP qubits, GKP qudits have a longer stabilizer length that is proportional to $\sqrt{d}$, such that they encode information further out in phase space, and a shorter distance between logical states that is proportional to $1/\sqrt{d}$.

In practice, we work with an approximate finite-energy version of this code, formally obtained by applying the Gaussian envelope operator $E_{\Delta} = \exp(-\Delta^{2}a^{\dagger}a)$ to both the operators and states of the ideal code \cite{noauthor_see_nodate, royer_stabilization_2020, grimsmo_quantum_2021}. The parameter $\Delta$ determines both the squeezing of individual quadrature peaks in the grid states as well as their overall extent in energy: for smaller $\Delta$, the peaks are more highly squeezed and the states have more energy. As we increase the qudit dimension $d$, we expect to require smaller $\Delta$, since the logical states are more closely spaced and contain information further out in phase space (i.e., at higher energies). With smaller $\Delta$, we in turn expect the lifetime of our GKP qudits to decrease, since having more energy amplifies the rate of oscillator photon loss, and having information stored further out in phase space amplifies the effects of oscillator dephasing.

To stabilize the finite-energy GKP qudit manifold, we adapt the small-Big-small (sBs) protocol \cite{royer_stabilization_2020} to the stabilizer length $\ell_{d} = \sqrt{\pi d}$, as shown in Fig. \ref{fig1}c. This circuit, consisting of echoed conditional displacement gates $\ECD(\beta) = D(\beta/2)|e\rangle\langle g| + D(-\beta/2)|g\rangle\langle e|$ and ancilla qubit rotations $R_{\phi}(\theta) = \exp[i(\sigma_{x}\cos\phi + \sigma_{y}\sin\phi)\theta/2]$, realizes an engineered dissipation onto the finite-energy GKP qudit manifold that removes the entropy associated with physical errors in the oscillator before they can accumulate into logical errors \cite{royer_stabilization_2020, de_neeve_error_2022, noauthor_see_nodate}. This protocol is autonomous, only requiring reset of the ancilla between rounds. We also update the reference phase of the cavity mode between rounds to stabilize both quadratures in phase space (see Methods). 

To verify that this generalized sBs protocol works as intended we run it for $300$ rounds, starting with the cavity in vacuum, which prepares the maximally mixed state of the finite-energy GKP qudit $\rho_{d}^{\mathrm{mix}} = \frac{1}{d}\sum_{n=0}^{d-1}|Z_{n}\rangle\langle Z_{n}|_{d}$. We perform characteristic function (CF) tomography \cite{fluhmann_direct_2020, barnett_methods_2002} of $\rho_{d}^{\mathrm{mix}}$ prepared in this way, the results of which are shown in Fig. \ref{fig1}d. As expected from its definition $\mathcal{C}(\beta) = \langle D(\beta)\rangle$, the CF of these states has peaks at the stabilizer lengths, which increase with $d$ according to $\ell_{d} = \sqrt{\pi d}$. The negative regions of $\mathrm{Re}[\mathcal{C}(\beta)]$ for odd $d$ are simply a consequence of the geometric phase associated with displacement operators, $D(e^{i\pi/4}\sqrt{2\pi d}) = (-1)^{d}D(\sqrt{\pi d})D(i\sqrt{\pi d})$. However, it is interesting to note that the states $\rho_{d}^{\mathrm{mix}}$ for odd $d$ have regions of Wigner negativity \cite{noauthor_see_nodate} and are therefore non-classical \cite{kenfack_negativity_2004}. 

\section*{Characterizing Quantum Memories}

To characterize the performance of our logical qudits as quantum memories, and establish the concept of QEC gain for qudits, we follow previous work \cite{sivak_real-time_2023} and use the average channel fidelity $\mathcal{F}_{d}(\mathcal{E},I)$, which quantifies how well a channel $\mathcal{E}$ realizes the identity $I$ \cite{nielsen_simple_2002}. Although $\mathcal{F}_{d}$ will have a non-exponential time evolution in general, it can always be expanded to short times $dt$ as $\mathcal{F}_{d}(\mathcal{E},I) \approx 1-\frac{d-1}{d}\Gamma dt$, where $\Gamma$ is the effective decay rate of the channel $\mathcal{E}$ at short times \cite{noauthor_see_nodate}. This rate $\Gamma$ enables us to compare different decay channels on the same footing. In particular, we want to compare the decay rate $\Gamma^{\mathrm{logical}}_{d}$ of our logical qudit to $\Gamma^{\mathrm{physical}}_{d}$ of the best physical qudit in our system. We define the QEC gain as their ratio $G_{d} = \Gamma^{\mathrm{physical}}_{d}/\Gamma^{\mathrm{logical}}_{d}$,
and the break-even point is when this gain is unity.

The average channel fidelity can be expressed in terms of the probabilities $\langle\psi|\mathcal{E}(|\psi\rangle\langle\psi|)|\psi\rangle_{d}$ that our error correction channel $\mathcal{E}$ preserves the qudit state $|\psi\rangle_{d}$, summed over a representative set of states $\{|\psi\rangle_{d}\}$ \cite{noauthor_see_nodate}. Each of these probabilities entails a separate experiment, in which we prepare the state $|\psi\rangle_{d}$, perform error correction, and measure our logical qudit in a basis containing $|\psi\rangle_{d}$. Herein lies the primary experimental challenge of the present work: devising ways of measuring our logical GKP qudit in bases containing each state in our representative set $\{|\psi\rangle_{d}\}$, using only binary measurements of our ancilla qubit.

For the qutrit in $d=3$, our representative set of states are the bases $\{|P_{n}\rangle_{3} : n=0,1,2\}$ of Pauli operators $P\in\mathcal{P}_{3} = \{X_{3},Z_{3},X_{3}Z_{3},X_{3}^{2}Z_{3}\}$, defined by $P|P_{n}\rangle_{3} = \omega^{n}|P_{n}\rangle_{3}$ for $\omega = \exp(2\pi i/3)$. The effective decay rate of our logical GKP qutrit can then be expressed
\begin{equation}\label{eq:qutrit_decay_rate}
\Gamma_{3}^{\mathrm{GKP}} = \frac{1}{12}\sum\limits_{P \in \mathcal{P}_{3}}\sum\limits_{n=0}^{2}\gamma_{P_{n}},
\end{equation}
where $\gamma_{P_{n}}$ is the rate at which the state $|P_{n}\rangle_{3}$ decays to the maximally mixed state $\rho_{3}^{\mathrm{mix}}$. For the ququart in $d=4$, our representative set of states consists of two types of bases. The first type are the bases $\{|P_{n}\rangle_{4} : n=0,1,2,3\}$ of Pauli operators $P\in\mathcal{P}_{4} = \{X_{4},Z_{4},\sqrt{\omega}X_{4}Z_{4},X_{4}^{2}Z_{4},\sqrt{\omega}X_{4}^{3}Z_{4},X_{4}Z_{4}^{2}\}$, defined by $P|P_{n}\rangle_{4} = \omega^{n}|P_{n}\rangle_{4}$ for $\omega = i$. The second type is what we call the ququart parity basis $\{|\pm,m\rangle_{4} : m=0,1\}$ consisting of the simultaneous eigenstates of $X_{4}^{2}$ and $Z_{4}^{2}$, such that $X_{4}^{2}|\pm,m\rangle_{4} = \pm |\pm,m\rangle_{4}$ and $Z_{4}^{2}|\pm,m\rangle_{4} = (-1)^{m} |\pm,m\rangle_{4}$. The effective decay rate of our logical GKP ququart can then be expressed
\begin{equation}\label{eq:ququart_decay_rate}
\Gamma_{4}^{\mathrm{GKP}} = \frac{1}{20}\Biggl[\sum\limits_{P \in \mathcal{P}_{4}}\sum\limits_{n=0}^{3}\gamma_{P_{n}} - \sum\limits_{\substack{s=\pm \\ m=0,1}}\gamma_{\pm,m} \Biggr],
\end{equation}
where $\gamma_{P_{n}}$ ($\gamma_{\pm,m}$) is the rate at which the Pauli eigenstate $|P_{n}\rangle_{4}$ (parity state $|\pm,m\rangle_{4}$) decays to the maximally mixed state $\rho_{4}^{\mathrm{mix}}$ \cite{noauthor_see_nodate}. 

As a basis of comparison, the best physical qudit in our system is the cavity Fock qudit spanned by the states $|0\rangle, |1\rangle, ..., |d-1\rangle$. The cavity hosting this qudit decoheres under both photon loss and pure dephasing at rates $\kappa_{1,c} = 1/T_{1,c}$ and $\kappa_{\phi,c} = 1/T_{2R,c} - 1/2T_{1,c}$. From these measured rates, we can extrapolate the effective decay rate $\Gamma_{d}^{\mathrm{Fock}}$ of the cavity Fock qudit under these decoherence channels. For $d=2$ through $d=4$ we obtain $\Gamma_{2}^{\mathrm{Fock}} = (851\pm 9 \:\mathrm{\mu s})^{-1}$, $\Gamma_{3}^{\mathrm{Fock}} = (488\pm 7 \:\mathrm{\mu s})^{-1}$, and $\Gamma_{4}^{\mathrm{Fock}} = (332\pm 6 \:\mathrm{\mu s})^{-1}$ \cite{noauthor_see_nodate}.

\section*{Logical Qutrit Beyond Break-even}

\begin{figure}[t!]
\includegraphics[width=\columnwidth]{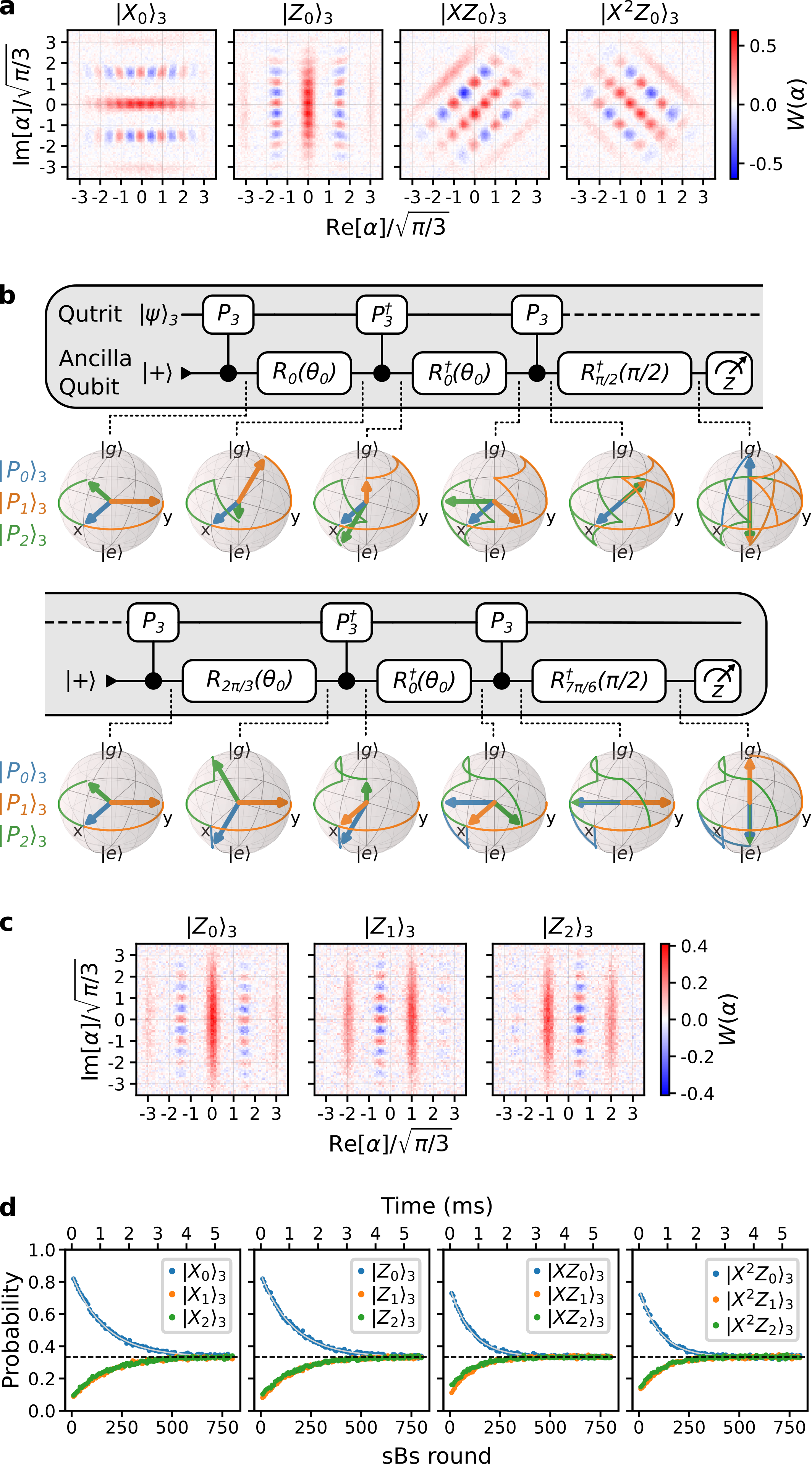}
\caption{
\textbf{Realization of a logical GKP qutrit.} \textbf{a}, State preparation of qutrit Pauli eigenstates $|P_{0}\rangle_{3}$ with $\Delta = 0.32$. \textbf{b}, Circuit for measuring a qutrit in the basis of Pauli operator $P_3$ using an ancilla qubit, where $\theta_{0} = 2\arctan(1/\sqrt{2})$. The first measurement distinguishes between the state $|P_{0}\rangle_{3}$ and the subspace $\{|P_{1}\rangle_{3},|P_{2}\rangle_{3}\}$, while the second distinguishes between $|P_{1}\rangle_{3}$ and $\{|P_{0}\rangle_{3},|P_{2}\rangle_{3}\}$. The Bloch spheres depict the trajectories taken by the ancilla when the qutrit is in each Pauli eigenstate. \textbf{c}, Backaction of the qutrit Pauli measurement in the $Z_{3}$ basis, applied to the maximally mixed qutrit state. \textbf{d}, Decay of qutrit Pauli eigenstates $|P_{0}\rangle_{3}$ under the optimized QEC protocol. The dashed black lines indicate a probability of $1/3$. The solid grey lines are exponential fits. From left to right we find $\gamma_{X_{0}}^{-1} = 1153\pm 13 \mathrm{\mu s}$, $\gamma_{Z_{0}}^{-1} = 1120\pm 15 \mathrm{\mu s}$, $\gamma_{XZ_{0}}^{-1} = 743\pm 10 \mathrm{\mu s}$, and $\gamma_{X^{2}Z_{0}}^{-1} = 727\pm 11 \mathrm{\mu s}$.
}\label{fig2}
\end{figure}

In order to measure the effective decay rate of the logical GKP qutrit via Eq. \eqref{eq:qutrit_decay_rate}, we need to prepare all of the eigenstates of the Pauli operators in $\mathcal{P}_{3}$ and perform measurements in the basis of these Pauli operators. To prepare the eigenstates $|P_{n}\rangle_{3}$ we use interleaved sequences of ECD gates and transmon rotations, which enable universal control of the oscillator mode in the cavity \cite{eickbusch_fast_2022}. We optimize depth-$8$ ECD circuits to implement the unitary that maps the cavity vacuum state $|0\rangle$ to the desired state $|P_{n}\rangle_{3}$ with envelope size $\Delta = 0.32$. Measured Wigner functions of our prepared $|P_{0}\rangle_{3}$ states are shown in Fig. \ref{fig2}a (see \cite{noauthor_see_nodate} for the other eigenstates). In general, the eigenstates $|P_{n}\rangle_{d}$ are oriented in phase space in the direction of the displacement induced by $P_{d}$, where $X_{d}$ displaces rightward, $Z_{d}$ displaces upward, and $P_{d}^{d-1} = P_{d}^{-1}$.

We measure the GKP qutrit in Pauli basis $P_{3}$ using the circuit shown in Fig. \ref{fig2}b. The generalized ancilla-qubit-controlled Pauli operators $CP_{3} = |g\rangle\langle g|I_{3} + |e\rangle\langle e|P_{3}$ are realized via ECD gates. Intuitively, since generalized Pauli operators on GKP qudits are implemented via displacements, the conditional versions of these displacements implement $CP_{d}$ operations, with some technical caveats (see Methods). The idea of this circuit is to perform a projective measurement in the $P_{3}$ basis using two binary measurements of the ancilla qubit. First we measure whether the qutrit is in state $|P_{0}\rangle_{3}$ or the $\{|P_{1}\rangle_{3},|P_{2}\rangle_{3}\}$ subspace, then we measure whether the qutrit is in state $|P_{1}\rangle_{3}$ or the $\{|P_{0}\rangle_{3},|P_{2}\rangle_{3}\}$ subspace. These two binary measurements uniquely determine the ternary measurement result in the $P_{3}$ basis and collapse the qutrit state accordingly. Note that this circuit is constructed for the ideal code, and incurs infidelity when applied to the finite-energy code. To verify that this circuit realizes the desired projective measurement, we prepare the maximally mixed state of the GKP qutrit $\rho_{3}^{\mathrm{mix}}$, measure in the $Z_{3}$ basis, and perform Wigner tomography of the cavity postselected on the three measurement outcomes. The results of this measurement are shown in Fig. \ref{fig2}c (see \cite{noauthor_see_nodate} for measurements in the other Pauli bases).

With these techniques, we use a reinforcement learning agent \cite{sivak_model-free_2022, sivak_quantum_nodate} to optimize the logical GKP qutrit as a ternary quantum memory following the method in \cite{sivak_real-time_2023} (also, see Methods). We then evaluate the optimal QEC protocol by preparing each eigenstate $|P_{n}\rangle_{3}$ for each $P_{3}\in\mathcal{P}_{3}$, implementing the optimized QEC protocol for a variable number of rounds, and measuring the final state in the $P_{3}$ basis. Finally, we fit an exponential decay to each probability $\langle P_{n}| \mathcal{E}\left(|P_{n}\rangle\langle P_{n}|\right)|P_{n}\rangle$ to obtain $\gamma_{P_{n}}$. The results of this evaluation for the $|P_{0}\rangle_{3}$ states are shown in Fig. \ref{fig2}d (with the remaining results in \cite{noauthor_see_nodate}). As with the GKP qubit \cite{campagne-ibarcq_quantum_2020, sivak_real-time_2023}, we find longer lifetimes for the ``Cartesian'' eigenstates of $X_{3}$ and $Z_{3}$ than for the remaining ``diagonal'' eigenstates, since the latter are more susceptible to both cavity photon-loss errors and ancilla bit-flip errors \cite{campagne-ibarcq_quantum_2020}. Using Eq. \eqref{eq:qutrit_decay_rate} with our measured rates $\gamma_{P_{n}}$, we obtain $\Gamma_{3}^{\mathrm{GKP}} = \left(886\pm 3 \:\mathrm{\mu s}\right)^{-1}$. Comparing to $\Gamma_{3}^{\mathrm{Fock}}$, we obtain the QEC gain
\begin{equation}
G_{3} = \Gamma_{3}^{\mathrm{Fock}}/\Gamma_{3}^{\mathrm{GKP}} = 1.82\pm 0.03,
\end{equation}
well beyond the break-even point.

\begin{figure*}[t!]
\includegraphics[width=\textwidth]{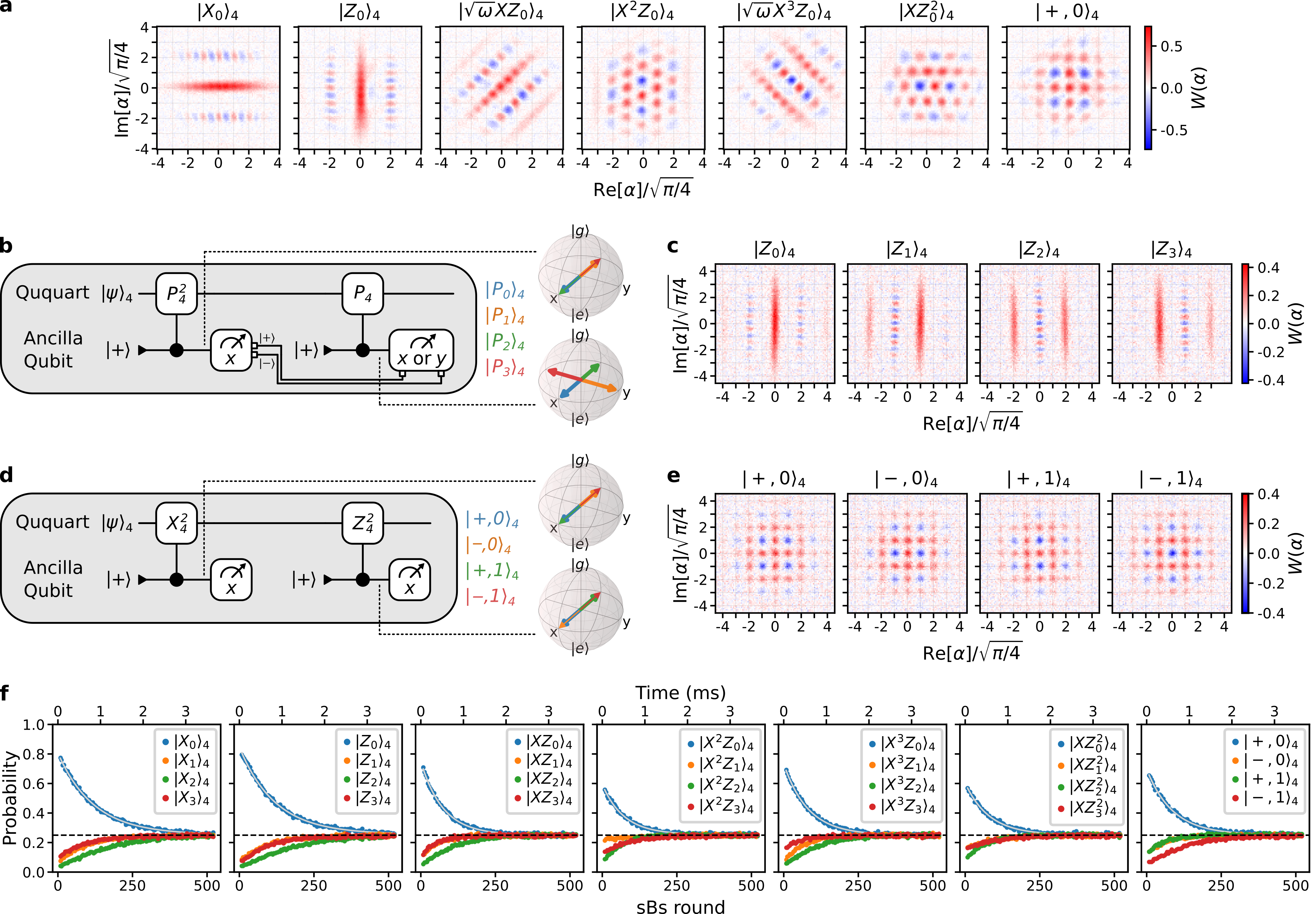}
\caption{
\textbf{Realization of a logical GKP ququart.} \textbf{a}, State preparation of ququart Pauli eigenstates $|P_{0}\rangle_{4}$ and parity state $|+,0\rangle_{4}$ with $\Delta = 0.32$. \textbf{b}, Circuit for measuring a ququart in the basis of Pauli operator $P_4$ using an ancilla qubit. The first measurement distinguishes between the even and odd states $|P_{\mathrm{even/odd}}\rangle_{4}$, and the second measurement distinguishes between the remaining two states. The Bloch spheres depict the trajectories taken by the ancilla when the ququart is in each Pauli eigenstate. \textbf{c}, Backaction of the GKP ququart Pauli measurement in the $Z_{4}$ basis, applied to the maximally mixed ququart state. \textbf{d}, Circuit for measuring a ququart in the parity basis $\{|\pm,m\rangle_{4} : m=0,1\}$, where $X_{4}^{2}|\pm,m\rangle_{4} = \pm |\pm,m\rangle_{4}$ and $Z_{4}^{2}|\pm,m\rangle_{4} = (-1)^{m} |\pm,m\rangle_{4}$. The first measurement determines the eigenvalue of $X_{4}^{2}$, and the second determines that of $Z_{4}^{2}$. \textbf{e}, Backaction of the GKP ququart parity measurement, applied to the maximally mixed ququart state. \textbf{f}, Decay of ququart Pauli eigenstates $|P_{0}\rangle_{4}$ and parity state $|+,0\rangle_{4}$ under the optimized QEC protocol. The dashed black lines indicate a probability of $1/4$. The solid grey lines are exponential fits. From left to right we find $\gamma_{X_{0}}^{-1} = 840\pm 8 \mathrm{\mu s}$, $\gamma_{Z_{0}}^{-1} = 836\pm 9 \mathrm{\mu s}$, $\gamma_{XZ_{0}}^{-1} = 519\pm 6 \mathrm{\mu s}$, $\gamma_{X^{2}Z_{0}}^{-1} = 507\pm 9 \mathrm{\mu s}$, $\gamma_{X^{3}Z_{0}}^{-1} = 571\pm 7 \mathrm{\mu s}$, $\gamma_{XZ^{2}_{0}}^{-1} = 562\pm 9 \mathrm{\mu s}$, and $\gamma_{+,0}^{-1} = 607\pm 8 \mathrm{\mu s}$.
}\label{fig3}
\end{figure*}

\section*{Logical Ququart Beyond Break-even}

We follow a similar procedure to measure the effective decay rate of our logical GKP ququart via Eq. \eqref{eq:ququart_decay_rate} as we did for the qutrit, the main difference being that we need to prepare and measure states in both the ququart parity basis and the Pauli bases $P\in\mathcal{P}_{4}$. We again use depth-8 ECD circuits \cite{eickbusch_fast_2022} to prepare the Pauli eigenstates $|P_{n}\rangle_{4}$ and parity states $|\pm,m\rangle_{4}$ with $\Delta = 0.32$. Measured Wigner functions of our prepared $|P_{0}\rangle_{4}$ states and $|+,0\rangle_{4}$ state are shown in Fig. \ref{fig3}a (see \cite{noauthor_see_nodate} for the remaining states). Again, the eigenstates $|P_{n}\rangle_{d}$ are oriented in phase space in the direction of the displacement induced by $P_{d}$. In contrast, the parity states $|\pm,m\rangle$ are uniform grids, equally oriented both horizontally and vertically.

We measure the GKP ququart in Pauli basis $P_{4}$ using the circuit shown in Fig. \ref{fig3}b. As with the qutrit, the generalized ancilla-qubit-controlled Pauli operators $CP_{4} = |g\rangle\langle g|I_{4} + |e\rangle\langle e|P_{4}$ are realized via ECD gates (see Methods). Note that the measurement circuit is simpler than the case of the qutrit because it is easier to sequentially split the $d=4$ Hilbert space of a ququart in half via binary measurements of an ancilla qubit. The first measurement distinguishes between the even subspace $\{|P_{0}\rangle,|P_{2}\rangle\}$ and odd subspace $\{|P_{1}\rangle,|P_{3}\rangle\}$ by measuring whether $P_{4}^{2}=\pm 1$, while the second measurement distinguishes between the remaining two states by measuring $P_{4}=\pm 1$ (if in the even subspace) or $P_{4}=\pm i$ (if in the odd subspace). To verify that this circuit realizes the desired projective measurement, we prepare the maximally mixed state of the GKP ququart $\rho_{4}^{\mathrm{mix}}$, measure in the $Z_{4}$ basis, and perform Wigner tomography of the cavity postselected on the four measurement outcomes. The results of this measurement are shown in Fig. \ref{fig3}c (see \cite{noauthor_see_nodate} for measurements in the other Pauli bases).

We measure the GKP ququart in the parity basis $\{|\pm,m\rangle_{4} : m=0,1\}$ using the circuit shown in Fig. \ref{fig3}d. The first circuit measures whether $X_{4}^{2}=\pm 1$, while the second circuit measures whether $Z_{4}^{2}=\pm 1$. To verify that this circuit realizes the desired projective measurement, we prepare the maximally mixed state of the GKP ququart $\rho_{4}^{\mathrm{mix}}$, measure in the parity basis, and perform Wigner tomography of the cavity postselected on the four measurement outcomes. The results of this measurement are shown in Fig. \ref{fig3}e. As with the qutrit, all of our logical measurements of the GKP ququart are constructed for the ideal code, and will incur infidelity when applied to the finite-energy code.

With these techniques, we again use a reinforcement learning agent \cite{sivak_model-free_2022, sivak_quantum_nodate} to optimize the logical GKP ququart as a quaternary quantum memory following the method in \cite{sivak_real-time_2023} (also, see Methods). We then evaluate the optimal QEC protocol by preparing each eigenstate $|P_{n}\rangle_{4}$ for each $P_{4}\in\mathcal{P}_{4}$ (plus the parity basis), implementing the optimized QEC protocol for a variable number of rounds, and measuring the final state in its corresponding basis. Finally, we fit an exponential decay to each probability $\langle P_{n}| \mathcal{E}\left(|P_{n}\rangle\langle P_{n}|\right)|P_{n}\rangle$ and $\langle \pm, m| \mathcal{E}\left(|\pm,m\rangle\langle \pm,m|\right)|\pm,m\rangle$ to obtain $\gamma_{P_{n}}$ and $\gamma_{\pm,m}$, respectively. The results of this evaluation for the $|P_{0}\rangle_{4}$ states and $|+,0\rangle_{4}$ state are shown in Fig. \ref{fig3}f (with the remaining results in \cite{noauthor_see_nodate}). Again, we find longer lifetimes for the ``Cartesian'' eigenstates of $X_{4}$ and $Z_{4}$ than for the remaining eigenstates. Using Eq. \eqref{eq:ququart_decay_rate} with our measured rates $\gamma_{P_{n}}$ and $\gamma_{\pm,m}$, we obtain $\Gamma_{4}^{\mathrm{GKP}} = \left(620\pm 2 \:\mathrm{\mu s}\right)^{-1}$. Comparing to $\Gamma_{4}^{\mathrm{Fock}}$, we obtain the QEC gain
\begin{equation}
G_{4} = \Gamma_{4}^{\mathrm{Fock}}/\Gamma_{4}^{\mathrm{GKP}} = 1.87\pm 0.03,
\end{equation}
again well beyond the break-even point.

\begin{figure}[t!]
\includegraphics[width=\columnwidth]{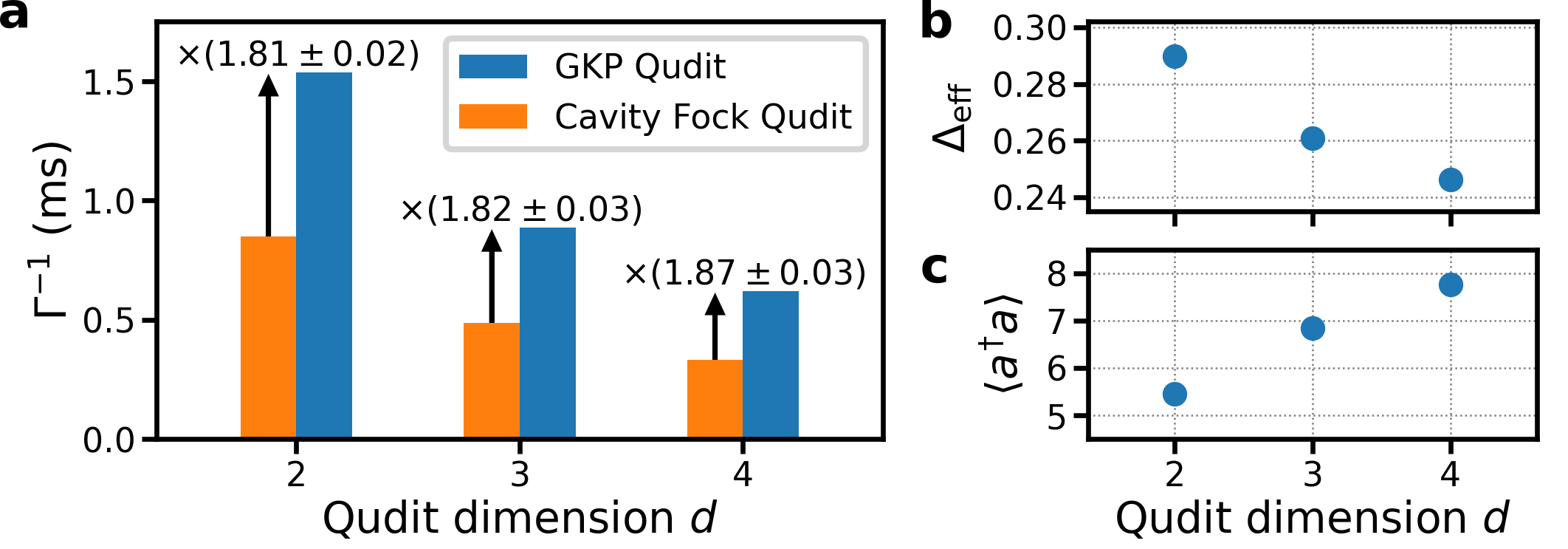}
\caption{
\textbf{a}, Effective lifetime of the physical cavity Fock qudit and logical GKP qudit for $d\in\{2,3,4\}$. The arrow indicates the QEC gain. \textbf{b} Effective envelope size $\Delta_{\mathrm{eff}}$ of the optimized GKP qudit for $d\in\{2,3,4\}$. \textbf{c}, Mean number of photons in the cavity for the optimized GKP qudit, for $d\in\{2,3,4\}$.
}\label{fig4}
\end{figure}

\section*{Discussion}

Remarkably, despite the increasing complexity of the code, we find that the QEC gain stays roughly constant at about $1.8$ as we increase the dimension of our logical GKP qudit from $2$ to $4$, as shown in Fig. \ref{fig4}a. Note that the gain $G_{2} = 1.81\pm 0.02$ we achieve with the GKP qubit is less than the gain of $2.3$ previously reported using the same device \cite{sivak_real-time_2023}, due to changes in both the device and experimental conditions \cite{noauthor_see_nodate}. Regardless, all of the measurements shown in Fig. \ref{fig4}a were taken under the same conditions, and they indicate that as we increase $d$ from $2$ to $4$, the lifetime of our logical GKP qudit decreases at about the same rate as that of our cavity Fock qudit. 

This decrease is due to the GKP qudit states being more closely spaced and containing information further out in phase space as we increase $d$, which should require smaller $\Delta$. To verify this, we prepare $\rho_{d}^{\mathrm{mix}}$ of our optimal GKP qudits and measure the central Gaussian peak of their characteristic functions for $d=2$ through $4$ \cite{noauthor_see_nodate}. The width $\Delta_{\mathrm{eff}}$ of this Gaussian is related to the parameter $\Delta$ and decreases with $d$, as shown in Fig. \ref{fig4}b, in agreement with our expectations. The average number of photons $\langle a^{\dagger}a\rangle$ in $\rho_{d}^{\mathrm{mix}}$ can also be inferred from this measurement of $\Delta_{\mathrm{eff}}$, and is presented in Fig. \ref{fig4}c. 

With a smaller $\Delta$, our logical GKP qudits have more energy and are more highly squeezed, which should amplify the rates of cavity photon loss and dephasing. To corroborate this, we simulate our optimal QEC protocols and isolate the relative contributions of different physical errors to our overall logical error rates \cite{noauthor_see_nodate}. We find that the three largest sources of logical errors are transmon bit-flips, whose relative contribution decreases as $d$ increases, cavity photon loss, whose relative contribution increases as $d$ increases, and cavity dephasing, which is the dominant source of error and whose relative contribution increases as $d$ increases. Since our cavity dephasing is primarily due to the thermal population $n_{\mathrm{th}} = 2.2\pm 0.1\%$ of the transmon \cite{reagor_quantum_2016}, the lifetimes of our logical GKP qudits could be substantially improved by either reducing $n_{\mathrm{th}}$ or using an ancilla that can be actively decoupled from the cavity when not in use \cite{rosenblum_fault-tolerant_2018, koottandavida_erasure_2024, ding_quantum_2024}.

In summary, we have demonstrated QEC of logical qudits with $d>2$, which represents a milestone achievement in the development of qudits for useful quantum technologies. Moreover, we have beaten the break-even point for QEC of quantum memories, a result few other experiments have accomplished \cite{ofek_extending_2016, ni_beating_2023, sivak_real-time_2023, acharya_quantum_2024}. This work builds on the promise of hardware efficiency offered by bosonic codes \cite{cai_bosonic_2021, gertler_protecting_2021, ofek_extending_2016, ni_beating_2023}, and represents a new way of leveraging the large Hilbert space of an oscillator. In exchange for a modest reduction in lifetime, we gain access to more logical quantum states in a single physical system. This could enable more efficient compilation of gates \cite{ralph_efficient_2007, fedorov_implementation_2012} and algorithms \cite{bocharov_factoring_2017, kiktenko_scalable_2020, gokhale_asymptotic_2019}, novel techniques for quantum communication \cite{schmidt_error-corrected_2024} and transduction \cite{wang_passive_2024}, and advantageous strategies for concatenation into an external multi-qudit code \cite{gottesman_stabilizer_1997, gottesman_fault-tolerant_1998, campbell_magic-state_2012, campbell_enhanced_2014, Anwar_2014}. With the realization of bosonic logical qudits we also establish a platform for concatenating codes \textit{internally}: by embedding a logical qubit within a bosonic logical qudit \cite{gottesman_encoding_2001,pirandola_minimal_2008,cafaro_quantum_2012,omanakuttan_fault-tolerant_2024, gross_designing_2021, gross_hardware-efficient_2024}, multiple layers of error correction could be implemented inside a single oscillator.

\section*{Methods}

\subsection*{Phase update between stabilization rounds}
The generalized sBs circuit in Fig. \ref{fig1}c realizes autonomous QEC of the finite-energy GKP code with respect to the ideal stabilizer $S_{X} = D(\sqrt{\pi d})$ \cite{royer_stabilization_2020}. The analogous circuit for $S_{Z} = D(i\sqrt{\pi d})$ is obtained by updating the phase of all subsequent cavity operations by $\pi/2$, which induces the transformation $q\rightarrow p$ and $p\rightarrow -q$ in the rotating frame of the cavity, where $q = (a+a^{\dagger})/\sqrt{2}$ is the position of the cavity and $p = i(a^{\dagger}-a)/\sqrt{2}$ is the momentum. To mitigate the effects of experimental imperfections, we symmetrize the protocol by also performing QEC with respect to stabilizers $S_{X}^{\dagger}$ and $S_{Z}^{\dagger}$, related to the circuit in Fig. \ref{fig1}c by phase updates of $\pi$ and $3\pi/2$, respectively. The full protocol is periodic with respect to four sBs rounds, where each stabilizer ($S_{X},S_{Z},S_{X}^{\dagger},S_{Z}^{\dagger}$) is measured once per period.

Ideally, we would measure the stabilizer $S_{X}$ by implementing the ancilla-controlled-stabilizer operation $CX_{d}^{d}$, but in practice we instead use the echoed conditional displacement $\ECD(\sqrt{\pi d}) = D(-\sqrt{\pi d}/2)CX_{d}^{d}\sigma_{x}$ \cite{terhal_encoding_2016, campagne-ibarcq_quantum_2020}. For even dimensions $d$ the additional displacement $D(-\sqrt{\pi d}/2)$ is the ideal Pauli operator $X_{d}^{d/2}$, the effect of which can be tracked in software (a similar result holds for the other stabilizers). In this case we are free to measure the stabilizers in any order. We choose to increment the cavity phase each round according to
\begin{equation}
\phi_{j}^{(d\:\mathrm{even})} = \pi/2,
\end{equation}
measuring the stabilizers in the order $S_{X},S_{Z},S_{X}^{\dagger},S_{Z}^{\dagger}$. However, for odd $d$ the displacement $D(-\sqrt{\pi d}/2)$ brings us outside the code space, an effect we have to reverse before moving on to measure $S_{Z}$. To do so we increment the cavity phase each round according to
\begin{equation}
\phi_{j}^{(d\:\mathrm{odd})} =
    \begin{cases}
        \pi, & \text{$j \equiv 0$ (mod $4$)} \\
        -\pi/2, & \text{$j \equiv 1$ (mod $4$)} \\
        \pi, & \text{$j \equiv 2$ (mod $4$)} \\
        \pi/2, & \text{$j \equiv 3$ (mod $d$)}
    \end{cases}
    ,
\end{equation}
measuring the stabilizers in the order $S_{X},S_{X}^{\dagger},S_{Z},S_{Z}^{\dagger}$. \\

\subsection*{Compiling generalized controlled-Pauli gates}

The ancilla-controlled version of generalized Pauli operator $P_{d} = e^{i\varphi}X_{d}^{n}Z_{d}^{m}$ on the ideal GKP code is given by $CP_{d} = |g\rangle\langle g| + |e\rangle\langle e|e^{i\varphi}D(\beta_{n})D(\beta_{m})$, where $\beta_{n} = n\sqrt{\pi/d}$ and $\beta_{m} = im\sqrt{\pi/d}$. We aim to compile $CP_{d}$ in terms of ancilla rotations and a single ECD gate
\begin{equation}
\ECD(\beta_{nm}) = D\left(-\beta_{nm}/2\right) \left(|g\rangle\langle e| + |e\rangle\langle g|D(\beta_{nm})\right),
\end{equation}
where $\beta_{nm} = \beta_{n}+\beta_{m}$. Using the fact that $D(\beta_{nm}) = \exp(inm\pi/d)D(\beta_{n})D(\beta_{m})$, this can be rewritten as
\begin{equation}
\ECD(\beta_{nm}) =  D\left(-\beta_{nm}/2\right)\sigma_{z}\left(\varphi_{nm}\right)CP_{d}\sigma_{x},
\end{equation}
where $\varphi_{nm} = nm\pi/d - \varphi$ and $\sigma_{z}(\theta) = |g\rangle\langle g| + |e\rangle\langle e|e^{i\theta}$. Rearranging terms, we obtain
\begin{equation}
CP_{d} = D\left(\beta_{nm}/2\right)\sigma_{z}(-\varphi_{nm})\ECD(\beta_{nm})\sigma_{x}.
\end{equation}

In our experiments we omit the unconditional displacement $D(\beta_{nm}/2)$ when compiling $CP_{d}$ gates, which affects the backaction of our GKP qudit logical measurements (see Figs. \ref{fig2} and \ref{fig3}, and \cite{noauthor_see_nodate}). In addition, we use the smallest amplitude $|\beta_{nm}|$ consistent with the Pauli operator $P_{d}$ (i.e., for $n=d-1$ and $d>2$ we use $\beta_{nm} = -\sqrt{\pi/d}+im\sqrt{\pi/d}$, since $X_{d}^{d-1} = X_{d}^{-1}$). We emphasize that this $CP_{d}$ gate is designed for the ideal GKP code, and will necessarily incur infidelity when applied to the finite-energy code, but it may be possible to adapt this construction to the finite-energy case \cite{royer_stabilization_2020, hastrup_measurement-free_2021, de_neeve_error_2022, rojkov_two-qubit_2024}.

\subsection*{Optimizing the QEC protocol}

To optimize our generalized sBs protocol (Fig. \ref{fig1}c), we follow the method described in Ref. \cite{sivak_real-time_2023}, parametrizing the sBs circuit using $45$ free parameters in total. Anticipating that the larger conditional displacements required for GKP qudit stabilization (nominally proportional to $\sqrt{\pi d}$) will take longer to execute, we fix the duration of each sBs round to be $7$ $\mathrm{\mu}$s (instead of $5$ $\mathrm{\mu}$s as in Ref. \cite{sivak_real-time_2023}).

We use a reinforcement learning agent to optimize our QEC protocol over these $45$ parameters in a model-free way \cite{sivak_model-free_2022, sivak_quantum_nodate}. Each training epoch, the agent sends a batch of $10$ parametrizations $\vec{p}_{i}$ to the experiment, collects a reward $R_{i}$ for each, and updates its policy in order to increase the reward. For our reward, we measure the probability that the QEC protocol keeps our logical qudit in its initial state, operationally quantified by
\begin{equation}
\begin{split}
R_{i} = \frac{1}{2}\Bigl[&\langle Z_{0}|\mathcal{E}_{\vec{p}_{i}}^{N}(|Z_{0}\rangle\langle Z_{0}|)|Z_{0}\rangle_{d} \\
&+ \langle X_{1}|\mathcal{E}_{\vec{p}_{i}}^{N}(|X_{1}\rangle\langle X_{1}|)|X_{1}\rangle_{d}\Bigr],
\end{split}
\end{equation}
where $\mathcal{E}_{\vec{p}_{i}}^{N}$ is the channel corresponding to $N$ rounds of the sBs protocol parametrized by $\vec{p}_{i}$. For our optimal GKP qubit we use $N=140$ and $200$ training epochs, for the qutrit $N=80$ and $200$ training epochs, and for the ququart $N=80$ and $300$ training epochs.

\bigskip
\noindent \textbf{Author contributions:}
BLB conceived the experiment, performed the measurements, and analyzed the results. BLB and SS developed the theory, with supervision from SMG. BLB devised the generalized Pauli measurement protocols, with input from AE and SS. VVS provided the experimental setup and wrote the reinforcement learning code. AE, VVS, AZD, and LF provided experimental support throughout the project. MHD supervised the project. BLB and MHD wrote the manuscript, with feedback from all authors.

\noindent \textbf{Ethics declarations:} BLB receives consulting fees from Nord Quantique. LF is a founder and shareholder of Quantum Circuits, Inc. SMG is an equity holder in, and receives consulting fees from, Quantum Circuits, Inc. The remaining authors declare no competing interests. 

\noindent \textbf{{Data availability:}} The data that support the findings of this study are available from the corresponding authors upon a reasonable request. 

\noindent \textbf{Code availability:} All computer code used in this study is available from the corresponding authors upon a reasonable request. 

\begin{acknowledgements}
We thank R.~G.~Corti\~{n}as, J.~Curtis, W.~Dai, S.~Hazra, A.~Koottandavida, A.~Miano, S.~Puri, K.~C.~Smith, and T.~Tsunoda for helpful discussions. This research was sponsored by the Army Research Office (ARO) under grant nos. W911NF-23-1-0051, and by the U.S. Department of Energy (DoE), Office of Science, National Quantum Information Science Research Centers, Co-design Center for Quantum Advantage (C2QA) under contract number DE-SC0012704. The views and conclusions contained in this document are those of the authors and should not be interpreted as representing the official policies, either expressed or implied, of the ARO, DoE or the US Government. The US Government is authorized to reproduce and distribute reprints for Government purposes notwithstanding any copyright notation herein. Fabrication facilities use was supported by the Yale Institute for Nanoscience and Quantum Engineering (YINQE) and the Yale SEAS Cleanroom.
\end{acknowledgements}

\let\oldaddcontentsline\addcontentsline
\renewcommand{\addcontentsline}[3]{}

\let\addcontentsline\oldaddcontentsline
\end{bibunit}

\let\oldaddcontentsline\addcontentsline
\renewcommand{\addcontentsline}[3]{}
\let\addcontentsline\oldaddcontentsline

\setcounter{equation}{0}
\setcounter{figure}{0}
\setcounter{table}{0}
\setcounter{section}{0}
\makeatletter

\crefname{section}{}{Sections}
\crefname{figure}{S}{Figures}
\crefname{equation}{}{Equations}
\crefname{table}{S}{Tables}
\renewcommand{\tablename}{Table S}
\renewcommand{\figurename}{FIG. S}
\renewcommand{\theequation}{S\arabic{equation}}

\newcommand{\subcell}[2]{\setlength\tabcolsep{0pt}\begin{tabular}{c} #1 \\[-6pt] #2 \\[3pt] \end{tabular}}

\newcommand{\fidsubcell}[2]{\setlength\tabcolsep{0pt}\begin{tabular}{l l} $\mathcal{F}_{\mathrm{exp}} = $ & #1 \\[-6pt] $\mathcal{F}_{\mathrm{thy}} = $ & #2 \\[3pt] \end{tabular}}

\title{%
  Supplementary Information \\
  \large ``Quantum Error Correction of Qudits Beyond Break-even''}

\clearpage
\maketitle

\onecolumngrid
\tableofcontents
\clearpage

\begin{bibunit}[apsrev4-2]

\section{Experimental Device and Methods}
\label{supp:sec:experimental_device_and_methods}

In this work we use the same experimental device as in \cite{sivak_real-time_2023}. It consists of a tantalum transmon fabricated on a sapphire chip \cite{place_new_2021, wang_towards_2022, ganjam_surpassing_2024}, which is inserted into a waveguide tunnel connected to a high-Q $\lambda/4$ 3D microwave cavity machined out of high purity aluminum \cite{reagor_quantum_2016, axline_architecture_2016}. We use the fundamental mode of the cavity as our oscillator (described by operator $a$, with Fock states $\{|0\rangle , |1\rangle, ...\}$), and the transmon's ground $|g\rangle$ and excited $|e\rangle$ states as our ancilla qubit (described by Pauli operators $\sigma_{x,y,z}$). The dipole coupling between the transmon and cavity gives rise to a dispersive coupling between these modes. Co-fabricated on the sapphire chip are a stripline readout resonator and Purcell filter, for dispersively measuring the state of the transmon qubit. We characterize the device and calibrate our control pulses using the methods described in detail in Ref. \cite{sivak_real-time_2023}. For the sake of brevity, in this section we limit ourselves to describing our experimental methods when they either differ from those used previously, or are important for contextualizing our results.

\begin{figure}[t!]
\includegraphics{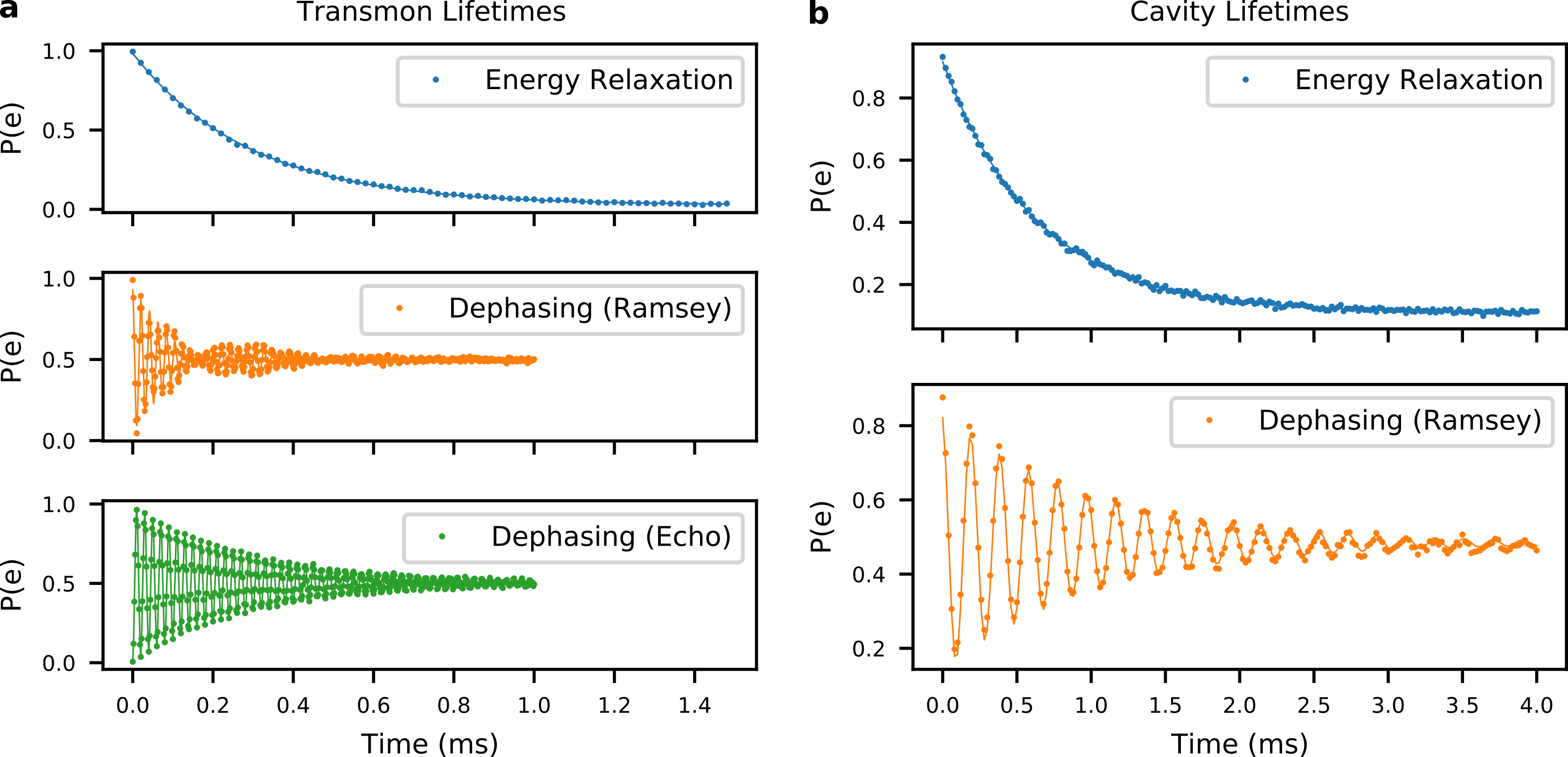}
\caption{
\textbf{a}, Transmon and \textbf{b}, Cavity lifetimes. The fit values are presented in Table \cref{supp:tab:experimental_parameters}.
}\label{supp:fig:system_lifetimes}
\end{figure}

We model the idling Hamiltonian of the transmon-cavity system, in the rotating frame of both the cavity (frequency $\omega_{c}$) and transmon qubit (frequency $\omega_{q}$), as
\begin{equation}\label{supp:eq:idling_hamiltonian}
H_{0}/\hbar = \frac{1}{2}\chi a^{\dagger}a \sigma_{z} + \frac{1}{2}K a^{\dagger 2}a^{2} + \frac{1}{4}\chi^{\prime}a^{\dagger 2}a^{2}\sigma_{z},
\end{equation}
where $\chi$ is the dispersive shift, $\chi^{\prime}$ is the second-order dispersive shift, and $K$ is the Kerr nonlinearity. In Table \cref{supp:tab:experimental_parameters} we present values for these parameters, measured in the same time-frame as the rest of our reported results. Our measurements of the transmon and cavity lifetimes are presented in Fig. \cref{supp:fig:system_lifetimes}. Due to the presence of a two-level system (TLS) coupled to our transmon we find beats in the Ramsey coherence measurement. Note that our measured value of the cavity pure-dephasing rate is
\begin{equation}
\kappa_{\phi,c} = \frac{1}{T_{2R,c}} - \frac{1}{2T_{1,c}} = \left(5.6\pm 0.6\: \mathrm{ms}\right)^{-1},
\end{equation}
which differs from from the rate of dephasing $\kappa_{\phi,c}^{(n_{\mathrm{th}})}$ induced by transmon thermal fluctuations 
\begin{equation}
\kappa_{\phi,c}^{(n_{\mathrm{th}})} \approx \frac{n_{\mathrm{nth}}}{T_{1,q}} = \left(13.4\pm 0.6\: \mathrm{ms}\right)^{-1},
\end{equation}
where the approximation we used for $\kappa_{\phi,c}^{(n_{\mathrm{th}})}$ is valid in the regime $\chi\gg T_{1,q}^{-1}$ and $n_{\mathrm{th}}\ll 1$ \cite{clerk_using_2007,wang_cavity_2019}. A similar discrepancy was found in Ref. \cite{sivak_real-time_2023}. Since we expect our 3D cavity to have negligible intrinsic dephasing \cite{reagor_quantum_2016, rosenblum_fault-tolerant_2018, koottandavida_erasure_2024, ding_quantum_2024}, we attribute this discrepancy to systematic error in our $T_{2R,c}$ measurement. In particular, we expect that errors during the echoed conditional displacement (ECD) circuit used in this measurement \cite{sivak_real-time_2023} cause leakage outside the $\{|0\rangle,|1\rangle\}$ subspace of the cavity, leading to an apparent reduction of $T_{2R,c}$.

The stability of our system is well-captured by the measurements reported in \cite{sivak_real-time_2023}. The primary source of fluctuations are spurious two-level-systems coming into and out of resonance with the transmon, causing fluctuations in the transmon lifetimes. These fluctuations typically occur on a week-by-week timescale. Our training and evaluation of the optimal GKP qubit, qutrit, and ququart were performed during periods of stability well-described by the reported experimental parameters.

\begin{center}
\begin{table}[t!]
\setlength\tabcolsep{6pt}
\begin{tabular}{ | c | l | l | } 
\hline 
\multirow{6}{*}{\textbf{Cavity mode}} & Frequency & $\omega_{c} = 2\pi\times 4.479$ GHz \\
& 1st order dispersive shift & $\chi = 2\pi\times 41.8$ kHz \\
& 2nd order dispersive shift & $\chi^{\prime} = 2\pi\times 4.3$ Hz\\
& Kerr nonlinearity & $K = -2\pi\times 3.7$ Hz \\
& Energy Relaxation & $T_{1,c} = 631\pm 3$ $\mathrm{\mu}$s \\
& Dephasing (Ramsey) & $T_{2R,c} = 1030\pm 20$ $\mathrm{\mu}$s \\
\hline
\multirow{6}{*}{\textbf{Ancilla transmon}} & Frequency & $\omega_{q} = 2\pi\times 6.009$ GHz \\
& Anharmonicity & $\alpha_{q} = -2\pi\times 221$ MHz \\
& Thermal Population & $n_{\mathrm{th}} = 2.2\pm 0.1 \%$ \\
& Energy Relaxation & $T_{1,q} = 295\pm 1$ $\mathrm{\mu}$s \\
& Dephasing (Ramsey) & $T_{2R,q} = 181\pm 5$ $\mathrm{\mu}$s \\
& Dephasing (Echo) & $T_{2E,q} = 286\pm 1$ $\mathrm{\mu}$s \\
\hline
\multirow{4}{*}{\textbf{Readout resonator}} & Frequency & $\omega_{r} = 2\pi\times 9.110$ GHz \\
& Dispersive shift & $\chi_{qr} = 2\pi\times 0.60$ MHz \\
& Coupling strength & $\kappa_{r,\mathrm{ext}} = 2\pi\times 0.47$ MHz \\
& Internal loss & $\kappa_{r,\mathrm{int}} = 2\pi\times 0.03$ MHz \\
\hline
\end{tabular}
\caption{Measured system parameters.}
\label{supp:tab:experimental_parameters}
\end{table}
\end{center}

\subsection{ECD Pulse Compilation}
\label{supp:sec:ecd_pulse_calibration}

We compile our echoed conditional displacement (ECD) gates \cite{eickbusch_fast_2022} in terms of displacements $D(\alpha) = \exp(\alpha a^{\dagger} - \alpha^{*}a)$, ancilla rotations, and evolution $U_{0}(\tau)=e^{-iH_{0}\tau/\hbar}$ under the idling Hamiltonian of Eq. \eqref{supp:eq:idling_hamiltonian} for time $\tau$. Our compilation takes the form
\begin{equation}
\ECD(\beta ; \alpha, \tau) = D(i\alpha e^{i\theta_{\beta}}\cos 2\phi)~U_{0}(\tau)~D(-i\alpha e^{i\theta_{\beta}}\cos\phi)~\sigma_{x}~D(-i\alpha e^{i\theta_{\beta}}\cos\phi)~U_{0}(\tau)~D(i\alpha e^{i\theta_{\beta}}), 
\end{equation}
where $\beta = |\beta|e^{i\theta_{\beta}}$ is the conditional displacement amplitude, $\alpha\in\mathbb{R}_{\geq 0}$ is the intermediate displacement amplitude, $\phi = \chi(\tau + \tau_{\mathrm{disp}})/2$, $\tau_{\mathrm{disp}} = 48$ ns is the duration of our cavity displacement pulses, and $\tau$ is the wait time required to achieve the desired $|\beta|$ given the intermediate amplitude $\alpha$. Following Ref. \cite{sivak_real-time_2023}, we calibrate the wait time $\tau(\alpha,\beta)$ empirically using the model
\begin{equation}
\tau(\alpha,\beta) = p_{1}\left|\frac{\beta}{\alpha}\right| - p_{2},
\end{equation}
and find typical values of $p_{1} \approx 1566$~ns and $p_{2} \approx 48$~ns. We also calibrate the geometric phase accumulated by the ancilla qubit during each ECD gate following established methods \cite{eickbusch_fast_2022, sivak_real-time_2023}. Compiling our ECD gates in terms of these primitive pulses, rather than long array pulses, gives us the flexibility to realize the complicated control sequences required for manipulating GKP qudits.

\subsection{Oscillator State Preparation}
\label{supp:sec:state_preparation}

We prepare arbitrary oscillator states using ECD gates $\ECD(\beta) = D(\beta/2)|e\rangle\langle g| + D(-\beta/2)|g\rangle\langle e|$ and ancilla qubit rotations $R_{\phi}(\theta) = \exp[i(\sigma_{x}\cos\phi + \sigma_{y}\sin\phi)\theta/2]$, which together enable universal control of the combined qubit-oscillator system \cite{eickbusch_fast_2022}. In particular, we can approximate the unitary operator $U_{\psi}$ that maps the state $|g\rangle|0\rangle$ to the target state $|g\rangle|\psi\rangle$ with the interleaved sequence
\begin{equation}
U_{\psi} \approx U_{\psi}(\vec{\beta},\vec{\phi},\vec{\theta}) =  \ECD(\beta_{N})R_{\phi_{N}}(\theta_{N})\ECD(\beta_{N-1})R_{\phi_{N-1}}(\theta_{N-1})\cdots\ECD(\beta_{1})R_{\phi_{1}}(\theta_{1}),
\end{equation}
where we refer to $N$ as the depth of the ECD circuit. We optimize this unitary offline \cite{eickbusch_ecd_nodate}, with a fixed circuit depth, finding the circuit parameters $\vec{\beta}$, $\vec{\phi}$, and $\vec{\theta}$ that maximize the fidelity $\mathcal{F} = |\langle g|\langle\psi| U_{\psi}(\vec{\beta},\vec{\phi},\vec{\theta})|g\rangle|0\rangle|^{2}$ in the absence of loss.

\subsection{Wigner and Characteristic Function Tomography}
\label{supp:sec:tomography_and_state_reconstruction}

We measure the state of our oscillator mode by performing either Wigner tomography or characteristic function tomography. The Wigner function can be defined as the expectation value of the displaced photon-number parity operator $\Pi = e^{i\pi a^{\dagger}a}$ according to
\begin{equation}
W(\alpha) = \langle D(\alpha)\Pi D(-\alpha) \rangle.
\end{equation}
We map the photon-number parity to the ancilla qubit by initializing the ancilla in $|+\rangle = (|g\rangle+|e\rangle)/\sqrt{2}$ and evolving under the dispersive Hamiltonian $H_{\mathrm{disp}} = \chi a^{\dagger}a\sigma_{z}/2$ for time $t = \pi/\chi$, such that
\begin{equation}
|+\rangle\sum\limits_{n=0}^{\infty}c_{n}|n\rangle \rightarrow e^{-i\pi a^{\dagger} a\sigma_{z}/2}|+\rangle\sum\limits_{n=0}^{\infty}c_{n}|n\rangle = |+\rangle\sum\limits_{n\: \mathrm{even}}^{\infty}i^{n}c_{n}|n\rangle + |-\rangle\sum\limits_{n\: \mathrm{odd}}^{\infty}i^{n}c_{n}|n\rangle,
\end{equation}
where $|-\rangle = (|g\rangle-|e\rangle)/\sqrt{2}$ \cite{vlastakis_deterministically_2013}. Afterwards, we rotate the ancilla around its $y$ axis by applying $R_{\pi/2}(-s\pi/2)$ and measuring $\sigma_{z}$ of the ancilla, such that the measurement result $\sigma_{z} = \pm 1$ corresponds to the parity result $\Pi = \pm s$, where $s=\pm 1$. In practice we symmetrize this measurement by using $s=1$ half the time and $s=-1$ half the time. To measure $W(\alpha)$, we simply prepend this parity measurement with a cavity displacement $D(-\alpha)$. To reconstruct density matrices from Wigner tomography measurements we follow the method described in Ref. \cite{sivak_real-time_2023}.

The characteristic function $\mathcal{C}(\beta)$ is defined as the expectation value of the displacement operator $\langle D(\beta) \rangle$ and it is also the Fourier transform of the Wigner function, which is why we often call $\beta$-space reciprocal phase space. Following previous work \cite{fluhmann_direct_2020, campagne-ibarcq_quantum_2020, eickbusch_fast_2022}, we map the characteristic function of the cavity state $|\psi\rangle$ to our ancilla qubit by performing an ECD gate starting with our ancilla in the $|+\rangle$ state. This entangles the cavity with the ancilla according to
\begin{equation}
\ECD(\beta)|+\rangle|\psi\rangle = \frac{1}{\sqrt{2}}\Bigl[|g\rangle D(-\beta/2)|\psi\rangle + |e\rangle D(\beta/2)|\psi\rangle\Bigr].
\end{equation}
Tracing out the cavity, the reduced density matrix on our qubit can be written
\begin{equation}
\rho_{q} = \frac{1}{2}
\begin{pmatrix}
1 & \langle D(-\beta)\rangle \\
\langle D(\beta)\rangle & 1
\end{pmatrix},
\end{equation}
such that the characteristic function is encoded in expectation values of the qubit with respect to the equator of its Bloch sphere according to
\begin{equation}
\langle D(\beta)\rangle = \langle\sigma_{x}\rangle + i\langle\sigma_{y}\rangle.
\end{equation}
The derivatives of the characteristic function at the origin of reciprocal phase space enable us to measure photon statistics of the field in the cavity according to \cite{fluhmann_direct_2020}
\begin{equation}
\langle a^{\dagger m}a^{n}\rangle = -\frac{\partial^{m+n}}{\partial \beta^{m}\partial(\beta^{*})^{n}}\mathcal{C}(\beta)\Bigr|_{\beta=0}.
\end{equation}
In particular, for $\beta = x+iy$ the average number of photons in the cavity can be written
\begin{equation}\label{supp:eq:cf_avg_photons}
\langle a^{\dagger} a\rangle = - \frac{1}{2} - \frac{1}{4}\left[\frac{\partial^{2}\mathcal{C}}{\partial x^{2}} + \frac{\partial^{2}\mathcal{C}}{\partial y^{2}}\right]_{\substack{x=0 \\ y=0}}.
\end{equation}

\section{Gottesman-Kitaev-Preskill Qudits}
\label{supp:sec:gkp_qudits}

In its most general form, the Gottesman-Kitaev-Preskill (GKP) code can be used to encode a logical qudit within $N$ oscillator modes \cite{gottesman_encoding_2001}. As discussed in the main text, we restrict ourselves to the single-mode square code with ideal stabilizer generators
\begin{alignat}{2}
S_{X} &= D(\ell_{d}) &&= e^{-i\ell_{d}\sqrt{2}p} ,\\
S_{Z} &= D(i\ell_{d}) &&= e^{i\ell_{d}\sqrt{2}q} ,
\end{alignat}
ideal codewords
\begin{equation}
|Z_{n}\rangle_{d} \propto \sum_{k=-\infty}^{\infty}|n\sqrt{2\pi/d} + k\sqrt{2\pi d}\rangle_{q}
\end{equation}
with $n=0,1,...,d-1$, and ideal logical Pauli operators
\begin{alignat}{2}
X_{d} &= D(\ell_{d}/d) &&= e^{-i\ell_{d}\sqrt{2}p/d} ,\\
Z_{d} &= D(i\ell_{d}/d) &&= e^{i\ell_{d}\sqrt{2}q/d} ,
\end{alignat}
where $\ell_{d} = \sqrt{\pi d}$ and $d\in\mathbb{Z}_{>0}$ is the dimension of the logical qudit. Note that with our choice of phase-space units, translations in position and displacements along the real axis of phase space differ in amplitude by a factor of $\sqrt{2}$. The Pauli operators act on the $Z$-codewords according to
\begin{equation}
    \begin{split}
    Z_{d}|Z_{n}\rangle_{d} &= \omega_{d}^{n}|Z_{n}\rangle_{d}, \\
    X_{d}|Z_{n}\rangle_{d} &= |Z_{(n+1)\:\mathrm{mod}\:d}\rangle_{d}, \\
    Z_{d}X_{d} &= \omega_{d} X_{d}Z_{d},
    \end{split}
\end{equation}
where $\omega_{d} = e^{2\pi i/d}$ is the primitive $d$th root of unity. Here the oscillator is described by mode operator $a$, position operator $q = (a+a^{\dagger})/\sqrt{2}$, momentum operator $p = i(a^{\dagger} - a)/\sqrt{2}$, and displacement operator $D(\alpha) = \exp(\alpha a^{\dagger} - \alpha^{*}a)$. 

The eigenstates of the generalized Pauli operators play a key role in our experiment. For Pauli operator $P_{d}$ with non-degenerate spectrum $\{1,\omega_{d}, \omega_{d}^{2},...,\omega_{d}^{d-1}\}$, we define the $n$th eigenstate $|P_{n}\rangle_{d}$ according to
\begin{equation}
P_{d}|P_{n}\rangle_{d} = \omega_{d}^{n}|P_{n}\rangle_{d}.
\end{equation}
To ensure $P_{d}$ has the desired spectrum we either define it as $X_{d}^{n}Z_{d}^{m}$ or $\sqrt{\omega_{d}}X_{d}^{n}Z_{d}^{m}$, for integers $n$ and $m$. Note that we sometimes omit subscripts $d$ and/or the factor of $\sqrt{\omega_{d}}$ for convenience, when the meaning remains contextually clear. We present the Pauli eigenstates of the qubit, qutrit, and ququart used throughout this work in terms of the Pauli $Z_{d}$ basis states in Table \cref{supp:tab:qudit_states}. We find these representations by numerically diagonalizing the relevant operators. In this table we also present the ququart parity basis $\{|\pm,n\rangle_{4} ; s=\pm , n=0,1\}$, which are the simultaneous eigenstates of $X_{4}^{2}$ and $Z_{4}^{2}$ according to $X_{4}^{2}|\pm,n\rangle_{4} = \pm |\pm,n\rangle_{4}$ and $Z_{4}^{2}|\pm,n\rangle_{4} = (-1)^{n}|\pm,n\rangle_{4}$, in terms of the $Z_{d}$ basis.

\begin{table}[p]
\setlength\tabcolsep{6pt}
\renewcommand{\arraystretch}{2}
\begin{tabular}{ | c | c | c | } 
 \cline{2-3} 
  \multicolumn{1}{c|}{\subcell{Qubit Pauli}{States}} & $n=0$ & $n=1$ \\[3pt]
\hline
 $|X_{n}\rangle_{2}$ & $|Z_{0}\rangle_{2}+|Z_{1}\rangle_{2}$ & $|Z_{0}\rangle_{2}-|Z_{1}\rangle_{2}$ \\ [3pt]
\hline
 $|\sqrt{\omega}XZ_{n}\rangle_{2}$ & $|Z_{0}\rangle_{2}+\sqrt{\omega}|Z_{1}\rangle_{2}$ & $|Z_{0}\rangle_{2}-\sqrt{\omega}|Z_{1}\rangle_{2}$ \\ [3pt]
 \hline
 \multicolumn{3}{c}{}\\[3pt]
\end{tabular}
\begin{tabular}{ | c | c | c | c |} 
 \cline{2-4} 
  \multicolumn{1}{c|}{\subcell{Qutrit Pauli}{States}} & $n=0$ & $n=1$ & $n=2$ \\[3pt]
\hline
 $|X_{n}\rangle_{3}$ & $|Z_{0}\rangle_{3}+|Z_{1}\rangle_{3}+|Z_{2}\rangle_{3}$ & $|Z_{0}\rangle_{3}+\omega^{2}|Z_{1}\rangle_{3}+\omega|Z_{2}\rangle_{3}$ & $|Z_{0}\rangle_{3}+\omega|Z_{1}\rangle_{3}+\omega^{2}|Z_{2}\rangle_{3}$ \\ [3pt]
\hline
 $|XZ_{n}\rangle_{3}$ & $|Z_{0}\rangle_{3}+|Z_{1}\rangle_{3}+\omega|Z_{2}\rangle_{3}$ & $|Z_{0}\rangle_{3}+\omega^{2}|Z_{1}\rangle_{3}+\omega^{2}|Z_{2}\rangle_{3}$ & $|Z_{0}\rangle_{3}+\omega|Z_{1}\rangle_{3}+|Z_{2}\rangle_{3}$ \\ [3pt]
 \hline
 $|X^{2}Z_{n}\rangle_{3}$ & $|Z_{0}\rangle_{3}+\omega^{2}|Z_{1}\rangle_{3}+|Z_{2}\rangle_{3}$ & $|Z_{0}\rangle_{3}+|Z_{1}\rangle_{3}+\omega^{2}|Z_{2}\rangle_{3}$ & $|Z_{0}\rangle_{3}+\omega|Z_{1}\rangle_{3}+\omega|Z_{2}\rangle_{3}$ \\ [3pt]
 \hline
 \multicolumn{4}{c}{}\\[3pt]
\end{tabular}
\begin{tabular}{ c | c | c | c | c |} 
 \cline{2-5}
  \multicolumn{1}{c|}{\subcell{Ququart Pauli}{States}} & $n=0$ & $n=1$ & $n=2$ & $n=3$ \\[3pt]
\hline
 \multicolumn{1}{|c|}{$|X_{n}\rangle_{4}$} & 
 \subcell{$|Z_{0}\rangle_{4}+|Z_{1}\rangle_{4}$}{$+|Z_{2}\rangle_{4}+|Z_{3}\rangle_{4}$} &
 \subcell{$|Z_{0}\rangle_{4}-\omega|Z_{1}\rangle_{4}$}{$-|Z_{2}\rangle_{4}+\omega|Z_{3}\rangle_{4}$} &
 \subcell{$|Z_{0}\rangle_{4}-|Z_{1}\rangle_{4}$}{$+|Z_{2}\rangle_{4}-|Z_{3}\rangle_{4}$} & 
 \subcell{$|Z_{0}\rangle_{4}+\omega|Z_{1}\rangle_{4}$}{$-|Z_{2}\rangle_{4}-\omega|Z_{3}\rangle_{4}$} \\ [3pt]
\hline
 \multicolumn{1}{|c|}{$|\sqrt{\omega}XZ_{n}\rangle_{4}$} & 
  \subcell{$|Z_{0}\rangle_{4}+\sqrt{\omega}|Z_{1}\rangle_{4}$}{$-|Z_{2}\rangle_{4}+\sqrt{\omega}|Z_{3}\rangle_{4}$} &
 \subcell{$\sqrt{\omega}|Z_{0}\rangle_{4}+|Z_{1}\rangle_{4}$}{$+\sqrt{\omega}|Z_{2}\rangle_{4}-|Z_{3}\rangle_{4}$} &
  \subcell{$|Z_{0}\rangle_{4}-\sqrt{\omega}|Z_{1}\rangle_{4}$}{$-|Z_{2}\rangle_{4}-\sqrt{\omega}|Z_{3}\rangle_{4}$} &
 \subcell{$\sqrt{\omega}|Z_{0}\rangle_{4}-|Z_{1}\rangle_{4}$}{$+\sqrt{\omega}|Z_{2}\rangle_{4}+|Z_{3}\rangle_{4}$} \\ [3pt]
 \hline
\multicolumn{1}{|c|}{$|X^{2}Z_{n}\rangle_{4}$} & 
$|Z_{1}\rangle_{4}+\omega|Z_{3}\rangle_{4}$ &
$|Z_{0}\rangle_{4}-\omega|Z_{2}\rangle_{4}$ &
$|Z_{1}\rangle_{4}-\omega|Z_{3}\rangle_{4}$ &
$|Z_{0}\rangle_{4}+\omega|Z_{2}\rangle_{4}$\\ [3pt]
 \hline
 \multicolumn{1}{|c|}{$|\sqrt{\omega}X^{3}Z_{n}\rangle_{4}$} & 
  \subcell{$|Z_{0}\rangle_{4}-\sqrt{\omega}|Z_{1}\rangle_{4}$}{$+|Z_{2}\rangle_{4}+\sqrt{\omega}|Z_{3}\rangle_{4}$} &
 \subcell{$\sqrt{\omega}|Z_{0}\rangle_{4}+|Z_{1}\rangle_{4}$}{$-\sqrt{\omega}|Z_{2}\rangle_{4}+|Z_{3}\rangle_{4}$} &
  \subcell{$|Z_{0}\rangle_{4}+\sqrt{\omega}|Z_{1}\rangle_{4}$}{$+|Z_{2}\rangle_{4}-\sqrt{\omega}|Z_{3}\rangle_{4}$} &
 \subcell{$\sqrt{\omega}|Z_{0}\rangle_{4}-|Z_{1}\rangle_{4}$}{$-\sqrt{\omega}|Z_{2}\rangle_{4}-|Z_{3}\rangle_{4}$} \\ [3pt]
 \hline
 \multicolumn{1}{|c|}{$|XZ^{2}_{n}\rangle_{4}$} & 
 \subcell{$|Z_{0}\rangle_{4}+|Z_{1}\rangle_{4}$}{$-|Z_{2}\rangle_{4}-|Z_{3}\rangle_{4}$} &
 \subcell{$|Z_{0}\rangle_{4}-\omega|Z_{1}\rangle_{4}$}{$+|Z_{2}\rangle_{4}-\omega|Z_{3}\rangle_{4}$} &
 \subcell{$|Z_{0}\rangle_{4}-|Z_{1}\rangle_{4}$}{$-|Z_{2}\rangle_{4}+|Z_{3}\rangle_{4}$} & 
 \subcell{$|Z_{0}\rangle_{4}+\omega|Z_{1}\rangle_{4}$}{$+|Z_{2}\rangle_{4}+\omega|Z_{3}\rangle_{4}$} \\ [3pt]
 \hline
\multicolumn{5}{c}{} \\[3pt]
\cline{2-5}
\multirow{2}{*}{\subcell{Ququart Parity}{States}} & $|+,0\rangle_{4}$ & $|-,0\rangle_{4}$ & $|+,1\rangle_{4}$ & $|-,1\rangle_{4}$ \\[3pt]
\cline{2-5}
 & $|Z_{0}\rangle_{4}+|Z_{2}\rangle_{4}$ &
 $|Z_{0}\rangle_{4}-|Z_{2}\rangle_{4}$ &
 $|Z_{1}\rangle_{4}+|Z_{3}\rangle_{4}$ &
 $|Z_{1}\rangle_{4}-|Z_{3}\rangle_{4}$ \\[3pt]
\cline{2-5}
\end{tabular}
\caption{Pauli eigenstates of the qubit, qutrit, and ququart expressed in the $Z_{d}$ basis, and the ququart parity states expressed in the $Z_{4}$ basis. For each qudit, $\omega = e^{2\pi i/d}$ is the primitive $d$th root of unity and $\sqrt{\omega} = e^{i\pi/d}$. The overall normalization constants are omitted for simplicity.}
\label{supp:tab:qudit_states}
\end{table}

In practice, we work with an approximate finite-energy version of the GKP code, obtained by applying the Gaussian envelope operator $E_{\Delta} = \exp(-\Delta^{2}a^{\dagger}a)$ to both the operators $O$ and states $|\psi\rangle$ of the ideal code according to\cite{gottesman_encoding_2001, grimsmo_quantum_2021}
\begin{equation}
\begin{split}
|\psi\rangle_{\Delta} &= E_{\Delta}|\psi\rangle, \\
O_{\Delta} &= E_{\Delta}OE_{\Delta}^{-1},
\end{split}
\end{equation}
up to normalization. The parameter $\Delta$ determines both the squeezing of individual quadrature peaks in the grid states as well as their overall extent in energy. 

To find the finite-energy Pauli eigenstates numerically, we extend the approach of Ref. \cite{eickbusch_fast_2022} to the case of GKP qudits. We find the finite-energy $|Z_{0}\rangle_{d,\Delta}$ state by computing the ground state of the fictitious Hamiltonian
\begin{equation}
H = -\frac{1}{2}E_{\Delta}\left(S_{X} + S_{X}^{\dagger}\right)E_{\Delta}^{-1} -\frac{1}{2}E_{\Delta}\left(S_{Z} + S_{Z}^{\dagger}\right)E_{\Delta}^{-1} - dE_{\Delta}\left(Z_{d} + Z_{d}^{\dagger}\right)E_{\Delta}^{-1},
\end{equation}
and then obtain the remaining $|Z_{n}\rangle_{d,\Delta}$ states according to
\begin{equation}
|Z_{n}\rangle_{d,\Delta} = E_{\Delta}X_{d}^{n}E_{\Delta}^{-1}|Z_{0}\rangle_{d,\Delta}.
\end{equation}
Finally, we obtain the remaining states using the expressions in Table \cref{supp:tab:qudit_states}.

\subsection{Stabilizing Finite-energy GKP Qudits}
\label{supp:sec:stabilizing_finite_energy_gkp_qudits}

We follow Ref.~\cite{royer_stabilization_2020} to generalize the sBs protocol to stabilize finite-energy GKP qudits. The stabilizers ($S$) of the ideal square GKP qudit are given by $S_Z=e^{i\sqrt{2\pi d} q}, S_X=e^{-i\sqrt{2\pi d}p}$. The finite-energy analog of these stabilizers can be expressed as
\begin{align}
S_{Z,\Delta}&=e^{i\sqrt{2\pi d}(q\cosh{\Delta^2}+ip\sinh{\Delta^2})},\\
S_{X,\Delta}&=e^{-i\sqrt{2\pi d}(p\cosh{\Delta^2}-iq\sinh{\Delta^2})}.\end{align}
To construct the engineered dissipation onto the codespace of these finite-energy stabilizers, we introduce an effective mode operator $a_{X,Z}\propto\log(S_{X,Z})$ which annihilates the GKP qudit codewords. These mode operators can be written
\begin{align}
    a_{Z} &= (q_{[\sqrt{2\pi/d}]}+ip\tanh{\Delta^{2}})/\sqrt{2\tanh{\Delta^{2}}},\\
    a_{X} &= (p_{[\sqrt{2\pi/d}]}-iq\tanh{\Delta^{2}})/\sqrt{2\tanh{\Delta^{2}}}.
\end{align}
Here $v_{[m]}$ denotes the symmetric version of the modular quadrature $v$ (mod $m$), also known as the Zak-basis~\cite{zak_finite_1967}. 

To realize our engineered dissipation, we siphon energy from the operator $a_{X,Z}$ into an ancilla qubit by evolving under 
\begin{equation}
H \propto a_{X,Z}(\sigma_{x}-i\sigma_{y}) + a_{X,Z}^{\dagger}(\sigma_{x}+i\sigma_{y})
\end{equation}
for a short time, resetting the ancilla, and repeating. Resetting the ancilla extracts entropy from the oscillator in a way that stabilizes the ground state manifold of $a_{X,Z}$, which is designed to be the finite-energy GKP qudit codespace. Starting with the ancilla in $|g\rangle$, the desired entangling unitary associated with $a_{Z}$ can be expressed
\begin{equation}
U_{Z} = \exp[-i\sqrt{\frac{\gamma\delta t}{\tanh\Delta^2}}\left(q_{[\sqrt{2\pi/d}]}\sigma_x +p\sigma_y\tanh\Delta^2\right)],
\end{equation}
where $\gamma$ is the rate of dissipation onto the ground state of $a_{Z}$. We can now use second-order trotterization  of $U_{Z}$ to derive the circuit that stabilizes the logical $Z$ subspace,
\begin{equation}     
U_{Z}^{\mathrm{sBs}} = 
\exp(-i\sqrt{\frac{\gamma\delta t \tanh(\Delta^2)}{4}}p\sigma_y) 
\exp(-i\sqrt{\frac{\gamma\delta t}{\tanh(\Delta^2)}}q_{[\sqrt{2\pi/d}]}\sigma_x)
\exp(-i\sqrt{\frac{\gamma\delta t \tanh(\Delta^2)}{4}}p\sigma_y)+\mathcal{O}[(\gamma\delta t)^{3/2}].
\end{equation}
Since the modular quadratures are translation invariant we impose the constraint $q_{[\sqrt{2\pi/d}]}\rightarrow q$ on the trotterized unitary, and choose
\begin{equation}
\gamma\delta t=\frac{\pi d}{2}\tanh{\Delta^2}\left(\cosh\Delta^{2}\right)^{2}.
\end{equation}
Plugging back in, we obtain
\begin{equation}     
U_{Z}^{\mathrm{sBs}} = 
\exp(-i\varepsilon_{d} p\sigma_{y}/2\sqrt{2})
\exp(-i\ell_{d,\Delta}q\sigma_x/\sqrt{2})
\exp(-i\varepsilon_{d}p\sigma_y/2\sqrt{2}),
\end{equation}
where $\varepsilon_{d} = \sqrt{\pi d}\sinh{\Delta^2}$ and $\ell_{d,\Delta} = \sqrt{\pi d}\cosh{\Delta^{2}}$. We next introduce Hadamard gates $H$ and ancilla rotations to express the circuit in terms of conditional displacement gates conditioned on the $\sigma_z$ axis of the ancilla Bloch sphere, as used in the main text. Doing so, we obtain
\begin{equation}
U_{Z}^{\mathrm{sBs}}=
\exp(-i\varepsilon_{d} q\sigma_{z}/2\sqrt{2})
R_0^\dagger(\pi/2)
\exp(-i\ell_{d,\Delta}p\sigma_z/\sqrt{2})
R_0(\pi/2)
\exp(-i\varepsilon_{d}q\sigma_z/2\sqrt{2}),
\end{equation}
where we now assume that we begin with the ancilla in $|+\rangle$ rather than $|g\rangle$ (and we ignore ancilla gates at the end since we follow the last entangling operation with a reset). This circuit can be expressed in terms of ECD gates as
\begin{equation}
U_{Z}^{\mathrm{sBs}} = \ECD(\varepsilon_{d}/2)~R_{0}(\pi/2)~\ECD(-i\ell_{d,\Delta})~R_{0}^{\dagger}(\pi/2)~\ECD(\varepsilon_{d}/2).
\end{equation}
If we update the cavity reference phase by a factor of $i$, we obtain the circuit in Fig. 1c of the main text.

\subsection{Maximally-mixed GKP Qudit States}
\label{supp:sec:maximally_mixed_gkp_qudit_states}

\begin{figure}[t!]
\includegraphics[width=0.75\textwidth]{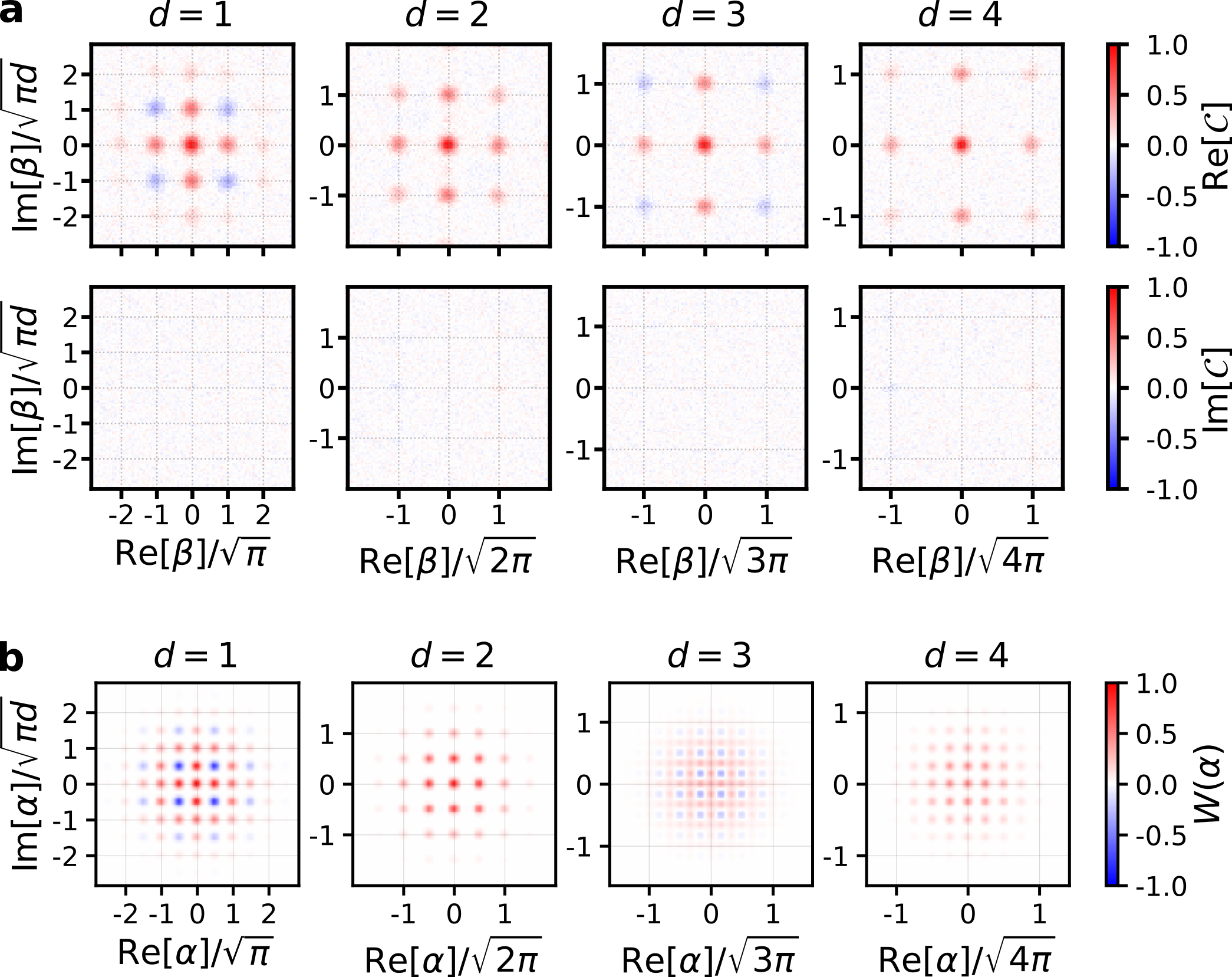}
\caption{
\textbf{Maximally-mixed GKP qudit states.} \textbf{a}, Measured characteristic functions of the maximally-mixed states of GKP qudits. This is the same dataset presented in Fig. 1d of the main text, except here we include our measurement of $\mathrm{Im}[\mathcal{C}]$ for completeness. \textbf{b}, Simulated Wigner functions of the maximally-mixed states of GKP qudits.
}\label{supp:fig:gkpd_mixed_states}
\end{figure}

Since the codespace of the GKP qudit is the steady state of the engineered dissipation realized by the sBs protocol, we can prepare the maximally mixed state $\rho_{d}^{\mathrm{mix}}$ by turning on sBs and waiting, as discussed in the main text. In Fig. \cref{supp:fig:gkpd_mixed_states}a we reproduce the dataset presented in Fig. 1d of the main text, but here we include our measurement of $\mathrm{Im}[\mathcal{C}]$ as well, which is zero as we expect. In Fig. \cref{supp:fig:gkpd_mixed_states}b we plot the simulated Wigner functions of these mixed states. 

Interestingly, the maximally mixed GKP states in odd dimension $d$ have regions of Wigner negativity, indicating non-classicality. This can be viewed as a consequence of the interference fringes of the $|Z_{n}\rangle_{d}$ basis states in phase space, which occur at half lattice spacings (see Figs. \cref{supp:fig:gkp2_ecd_prep},\cref{supp:fig:gkp3_ecd_prep}, and \cref{supp:fig:gkp4_ecd_prep}). In even dimensions this half lattice spacing is commensurate with the lattice, and when the states mix incoherently the negative peaks get canceled out by the positive peaks from other states. In odd dimensions, however, this half lattice spacing is not commensurate with the lattice, and these negative peaks remain even after incoherent mixing of the states. For our measurement of these mixed states, we opted to perform characteristic function tomography rather than Wigner tomography because of the inherently low contrast of the Wigner function of $\rho_{d}^{\mathrm{mix}}$ for increasing $d$; achieving the same signal-to-noise ratio with Wigner tomography would have taken a prohibitively long time.

\section{Characterizing Quantum Memories}
\label{supp:sec:characterizing_quantum_memories}

As discussed in the main text, we characterize the performance of our logical GKP qudits as non-binary quantum memories via the average channel fidelity
\begin{equation}\label{supp:eq:avg_channel_fidelity}
\mathcal{F}_{d}(\mathcal{E},I) = \frac{1}{d+1} + \frac{1}{d^{2}(d+1)}\sum\limits_{n,m=0}^{d-1}\mathrm{Tr}\left[U_{nm}^{\dagger}\mathcal{E}(U_{nm})\right],
\end{equation}
which compares the channel $\mathcal{E}$ to the identity $I$ \cite{nielsen_simple_2002}. Here the $U_{nm}$'s are elements of a unitary operator basis satisfying the orthogonality relation $\mathrm{Tr}[U_{nm}^{\dagger}U_{kl}]=\delta_{nk}\delta_{ml}d$, which in our case we take to be the generalized Pauli operators $U_{nm} = X_{d}^{n}Z_{d}^{m}$ (up to a phase factor, which will be important in our analysis later). Although $\mathcal{F}_{d}$ will have a complicated time evolution in general, it can always be expanded to short times $dt$ as
\begin{equation}\label{supp:eq:eff_depolarizing_rate}
\mathcal{F}_{d}(\mathcal{E},I) \approx 1-\frac{d-1}{d}\Gamma dt,
\end{equation}
where $\Gamma$ can be interpreted as the effective depolarization rate of the channel $\mathcal{E}$ at short times. This interpretation comes from considering the qudit depolarizing channel
\begin{equation}
\mathcal{E}_{\mathrm{dep}}(\rho) = e^{-\gamma t}\rho + \left(1-e^{-\gamma t}\right)\frac{I}{d}\mathrm{Tr}[\rho],
\end{equation}
which transfers probability from $\rho$ to the maximally-mixed state $I/d$ at rate $\gamma$. Note that the factor of $\mathrm{Tr}[\rho]=1$ is included so this channel is explicitly trace-preserving, even when applied to operators $U_{nm}$ as in Eq. \eqref{supp:eq:avg_channel_fidelity}. The average channel fidelity of $\mathcal{E}_{\mathrm{dep}}$ relative to $I$ takes the form
\begin{equation}
\mathcal{F}_{d}(\mathcal{E}_{\mathrm{dep}},I) = \frac{1}{d} + \frac{d-1}{d}e^{-\gamma t} \approx 1-\frac{d-1}{d}\gamma dt,
\end{equation}
where in the last step we have expanded the expression to short times $dt$.

This rate $\Gamma$ enables us to compare different decay channels on the same footing. In particular, we want to compare the decay rate $\Gamma^{\mathrm{logical}}_{d}$ of our logical qudit to $\Gamma^{\mathrm{physical}}_{d}$ of the best physical qudit in our system. We define the QEC gain as their ratio
\begin{equation}
G_{d} = \Gamma^{\mathrm{physical}}_{d}/\Gamma^{\mathrm{logical}}_{d},
\end{equation}
and the break-even point is when this gain is unity.

\subsection{Cavity Fock Qudit}
The best physical qudit in our system is the Fock qudit in the cavity, spanned by the states $\{|0\rangle, |1\rangle, ..., |d-1\rangle \}$. The cavity itself decays under both photon loss and dephasing, described by the Kraus operators
\begin{equation}
\begin{split}
K_{0} =& \, I - \frac{\kappa_{1,c}dt}{2}a^{\dagger}a - \kappa_{\phi,c}dt (a^{\dagger}a)^{2} ,\\
K_{1} =& \sqrt{\kappa_{1,c}dt} \, a ,\\
K_{2} =& \sqrt{2\kappa_{\phi,c}dt} \, a^{\dagger}a ,
\end{split}
\end{equation}
where $\kappa_{1,c} = 1/T_{1,c}$ and $\kappa_{\phi,c} = 1/T_{2R,c} - 1/2T_{1,c}$. The overall decay channel of the cavity is then given by
\begin{equation}
\mathcal{E}_{c}(\rho) = K_{0}\rho K_{0}^{\dagger} + K_{1}\rho K_{1}^{\dagger} + K_{2}\rho K_{2}^{\dagger}.
\end{equation}
To express the effective short-time depolarization rate $\Gamma_{d}^{\mathrm{Fock}}$ of the cavity Fock qudit in terms of $\kappa_{1,c}$ and $\kappa_{\phi,c}$, we truncate the channel $\mathcal{E}_{c}$ to the first $d$ Fock states, plug into Eq. \eqref{supp:eq:avg_channel_fidelity}, expand to lowest order in $dt$, and compare the result to Eq. \eqref{supp:eq:eff_depolarizing_rate}. We perform this calculation in Mathematica and obtain
\begin{equation}
\begin{split}
\Gamma_{2}^{\mathrm{Fock}} &= \frac{2}{3}\kappa_{1,c} + \frac{2}{3}\kappa_{\phi,c} = \left(851\pm 9 \:\mathrm{\mu s}\right)^{-1} ,\\
\Gamma_{3}^{\mathrm{Fock}} &= \frac{9}{8}\kappa_{1,c} + \frac{3}{2}\kappa_{\phi,c} = \left(488\pm 7 \:\mathrm{\mu s}\right)^{-1} ,\\
\Gamma_{4}^{\mathrm{Fock}} &= \frac{8}{5}\kappa_{1,c} + \frac{8}{3}\kappa_{\phi,c} = \left(332\pm 6 \:\mathrm{\mu s}\right)^{-1} ,
\end{split}
\end{equation}
where in the last step we plugged in our measured values for $\kappa_{1,c}$ and $\kappa_{\phi,c}$ (see Table \cref{supp:tab:experimental_parameters}).

\subsection{Breaking Down the Average Channel Fidelity}
We next analyze the average channel fidelity of our logical GKP qudits by breaking down Eq. \eqref{supp:eq:avg_channel_fidelity} in terms of quantities we can measure experimentally. Our strategy is to partition the set of Pauli operators $\{U_{nm}\}$ into subsets $\{P,P^{2},...,P^{d-1}\}$, where $P\in\{U_{nm}\}$ is assumed to have a non-degenerate spectrum. Since we are free to choose an overall phase factor of our $U_{nm}$, we may assume without loss of generality that the eigenvalues of $P$ are $1,\omega_{d},\omega_{d}^{2},...,\omega_{d}^{d-1}$, where $\omega_{d} = \exp(2\pi i/d)$ is the primitive $d$th root of unity. The operator $P$ can therefore be spectrally decomposed as
\begin{equation}
P = \sum\limits_{k=0}^{d-1}\omega_{d}^{k}|P_{k}\rangle\langle P_{k}|,
\end{equation}
where $|P_{k}\rangle$ is the $k$th eigenket of $P$ satisfying $P|P_{k}\rangle = \omega_{d}^{k}|P_{k}\rangle$. We can also express the projectors $|P_{k}\rangle\langle P_{k}|$ in terms of the operators $P^{k}$ by considering the expression
\begin{equation}
\begin{split}
\frac{1}{d}\sum\limits_{\ell=0}^{d-1}\omega_{d}^{-\ell k} P^{\ell} =& \frac{1}{d}\sum\limits_{\ell=0}^{d-1}\omega_{d}^{-\ell k}\sum\limits_{j=0}^{d-1}\omega_{d}^{\ell j}|P_{j}\rangle\langle P_{j}| ,\\
=&  \frac{1}{d}\sum\limits_{j=0}^{d-1}|P_{j}\rangle\langle P_{j}|\sum\limits_{\ell=0}^{d-1}\omega_{d}^{\ell (j-k)} ,\\
=& |P_{k}\rangle\langle P_{k}| ,
\end{split}
\end{equation}
where in the last step we used the identity
\begin{equation}\label{supp:eq:geometric_sum_identity}
\sum\limits_{\ell=0}^{d-1}\omega_{d}^{\ell (j-k)} = d\delta_{jk}.
\end{equation}
We next consider terms of the form $\mathrm{Tr}\left[(P^{j})^{\dagger}\mathcal{E}(P^{j})\right]$, as appear in Eq. \eqref{supp:eq:avg_channel_fidelity}. Using the above results and summing over $j$, we can write
\begin{equation}\label{supp:eq:derive_sum_over_trace_terms}
\begin{split}
\sum\limits_{j=0}^{d-1}\mathrm{Tr}\left[(P^{j})^{\dagger}\mathcal{E}(P^{j})\right] =& \sum\limits_{j,k=0}^{d-1}\langle P_{k}|(P^{j})^{\dagger}\mathcal{E}(P^{j}) |P_{k}\rangle ,\\
=& \sum\limits_{j,k,\ell,m=0}^{d-1}\omega_{d}^{j(m-\ell)}\langle P_{k}|P_{\ell}\rangle\langle P_{\ell}|\mathcal{E}(|P_{m}\rangle\langle P_{m}|) |P_{k}\rangle ,\\
=& \sum\limits_{j,k,m=0}^{d-1}\omega_{d}^{j(m-k)}\langle P_{k}|\mathcal{E}(|P_{m}\rangle\langle P_{m}|) |P_{k}\rangle ,\\
=& d\sum\limits_{k=0}^{d-1}\langle P_{k}|\mathcal{E}(|P_{k}\rangle\langle P_{k}|) |P_{k}\rangle ,\\
\end{split}
\end{equation}
where in the last step we again used Eq. \eqref{supp:eq:geometric_sum_identity}. Finally, we want to remove the contribution of the identity to the left hand side of Eq. \cref{supp:eq:derive_sum_over_trace_terms}. Assuming the steady state of the channel $\mathcal{E}$ is the maximally mixed qudit state, and therefore that $\mathcal{E}(I) = I$, we obtain
\begin{equation}\label{supp:eq:cycle_subset_sum}
\sum\limits_{j=1}^{d-1}\mathrm{Tr}\left[(P^{j})^{\dagger}\mathcal{E}(P^{j})\right] = -d + d\sum\limits_{k=0}^{d-1}\langle P_{k}|\mathcal{E}(|P_{k}\rangle\langle P_{k}|) |P_{k}\rangle. 
\end{equation}
The right hand side of this equation is now in terms of experimentally accessible quantities: the terms in the sum can be measured by preparing the state $|P_{k}\rangle$, applying the channel $\mathcal{E}$, and measuring the probability that the final state is still $|P_{k}\rangle$.

\subsection{Average Channel Fidelity for GKP Qubits}
In $d=2$, we partition the set $\{U_{nm}\}$ as
\begin{equation}
\{U_{nm}\} = \{I\}\cup\{X_{2}\}\cup\{Z_{2}\}\cup\{\sqrt{\omega_{2}}X_{2}Z_{2}\},
\end{equation}
which is somewhat trivial since each constituent set has only one element, but serves to illustrate our approach nonetheless. We break up Eq. \eqref{supp:eq:avg_channel_fidelity} into sums over each subset of the partition and use Eq. \eqref{supp:eq:cycle_subset_sum} to obtain the average channel fidelity
\begin{equation}\label{supp:eq:gkp2_avg_channel_fidelity}
\mathcal{F}_{2}(\mathcal{E},I) = \frac{1}{6}\sum\limits_{P\in\mathcal{P}_{2}}\sum\limits_{k=0}^{1} \langle P_{k}|\mathcal{E}(|P_{k}\rangle\langle P_{k}|) |P_{k}\rangle,
\end{equation}
where $\mathcal{P}_{2} = \{X_{2}, Z_{2}, \sqrt{\omega_{2}}X_{2}Z_{2}\}$. Assuming each Pauli eigenstate $|P_{k}\rangle$ of the GKP qubit decays exponentially toward the maximally mixed state $I/2$ at rate $\gamma_{P_{k}}$, the effective depolarization rate $\Gamma_{2}^{\mathrm{GKP}}$ can be expressed as
\begin{equation}\label{supp:eq:gkp2_eff_depolarization_rate}
\Gamma_{2}^{\mathrm{GKP}} = \frac{1}{6}\sum\limits_{P \in \mathcal{P}_{2}}\sum\limits_{k=0}^{1}\gamma_{P_{k}},
\end{equation}
in agreement with previous work \cite{sivak_real-time_2023}.

\subsection{Average Channel Fidelity for GKP Qutrits}
In $d=3$, we partition the set $\{U_{nm}\}$ as
\begin{equation}
\{U_{nm}\} = \{I\}\cup\{X_{3},X_{3}^{2}\}\cup\{Z_{3},Z_{3}^{2}\}\cup\{X_{3}Z_{3},X_{3}^{2}Z_{3}^{2}\}\cup\{X_{3}^{2}Z_{3},X_{3}Z_{3}^{2}\}.
\end{equation}
We again break up Eq. \eqref{supp:eq:avg_channel_fidelity} into sums over each subset of the partition and use Eq. \eqref{supp:eq:cycle_subset_sum} to obtain the average channel fidelity
\begin{equation}\label{supp:eq:gkp3_avg_channel_fidelity}
\mathcal{F}_{3}(\mathcal{E},I) = \frac{1}{12}\sum\limits_{P\in\mathcal{P}_{3}}\sum\limits_{k=0}^{2} \langle P_{k}|\mathcal{E}(|P_{k}\rangle\langle P_{k}|) |P_{k}\rangle,
\end{equation}
where $\mathcal{P}_{3} = \{X_{3},Z_{3},X_{3}Z_{3}, X_{3}^{2}Z_{3}\}$. We note that the bases of Pauli operators $P\in\mathcal{P}_{3}$ form a maximal set of mutually unbiased bases in $d=3$ \cite{wootters_optimal_1989}. Assuming each Pauli eigenstate $|P_{k}\rangle$ of the GKP qutrit decays exponentially toward the maximally mixed state $I/3$ at rate $\gamma_{P_{k}}$, the effective depolarization rate $\Gamma_{3}^{\mathrm{GKP}}$ can be expressed as
\begin{equation}\label{supp:eq:gkp3_eff_depolarization_rate}
\Gamma_{3}^{\mathrm{GKP}} = \frac{1}{12}\sum\limits_{P \in \mathcal{P}_{3}}\sum\limits_{k=0}^{2}\gamma_{P_{k}}.
\end{equation}

\subsection{Average Channel Fidelity for GKP Ququarts}
In $d=4$ the situation is more complex, since $\{U_{nm}\}$ can't be partitioned into disjoint subsets $\{P,P^{2},...,P^{d-1}\}$. The best we can do is
\begin{equation}\label{supp:eq:ququart_pauli_partition}
\begin{split}
\{U_{nm}\} = \{I\}
&\cup 
\{X_{4},X_{4}^{2},X_{4}^{3}\} \\
&\cup 
\{Z_{4},Z_{4}^{2},Z_{4}^{3}\} \\
&\cup\{\sqrt{\omega_{4}}X_{4}Z_{4},\omega_{4}X_{4}^{2}Z_{4}^{2},\omega_{4}^{3/2}X_{4}^{3}Z_{4}^{3}\} \\
&\cup
\{X_{4}^{2}Z_{4},Z_{4}^{2},X_{4}^{2}Z_{4}^{3}\} \\
&\cup
\{\sqrt{\omega_{4}}X_{4}^{3}Z_{4},\omega_{4}X_{4}^{2}Z_{4}^{2},\omega_{4}^{3/2}X_{4}Z_{4}^{3}\} \\
&\cup
\{X_{4}Z_{4}^{2},X_{4}^{2},X_{4}^{3}Z_{4}^{2}\}.
\end{split}
\end{equation}
If we apply the same procedure here as in $d=2$ and $d=3$ we would end up double-counting the contribution of Pauli operators $X_{4}^{2}$, $Z_{4}^{2}$, and $X_{4}^{2}Z_{4}^{2}$ to Eq. \eqref{supp:eq:avg_channel_fidelity}. As it turns out, however, $[X_{4}^{2},Z_{4}^{2}] = 0$, so all three of these operators can be simultaneously diagonalized, enabling us to effectively subtract off the effect of this double counting. To this end, we introduce what we call the ququart parity basis $\{|\pm,n\rangle_{4} : n=0,1\}$, which are the simultaneous eigenstates of $X_{4}^{2}$ and $Z_{4}^{2}$ according to
\begin{equation}
\begin{split}
X_{4}^{2}|\pm,n\rangle_{4} &= \pm|\pm,n\rangle_{4}, \\
Z_{4}^{2}|\pm,n\rangle_{4} &= (-1)^{n}|\pm,n\rangle_{4}.
\end{split}
\end{equation}
Using the spectral decomposition of the operators in $\mathcal{P}_{\mathrm{parity}} = \{X_{4}^{2},Z_{4}^{2},X_{4}^{2}Z_{4}^{2}\}$
\begin{equation}
\begin{split}
X_{4}^{2} =& \sum\limits_{\substack{s = \pm \\ n = 0,1}}s|s,n\rangle\langle s,n| ,\\
Z_{4}^{2} =& \sum\limits_{\substack{s = \pm \\ n = 0,1}}(-1)^{n}|s,n\rangle\langle s,n| ,\\
X_{4}^{2}Z_{4}^{2} =& \sum\limits_{\substack{s = \pm \\ n = 0,1}}s(-1)^{n}|s,n\rangle\langle s,n| ,\\
\end{split}
\end{equation}
we can calculate
\begin{equation}\label{supp:eq:parity_contribution}
\begin{split}
\sum\limits_{P\in\mathcal{P}_{\mathrm{parity}}} \mathrm{Tr}\left[P^{\dagger}\mathcal{E}(P)\right] =& \sum\limits_{\substack{s,t = \pm \\ n,m = 0,1}}\left(st+st(-1)^{n+m}+(-1)^{n+m}\right)\langle s,n|\mathcal{E}(|t,m\rangle\langle t,m|)|s,n\rangle ,\\
=& -4 + \sum\limits_{\substack{s,t = \pm \\ n,m = 0,1}}\left(1+st+st(-1)^{n+m}+(-1)^{n+m}\right)\langle s,n|\mathcal{E}(|t,m\rangle\langle t,m|)|s,n\rangle ,\\
=& -4 + 4\sum\limits_{\substack{s = \pm \\ n = 0,1}}\langle s,n|\mathcal{E}(|s,n\rangle\langle s,n|)|s,n\rangle ,
\end{split}
\end{equation}
where in the last step we used $1+st+st(-1)^{n+m}+(-1)^{n+m} = 4\delta_{nm}\delta_{st}$ for $s,t\in\{-1,1\}$ and $n,m\in\{0,1\}$. We can now once again break up Eq. \eqref{supp:eq:avg_channel_fidelity} into sums over each subset of the non-disjoint partition from Eq. \cref{supp:eq:ququart_pauli_partition}, evaluate these sums using Eq. \eqref{supp:eq:cycle_subset_sum}, and subtract off the contribution of the set $\mathcal{P}_{\mathrm{parity}}$ that is double-counted due to the non-disjointness of the partitioning. Doing so, we obtain the average channel fidelity
\begin{equation}\label{supp:eq:gkp4_avg_channel_fidelity}
\mathcal{F}_{4}(\mathcal{E},I) = \frac{1}{20}\Biggl[\sum\limits_{P \in \mathcal{P}_{4}}\sum\limits_{k=0}^{3}\langle P_{k}| \mathcal{E}\left(|P_{k}\rangle\langle P_{k}|\right)|P_{k}\rangle 
- \sum\limits_{\substack{s=\pm \\ n=0,1}}\langle s,n|\mathcal{E}\left(|s,n\rangle\langle s,n|\right)|s,n\rangle\Biggr].
\end{equation}
Finally, if each Pauli eigenstate $|P_{k}\rangle$ (parity eigenstate $|\pm,n\rangle$) of the GKP ququart decays exponentially to the maximally mixed state $I/4$ at rate $\gamma_{P_{k}}$ ($\gamma_{\pm,n}$), the effective depolarization rate $\Gamma_{4}^{\mathrm{GKP}}$ can be expressed as
\begin{equation}\label{supp:eq:gkp4_eff_depolarization_rate}
\Gamma_{4}^{\mathrm{GKP}} = \frac{1}{20}\Biggl[\sum\limits_{P \in \mathcal{P}_{4}}\sum\limits_{k=0}^{3}\gamma_{P_{k}} - \sum\limits_{\substack{s=\pm \\ n=0,1}}\gamma_{\pm,n} \Biggr].
\end{equation}

\clearpage
\section{Logical GKP Qubit}
\label{supp:sec:logical_gkp_qubit}
In this section we present the methods and results pertaining to our experimental realization of a logical GKP qubit beyond break-even, using established techniques from prior work on GKP qubits \cite{fluhmann_encoding_2019, campagne-ibarcq_quantum_2020, de_neeve_error_2022, sivak_real-time_2023}. Our goal is to provide a basis of comparison with our novel techniques and results for the GKP qutrit and ququart.

\subsection{State Preparation}\label{supp:sec:gkp2_state_prep}
\begin{figure}[b!]
{
\includegraphics{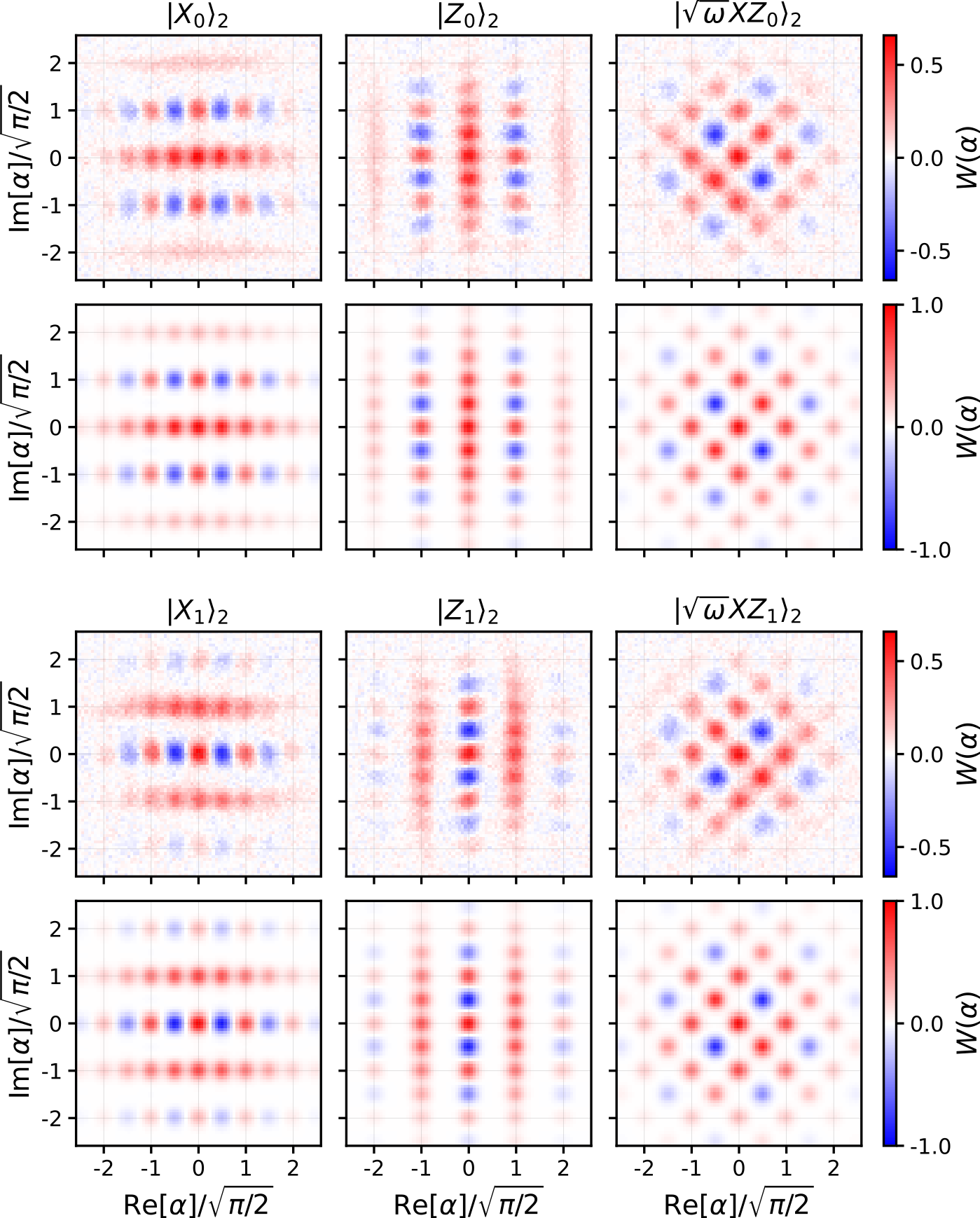}
}
\caption{
\textbf{GKP qubit state preparation.} For each state, the measured Wigner function of the prepared state is on top, and that of the target state is on the bottom. We use depth $8$ ECD circuits and a finite-energy envelope size $\Delta = 0.34$.
}\label{supp:fig:gkp2_ecd_prep}
\end{figure}

To prepare the Pauli eigenstates of the GKP qubit defined in Table \cref{supp:tab:qudit_states}, we use interleaved sequences of ECD gates and ancilla qubit rotations \cite{eickbusch_fast_2022}, as described in Sec. \cref{supp:sec:state_preparation}. We optimize depth $8$ ECD circuits for each state with envelope size $\Delta = 0.34$, approximated numerically using the method described in Sec. \cref{supp:sec:gkp_qudits}. To verify that we are preparing the desired states, we run these ECD circuits experimentally and perform Wigner tomography of the prepared states, the results of which are presented in Fig. \cref{supp:fig:gkp2_ecd_prep}. We then reconstruct the density matrices $\rho_{\mathrm{prep}}$ of the prepared states from their measured Wigner functions and compute their fidelity with respect to the target states $|\psi_{\mathrm{target}}\rangle$ according to $\mathcal{F}_{\mathrm{prep}} = \langle \psi_{\mathrm{target}}|\rho_{\mathrm{prep}} |\psi_{\mathrm{target}}\rangle$, the results of which are presented in Table \cref{supp:tab:gkp2_state_prep_fidelities}.

\begin{center}
\begin{table}[t!]
\setlength\tabcolsep{6pt}
\renewcommand{\arraystretch}{2}
\begin{tabular}{ | c | c | c | c |} 
\cline{2-4} 
\multicolumn{1}{c|}{$\mathcal{F}_{\mathrm{prep}}$ (\%)} & $|X_{n}\rangle_{2}$ & $|Z_{n}\rangle_{2}$ &  $|\sqrt{\omega}XZ_{n}\rangle_{2}$ \\[3pt]
\hline
$n=0$ & $90$ & $88$ & $89$ \\ [3pt]
\hline
$n=1$ & $89$ & $89$ & $90$ \\ [3pt]
\hline
\end{tabular}
\caption{Fidelity $\mathcal{F}_{\mathrm{prep}} = \langle \psi_{\mathrm{target}}|\rho_{\mathrm{prep}} |\psi_{\mathrm{target}}\rangle$ of prepared Pauli eigenstates of the GKP qubit, determined by reconstructing the prepared state from its measured Wigner function and comparing it to the target state.}
\label{supp:tab:gkp2_state_prep_fidelities}
\end{table}
\end{center}

\subsection{Pauli Measurements}\label{supp:sec:gkp2_pauli_msmts}
Following previous work \cite{terhal_encoding_2016, fluhmann_encoding_2019, campagne-ibarcq_quantum_2020}, we measure the GKP qubit in the basis of Pauli operator $P_{2}$ by initializing the ancilla in the $|+\rangle$ state and performing a controlled-Pauli operation $CP_{2}$. If the GKP qubit starts out in state $|\psi\rangle_{2} = c_{0}|P_{0}\rangle_{2} + c_{1}|P_{1}\rangle_{2}$, then this operation entangles the two systems as
\begin{equation}
CP_{2}|+\rangle|\psi\rangle_{2} = c_{0}|+\rangle|P_{0}\rangle_{2} + c_{1}|+\rangle|P_{1}\rangle_{2},
\end{equation}
such that a subsequent measurement of the ancilla along its $x$ axis constitutes a measurement of the GKP qubit in the basis of $P_{2}$, and collapses the state accordingly. In this work we use $CP_{d}$ operations designed for the ideal GKP code, which incur infidelity when applied to the finite-energy code. In the case of GKP qubits these $CP_{2}$ gates have been adapted to the finite-energy code \cite{royer_stabilization_2020, hastrup_measurement-free_2021, de_neeve_error_2022} but we choose not to use them in the present work, in the interest of comparison with the techniques we develop for the GKP qutrit and ququart. For $P_{2}\in\{X_{2},Z_{2},\sqrt{\omega}X_{2}Z_{2}\}$, we compile the controlled Pauli gate $CP_{2}$ in terms of an ECD gate and ancilla rotations according to
\begin{equation}
\begin{split}
CX_{2} =& D(\sqrt{\pi/2}/2)\ECD(\sqrt{\pi/2})\sigma_{x} , \\
CZ_{2} =& D(i\sqrt{\pi/2}/2)\ECD(i\sqrt{\pi/2})\sigma_{x} , \\
C(\sqrt{\omega}X_{2}Z_{2}) =& D(e^{i\pi/4}\sqrt{\pi}/2)\ECD(e^{i\pi/4}\sqrt{\pi})\sigma_{x} ,
\end{split}
\end{equation}
as described in the Methods section of the main text. In our experiments we do not implement the unconditional displacements $D(\beta_{nm}/2)$, which means that our measurement sequence will apply an overall displacement $D(-\beta_{nm}/2)$ that brings us out of the code space. However, it also means that the envelope of our finite-energy code remains centered, since our conditional displacements are symmetric about the origin. 

\begin{figure}[t!]
\includegraphics{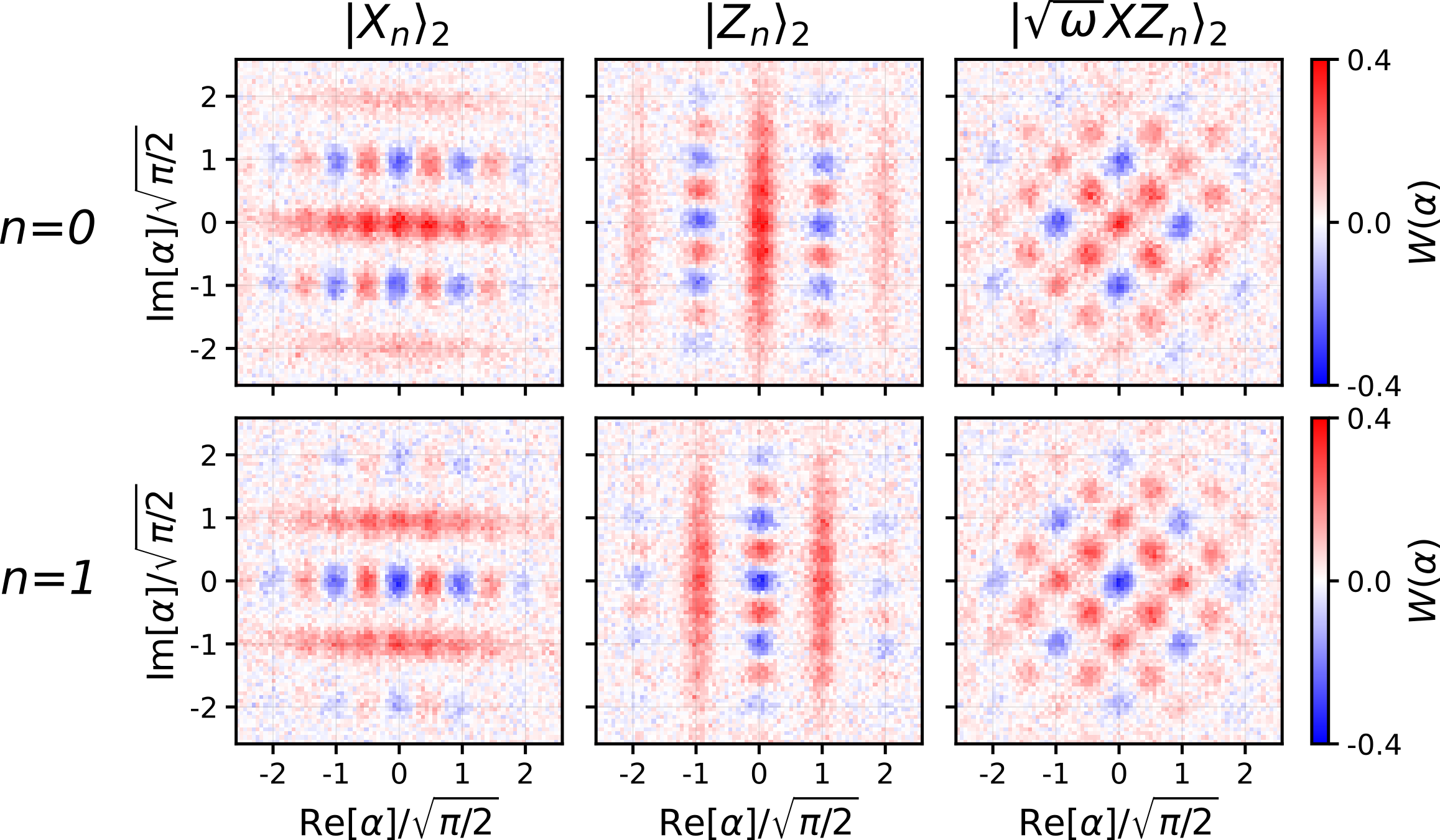}
\caption{
\textbf{GKP qubit Pauli measurement backaction.} In each column, we implement a projective measurement with respect to the indicated basis and then perform Wigner tomography post-selected on the measurement outcomes.
}\label{supp:fig:gkp2_pauli_msmt}
\end{figure}

To verify that our logical Pauli measurements work as intended, we prepare the maximally mixed state of the GKP qubit $\rho_{2}^{\mathrm{mix}}$ by performing $160$ rounds of sBs starting from the cavity in vacuum, implement the $CP_{2}$ gate, measure the ancilla along its $x$ axis, and perform Wigner tomography of the cavity post-selected on outcome of this measurement. The results of this experiment are shown in Fig. \cref{supp:fig:gkp2_pauli_msmt}, from which we can see that the Pauli measurement works as we expect: it projects onto the Pauli eigenstate $|P_{n}\rangle_{2}$ according to the measurement result, and implements a $D(-\beta_{nm}/2)$ operation on the underlying ideal GKP codewords while leaving the finite-energy envelope centered.

\subsection{Optimized Qubit Beyond Break-even}\label{supp:sec:gkp2_rl_optimization}

\begin{figure}[t!]
\includegraphics{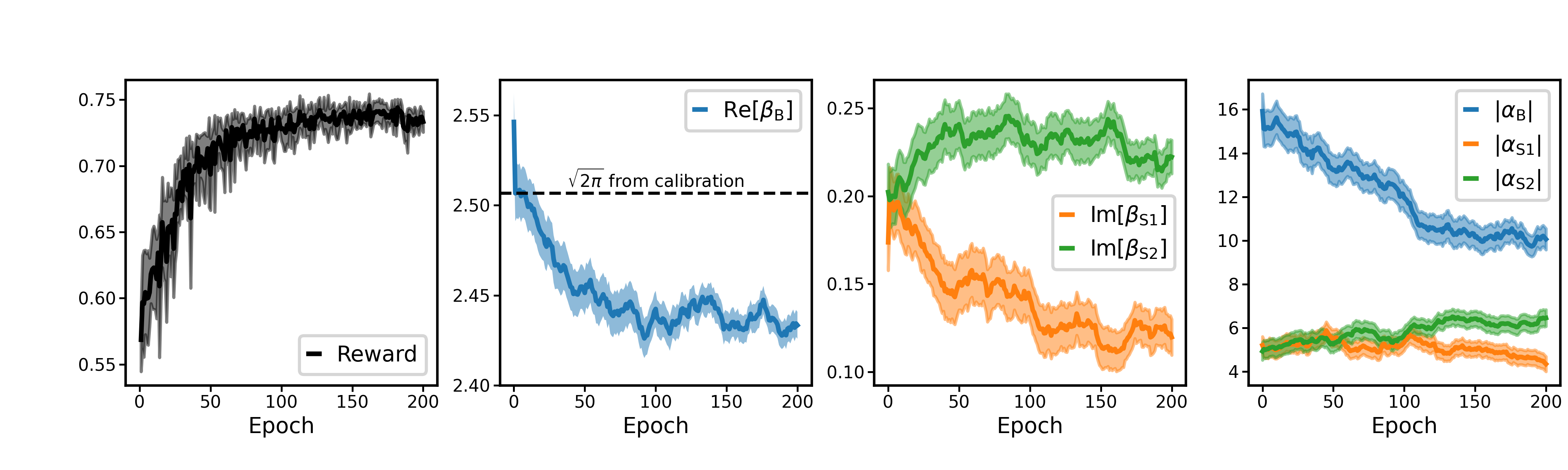}
\caption{
\textbf{GKP qubit RL training.} Lines indicate mean values while the shaded regions indicate the one-standard-deviation confidence interval. 
}\label{supp:fig:gkp2_rl_training}
\end{figure}

We optimize the QEC protocol of the GKP qubit using a reinforcement learning (RL) agent \cite{sivak_model-free_2022} following the method described in Ref. \cite{sivak_real-time_2023} (also, see the Methods section of the main text). In Fig. \cref{supp:fig:gkp2_rl_training} we present both the reward signal and key parameters of the sBs circuit as a function of training epoch. Our interpretation is that the RL training serves three main purposes \cite{sivak_real-time_2023}. First, it enables us to go beyond the constraints imposed by our model for the QEC protocol (for example, finding a better strategy where the second small conditional displacement amplitude $|\beta_{\mathrm{S2}}|$ of the sBs protocol is greater than $|\beta_{\mathrm{S1}}|$ of the first). Second, it mitigates experimental imperfections that are difficult to model (for example, finding intermediate displacement amplitudes $|\alpha_{\mathrm{S1},\mathrm{B},\mathrm{S2}}|$ that avoid bringing spurious two-level-systems onto resonance with our ancilla qubit). Third, it performs a meta-calibration of all pulse parameters in the midst of the complicated control sequences required for QEC (for example, finding a $3\%$ deviation between $|\beta_{B}|$ and $\sqrt{2\pi}$, and finding optimal DRAG parameters \cite{chen_measuring_2016} for each ancilla qubit pulse). The total duration of each sBs round is constrained to be $7$ $\mathrm{\mu}$s; the optimal protocol found by the RL training spends $3632$ ns measuring and resetting the ancilla qubit, $1792$ ns performing the sBs protocol, $1352$ ns idling, and $224$ ns executing FPGA sequencing instructions.

\begin{figure}[t!]
\includegraphics{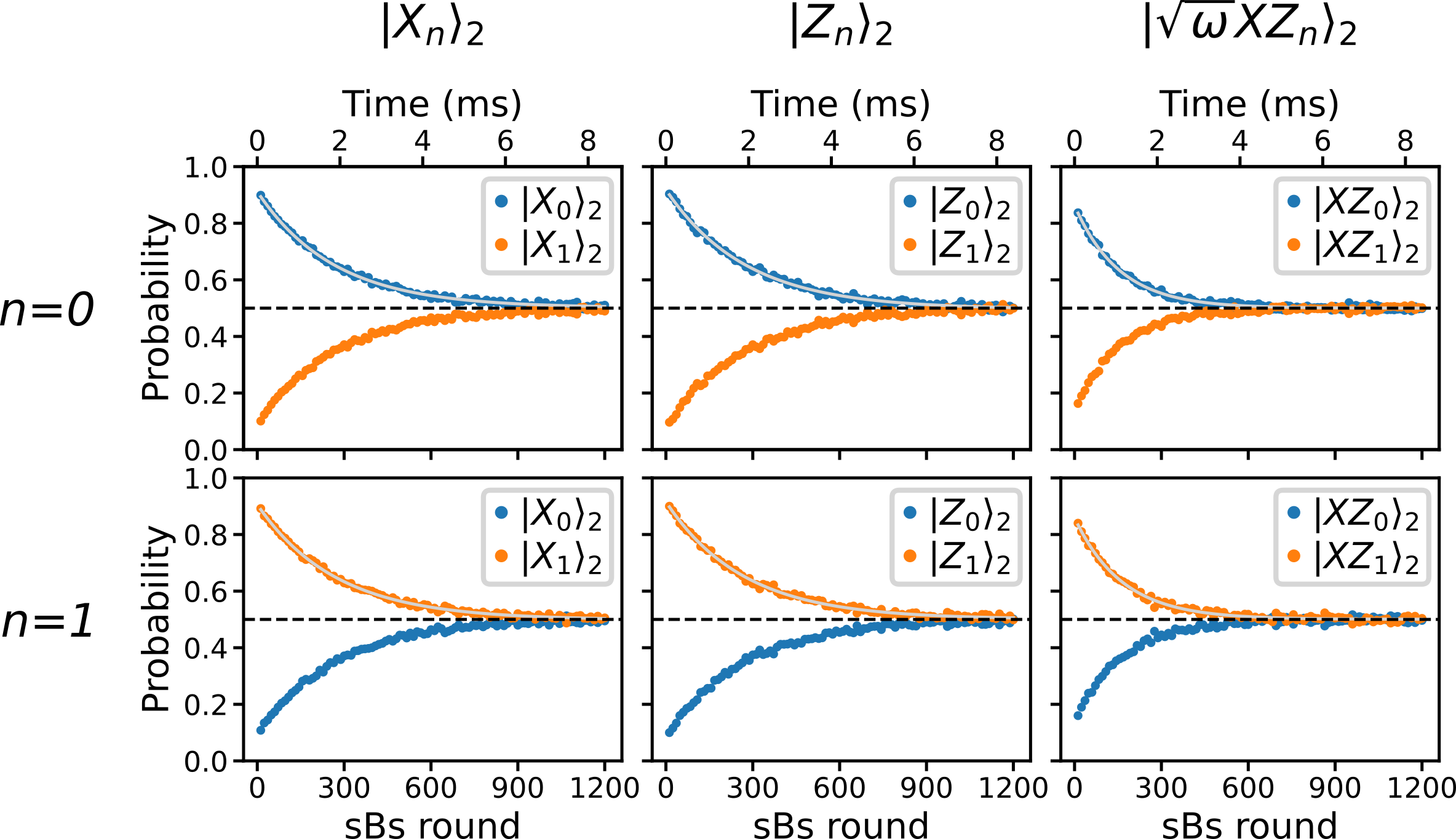}

\bigskip

\setlength\tabcolsep{6pt}
\renewcommand{\arraystretch}{2}
\begin{tabular}{ | c | c | c | c |} 
\cline{2-4} 
\multicolumn{1}{c|}{$\gamma_{P_{n}}^{-1}$ ($\mathrm{\mu s}$)} & $|X_{n}\rangle_{2}$ & $|Z_{n}\rangle_{2}$ &  $|\sqrt{\omega}XZ_{n}\rangle_{2}$ \\[3pt]
\hline
$n=0$ & $1863\pm 27$ & $1935\pm 35$ & $1086\pm 19$ \\ [3pt]
\hline
$n=1$ & $1902\pm 28$ & $1872\pm 32$ & $1150\pm 25$ \\ [3pt]
\hline
\end{tabular}
\caption{
\textbf{Lifetime of the optimized logical GKP qubit.} The solid grey lines are fits to an exponential decay. The dashed black line indicates a probability of $1/2$. 
}\label{supp:fig:gkp2_lifetime}
\end{figure}

To evaluate the average channel fidelity of the optimal QEC protocol via Eq. \eqref{supp:eq:gkp2_avg_channel_fidelity}, we measure the probabilities $\langle P_{n}|\mathcal{E}_{\mathrm{opt}}^{N}(|P_{n}\rangle\langle P_{n}|)|P_{n}\rangle_{2}$ as a function of $N$ for $n=0,1$ and $P\in\mathcal{P}_{2} = \{X_{2},Z_{2},\sqrt{\omega}X_{2}Z_{2}\}$, where $\mathcal{E}_{\mathrm{opt}}^{N}$ is the channel corresponding to $N$ rounds of the optimal QEC protocol. We measure these probabilities by preparing state $|P_{n}\rangle$, performing $N$ rounds of QEC with the optimal protocol (sweeping the value of $N$), and measuring the GKP qubit in the basis of Pauli operator $P$. The results of these six experiments are presented in Fig. \cref{supp:fig:gkp2_lifetime}. We then fit each probability $\langle P_{n}|\mathcal{E}_{\mathrm{opt}}^{N}(|P_{n}\rangle\langle P_{n}|)|P_{n}\rangle_{2}$ to an exponential decay and find the decay rate $\gamma_{P_{n}}$, which enables us to calculate the effective short-time depolarization rate of the optimized logical GKP qubit
\begin{equation}
\Gamma_{2}^{\mathrm{GKP}} = \left(1537\pm 12 \: \mathrm{\mu s}\right)^{-1},
\end{equation}
using Eq. \eqref{supp:eq:gkp2_eff_depolarization_rate}. Comparing with $\Gamma_{2}^{\mathrm{Fock}} = (851\pm 9 \: \mathrm{\mu s})^{-1}$, we obtain a QEC gain of
\begin{equation}
G_{2} = \Gamma_{2}^{\mathrm{Fock}}/\Gamma_{2}^{\mathrm{GKP}} = 1.81\pm 0.02,
\end{equation}
well beyond the break-even point. There are several possible reasons why this result is worse than the factor of $2.3$ previously reported for the same device \cite{sivak_real-time_2023}. First, we reclamped the chip containing the transmon, leading to lower transmon thermal population, less storage dephasing, a smaller $\Gamma_{2}^{\mathrm{Fock}}$, and thus a smaller $G_{2}$. Second, this reclamping and thermal cycling of the device likely changed the bath of two-level-systems coupling to the transmon \cite{shalibo_lifetime_2010, grabovskij_strain_2012}. Third, we use a longer sBs duration of $7$ $\mathrm{\mu}$s that may be suboptimal. 

\begin{figure}[t!]
\includegraphics{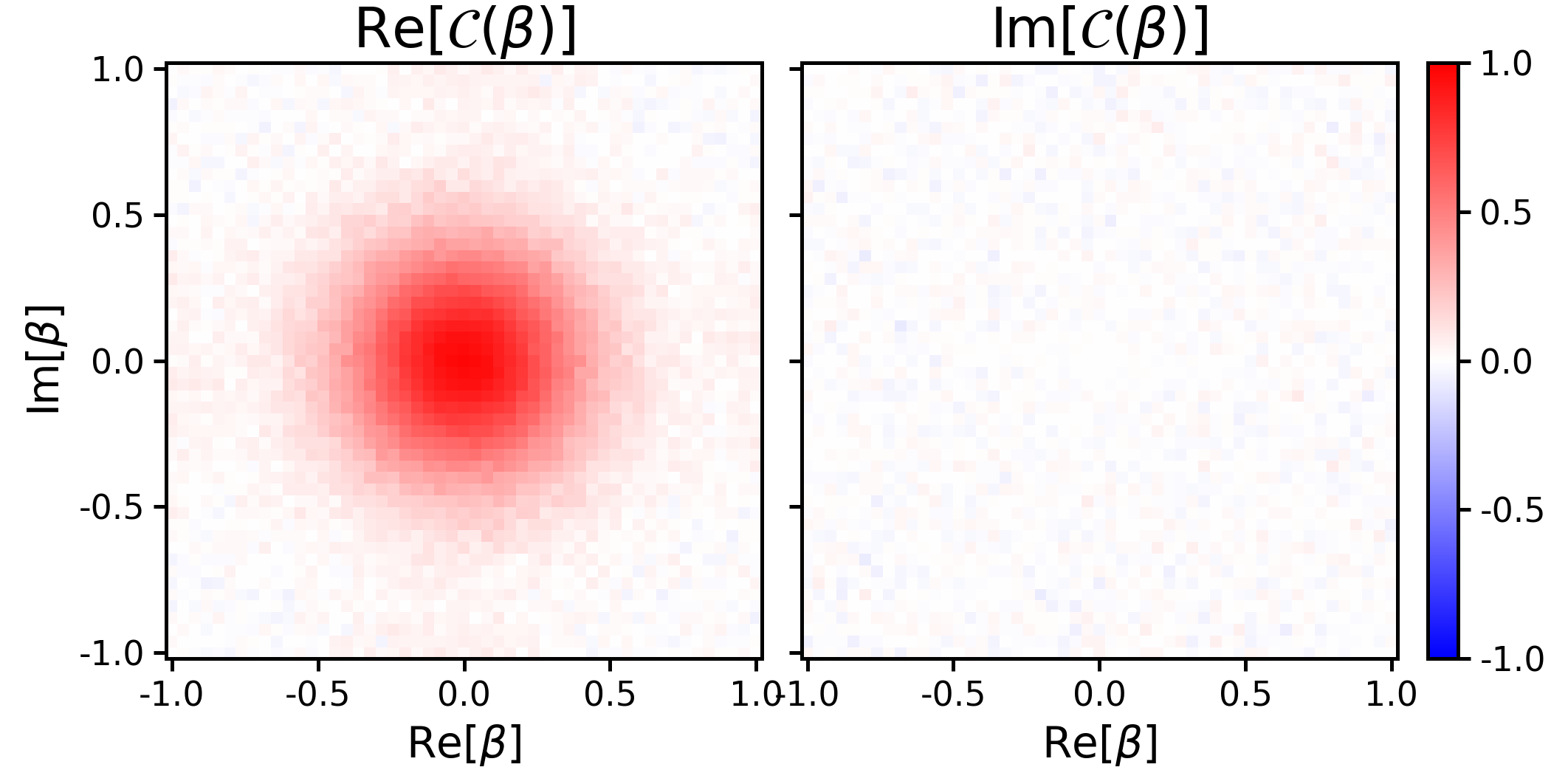}
\caption{
Steady-state characteristic function of the optimal GKP qubit QEC protocol, sampled near the origin of reciprocal phase space.
}\label{supp:fig:gkp2_steady_state_cf}
\end{figure}

Lastly, we find the effective envelope size of our optimal GKP qubit by measuring its steady-state characteristic function $\mathcal{C}(\beta)$ near the origin of reciprocal phase space. In this experiment we run the optimal protocol for $500$ rounds followed by characteristic function tomography, the result of which is shown in Fig. \cref{supp:fig:gkp2_steady_state_cf}. We fit the real part of the characteristic function to a Gaussian according to $\mathrm{Re}[\mathcal{C}(\beta)] = A\exp(-|\beta|^{2}/2\Delta_{\mathrm{eff}}^{2})+B$, and find $\Delta_{\mathrm{eff}} = 0.290\pm 0.001$. This is not exactly the same thing as the envelope size, since $\Delta_{\mathrm{eff}}$ will be affected by population in error spaces, such as photon-loss and photon-gain acting on the code space \cite{sivak_real-time_2023}. More precisely, this measurement of the characteristic function near the origin of reciprocal phase space enables us to calculate photon statistics of the state \cite{fluhmann_direct_2020} (see Sec. \cref{supp:sec:tomography_and_state_reconstruction}). For Gaussian $\mathrm{Re}[\mathcal{C}(\beta)]$ and null $\mathrm{Im}[\mathcal{C}(\beta)]$, we have $\langle a^{\dagger}a \rangle = (1/\Delta_{\mathrm{eff}}^{2} - 1)/2 = 5.45\pm 0.04$ photons.

\clearpage
\section{Logical GKP Qutrit}
In this section we present the methods and results pertaining to our experimental realization of a logical GKP qutrit ($d=3$) beyond break-even.

\subsection{State Preparation}
\begin{figure}[b!]
\includegraphics{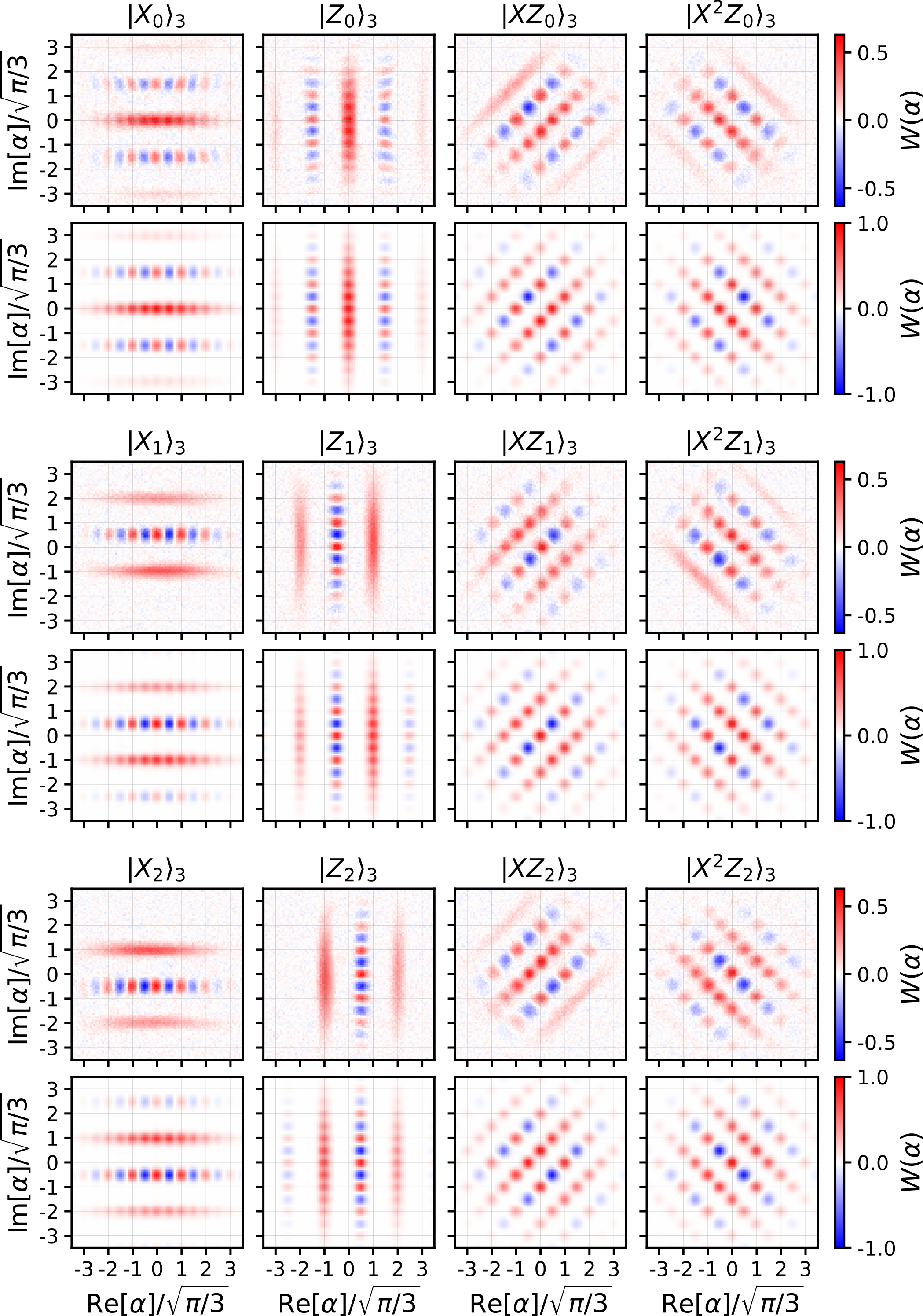}
\caption{
\textbf{GKP Qutrit state preparation.}  For each state, the measured Wigner function of the prepared state is on top, and that of the target state is on the bottom. We use depth $8$ ECD circuits and a finite-energy envelope size $\Delta = 0.32$.
}\label{supp:fig:gkp3_ecd_prep}
\end{figure}

To prepare the Pauli eigenstates of the GKP qutrit defined in Table \cref{supp:tab:qudit_states}, we again use interleaved sequences of ECD gates and ancilla qubit rotations \cite{eickbusch_fast_2022}, as described in Sec. \cref{supp:sec:state_preparation}. We optimize depth $8$ ECD circuits for each state with envelope size $\Delta = 0.32$, approximated numerically using the method described in Sec. \cref{supp:sec:gkp_qudits}. To verify that we are preparing the desired states, we run these ECD circuits experimentally and perform Wigner tomography of the prepared states, the results of which are presented in Fig. \cref{supp:fig:gkp3_ecd_prep}. We then reconstruct the density matrices $\rho_{\mathrm{prep}}$ of the prepared states from their measured Wigner functions and compute their fidelity with respect to the target states $|\psi_{\mathrm{target}}\rangle$ according to $\mathcal{F} = \langle \psi_{\mathrm{target}}|\rho_{\mathrm{prep}} |\psi_{\mathrm{target}}\rangle$, the results of which are presented in Table \cref{supp:tab:gkp3_state_prep_fidelities}.

\begin{center}
\begin{table}[t!]
\setlength\tabcolsep{6pt}
\renewcommand{\arraystretch}{2}
\begin{tabular}{ | c | c | c | c | c |} 
\cline{2-5}
  \multicolumn{1}{c|}{ $\mathcal{F}_{\mathrm{prep}}$ (\%)} & $|X_{n}\rangle_{3}$ & $|Z_{n}\rangle_{3}$ &  $|XZ_{n}\rangle_{3}$ & $|X^{2}Z_{n}\rangle_{3}$ \\[3pt]
\hline
$n=0$ & $86$ & $85$ & $85$ & $74$ \\ [3pt]
\hline
$n=1$ & $85$ & $86$ & $81$ & $85$ \\ [3pt]
\hline
$n=2$ & $87$ & $84$ & $83$ & $82$ \\ [3pt]
\hline
\end{tabular}
\caption{Fidelity $\mathcal{F}_{\mathrm{prep}} = \langle \psi_{\mathrm{target}}|\rho_{\mathrm{prep}} |\psi_{\mathrm{target}}\rangle$ of prepared Pauli eigenstates of the GKP qutrit, determined by reconstructing the prepared state from its measured Wigner function and comparing it to the target state.}
\label{supp:tab:gkp3_state_prep_fidelities}
\end{table}
\end{center}

\subsection{Pauli Measurements}

To measure our GKP qutrit in the basis of Pauli operator $P_{3}$ we devised the two-part circuit depicted in Fig. 2b of the main text, which measures the state of our logical three-level system via two binary measurements of our ancilla. If the GKP qutrit starts out in state $|\psi\rangle_{3} = c_{0}|P_{0}\rangle_{3} + c_{1}|P_{1}\rangle_{3} + c_{2}|P_{2}\rangle_{3}$ and the ancilla starts out in state $|+\rangle$, then the first part of the circuit entangles the two systems as
\begin{equation}
R_{\pi/2}^{\dagger}(\pi/2)~CP_{3}~R_{0}^{\dagger}(\theta_{0})~CP_{3}^{\dagger}~R_{0}(\theta_{0})~CP_{3}~|+\rangle|\psi_{3}\rangle = c_{0}|g\rangle|P_{0}\rangle_{3} + e^{-i\theta_{0}/2}|e\rangle\Bigl(c_{1}|P_{1}\rangle_{3} - c_{2}|P_{2}\rangle_{3}\Bigr),
\end{equation}
where $\theta_{0} = 2\arctan(1/\sqrt{2})$, such that a subsequent measurement of the ancilla along its $z$ axis constitutes a projective measurement on the qutrit distinguishing between the $|P_{0}\rangle_{3}$ state and the $\{|P_{1}\rangle_{3},|P_{2}\rangle_{3}\}$ subspace. Similarly, the second part of the circuit entangles the two systems as
\begin{equation}
R_{7\pi/6}^{\dagger}(\pi/2)~CP_{3}~R_{0}^{\dagger}(\theta_{0})~CP_{3}^{\dagger}~R_{2\pi/3}(\theta_{0})~CP_{3}~|+\rangle|\psi_{3}\rangle = c_{1}|g\rangle|P_{1}\rangle_{3} + e^{-i\theta_{0}/2}e^{2\pi i/3}|e\rangle\Bigl(c_{0}|P_{0}\rangle_{3} + c_{2}|P_{2}\rangle_{3}\Bigr),
\end{equation}
such that a subsequent measurement of the ancilla along its $z$ axis constitutes a projective measurement on the qutrit distinguishing between the $|P_{1}\rangle_{3}$ state and the $\{|P_{0}\rangle_{3},|P_{2}\rangle_{3}\}$ subspace. Performing these two measurements sequentially therefore projects the qutrit onto one of the three states, with probabilities determined by the initial amplitudes $\{c_{0},c_{1},c_{2}\}$. Note that the relative phases accumulated between the states $|P_{n}\rangle_{3}$ do not affect the overall projective measurement. 

Experimentally, we find that for finite-energy GKP qutrits the fidelity of this measurement scheme depends on the state $|P_{n}\rangle_{3}$. To mitigate this effect we symmetrize the measurement by adding a geometric phase to the ancilla via a $\sigma_{z}(2\pi k/3)$ gate after every $CP_{3}$ operation (and a $\sigma_{z}(-2\pi k/3)$ gate after every $CP_{3}^{\dagger}$ operation), where we call $k$ the symmetrization index. This modifies the measurement circuit such that the first part distinguishes between $|P_{k}\rangle_{3}$ and the subspace $\{|P_{n\neq k}\rangle_{3}\}$, and the second part distinguishes between $|P_{k+1}\rangle_{3}$ and the subspace $\{|P_{n\neq k+1}\rangle_{3}\}$. For all the GKP qutrit Pauli measurements presented in this work, we average this measurement scheme over $k=0,1,2$.

For $P_{3}\in\{X_{3},Z_{3},X_{3}Z_{3},X_{3}^{2}Z_{3}\}$, we compile the controlled Pauli gate $CP_{3}$ in terms of an ECD gate and ancilla rotations according to
\begin{equation}
\begin{split}
CX_{3} =& D(\sqrt{\pi/3}/2)~\ECD(\sqrt{\pi/3})~\sigma_{x} ,\\
CZ_{3} =& D(i\sqrt{\pi/3}/2)~\ECD(i\sqrt{\pi/3})~\sigma_{x} ,\\
CX_{3}Z_{3} =& D(e^{i\pi/4}\sqrt{2\pi/3}/2)~\sigma_{z}(-\pi/3)~\ECD(e^{i\pi/4}\sqrt{2\pi/3})~\sigma_{x} ,\\
CX_{3}^{2}Z_{3} =& D(e^{3\pi i/4}\sqrt{2\pi/3}/2)~\sigma_{z}(\pi/3)~\ECD(e^{3\pi i/4}\sqrt{2\pi/3})~\sigma_{x} ,\\
\end{split}
\end{equation}
as described in the Methods section of the main text. In our experiments we do not implement the unconditional displacements $D(\beta_{nm}/2)$ associated with each $CP_{3}$, but here they all commute through the entire circuit so their combined effect can be considered at the end of the two-part circuit. This means that our measurement sequence will apply an overall displacement $D(-\beta_{nm})$ that keeps us in the code space and has no effect on the measurement outcomes.

\begin{figure}[t!]
\includegraphics{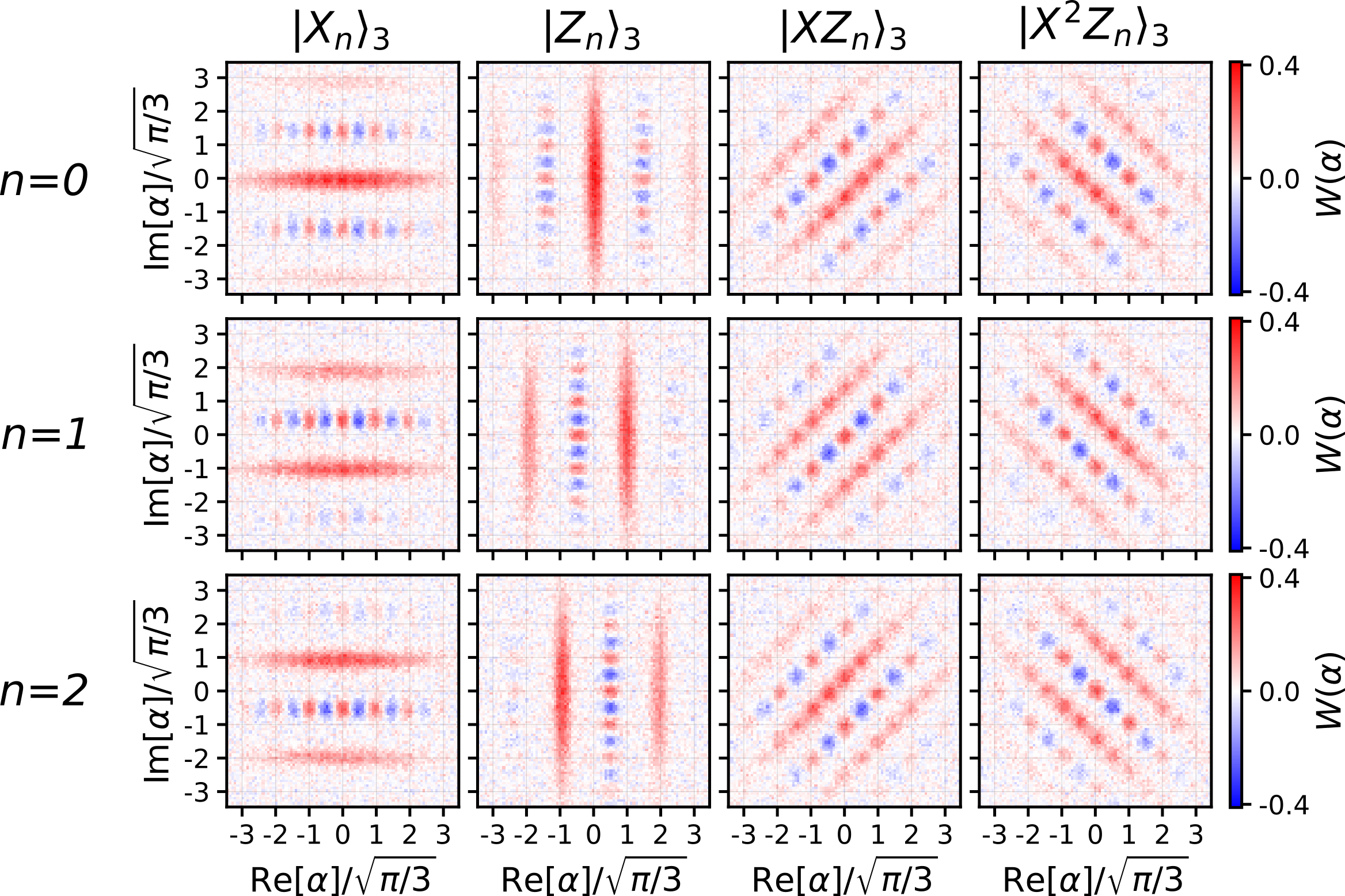}
\caption{
\textbf{GKP qutrit Pauli measurement backaction.} In each column, we implement a projective measurement with respect to the indicated basis and then perform Wigner tomography post-selected on the measurement outcomes.
}\label{supp:fig:gkp3_pauli_msmt}
\end{figure}

To verify that our logical Pauli measurements work as intended, we prepare the maximally mixed state of the GKP qutrit $\rho_{3}^{\mathrm{mix}}$ by performing $200$ rounds of sBs starting from the cavity in vacuum, implement the two-part measurement circuit, and perform Wigner tomography of the cavity post-selected on the outcomes of the two binary measurements. The results of this experiment are shown in Fig. \cref{supp:fig:gkp3_pauli_msmt}, from which we can see that the Pauli measurement works as we expect: it projects onto the Pauli eigenstate $|P_{n}\rangle_{3}$ according to measurement result, maintaining its centered envelope and remaining in the codespace.

\begin{figure}[t!]
\includegraphics{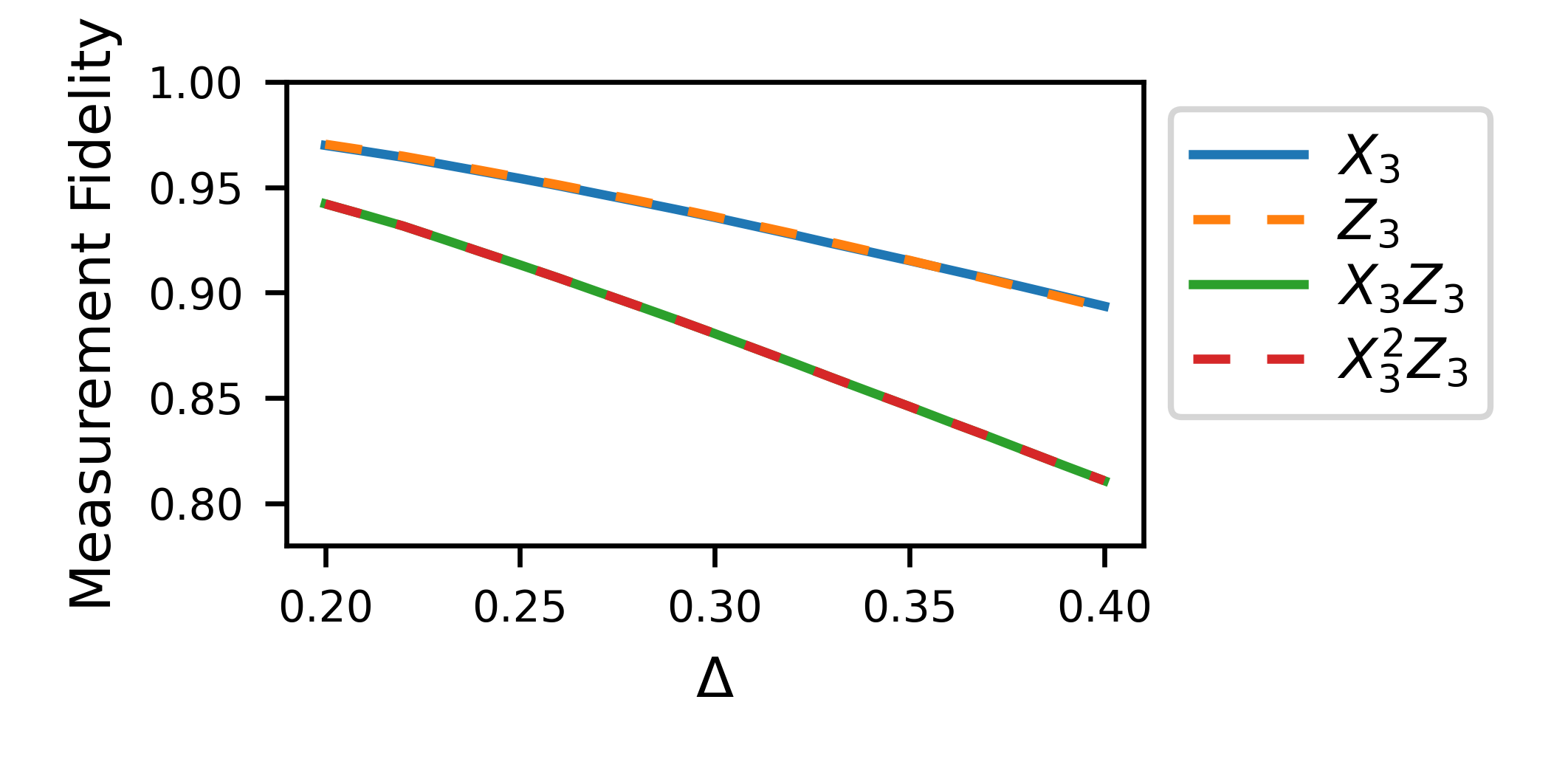}
\vspace{-0.25in}
\caption{
GKP qutrit Pauli measurement fidelity (simulation).
}\label{supp:fig:gkp3_measurement_fidelity}
\end{figure}

We simulate the fidelity of these logical Pauli measurements on the qutrit by preparing eigenstate $|P_{n}\rangle_{3}$ with finite-energy envelope size $\Delta$, performing the Pauli measurement in the basis of $P_{3}$, and finding the probability $p(P_{n} | P_{n})$ that we measure state $|P_{n}\rangle_{3}$ at the end. We quantify the average measurement fidelity by
\begin{equation}
    \mathcal{F}_{P_{3}} = \frac{1}{3}\left[p(P_{0} | P_{0}) + p(P_{1} | P_{1}) + p(P_{2} | P_{2})\right].
\end{equation}
The results of this simulation as a function of $\Delta$ are shown in Fig. \cref{supp:fig:gkp3_measurement_fidelity}. As noted in the main text, these measurements are designed for the ideal code, leading to limited fidelity when applied to finite-energy codewords. We expect it is possible to adapt these measurements to the finite-energy case \cite{royer_stabilization_2020, hastrup_measurement-free_2021, de_neeve_error_2022, rojkov_two-qubit_2024}, but we leave this for future work.

\subsection{Optimized Qutrit Beyond Break-even}

\begin{figure}[t!]
\includegraphics{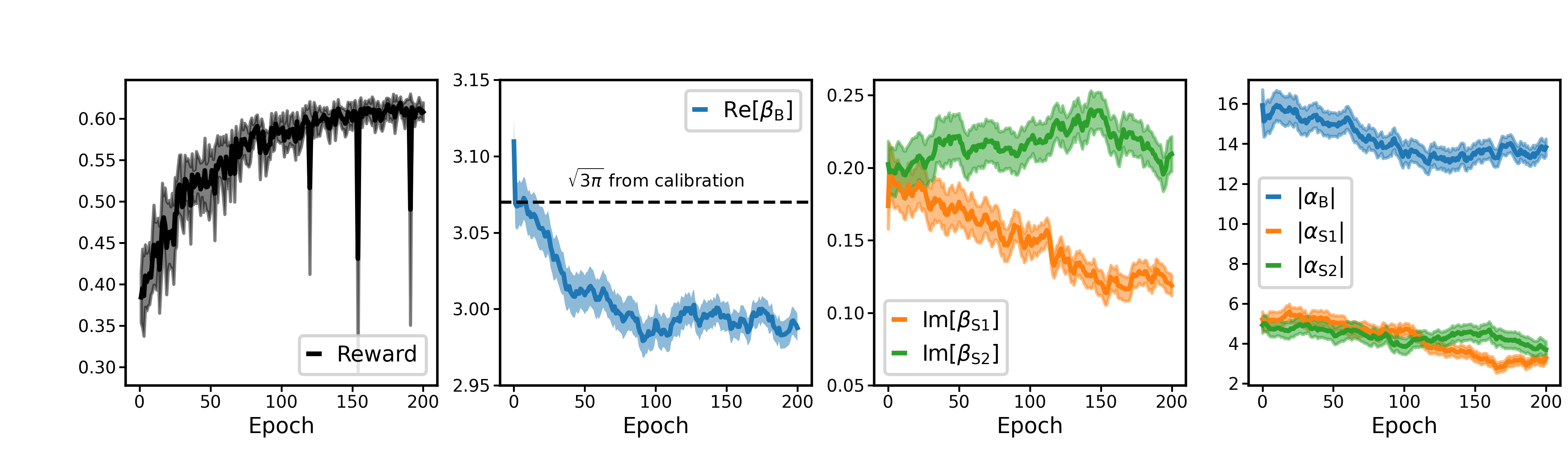}
\caption{
\textbf{GKP qutrit RL training.} Lines indicate mean values while the shaded regions indicate the one-standard-deviation confidence interval. 
}\label{supp:fig:gkp3_rl_training}
\end{figure}

We optimize the QEC protocol of the GKP qutrit using a reinforcement learning (RL) agent \cite{sivak_model-free_2022} following the method described in Ref. \cite{sivak_real-time_2023} (also, see the Methods section of the main text). In Fig. \cref{supp:fig:gkp3_rl_training} we present both the reward signal and key parameters of the sBs circuit as a function of training epoch. For our interpretation of these results, see our discussion in Sec. \cref{supp:sec:gkp2_rl_optimization}. The total duration of each sBs round is constrained to be $7$ $\mathrm{\mu}$s; the optimal protocol found by the RL training spends $3632$ ns measuring and resetting the ancilla qubit, $1808$ ns performing the sBs protocol, $1336$ ns idling, and $224$ ns executing FPGA sequencing instructions.

\begin{figure}[t!]
\includegraphics{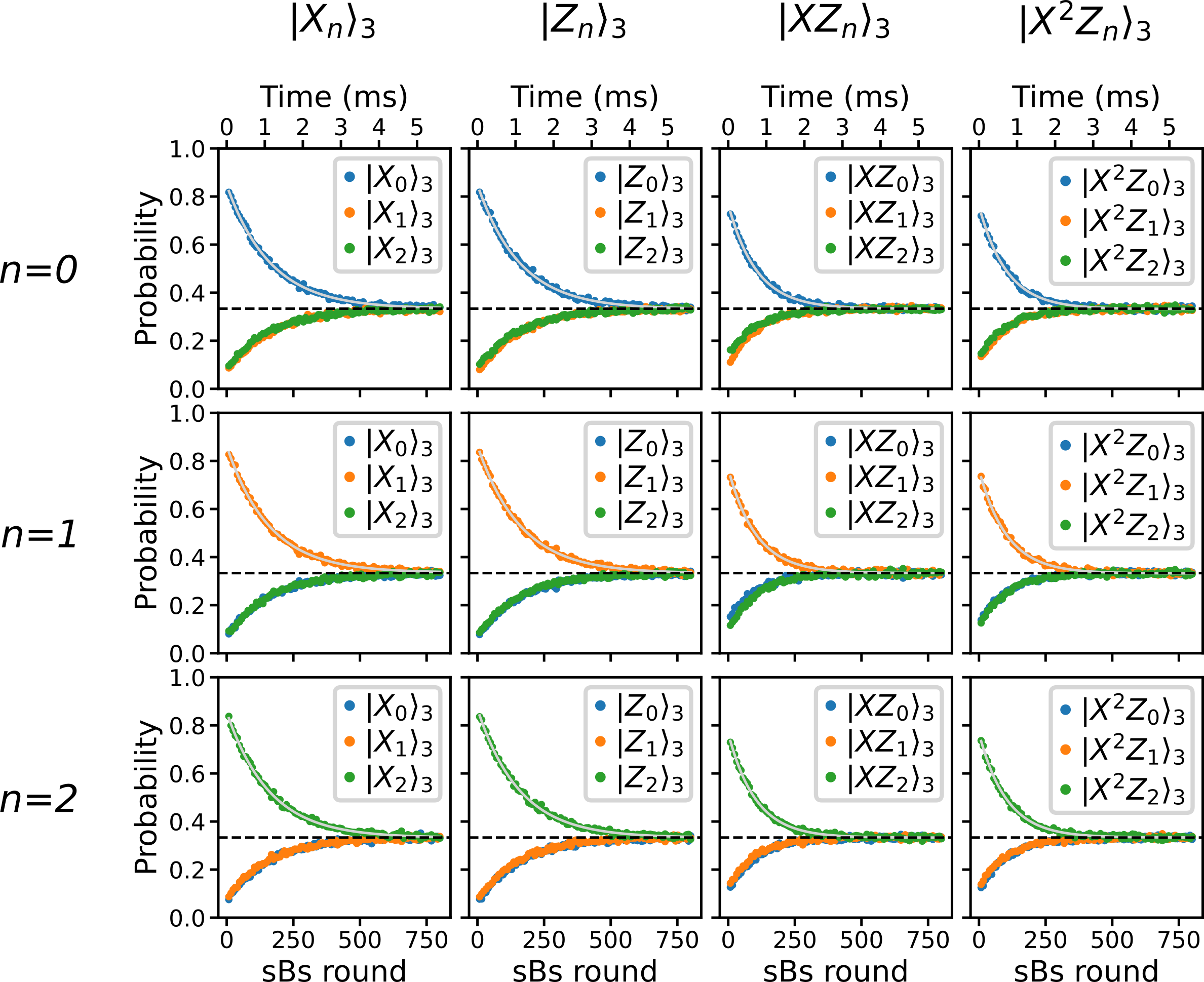}

\bigskip

\setlength\tabcolsep{6pt}
\renewcommand{\arraystretch}{2}
\begin{tabular}{ | c | c | c | c | c |} 
\cline{2-5}
  \multicolumn{1}{c|}{$\gamma_{P_{n}}^{-1}$ ($\mathrm{\mu s}$)} & $|X_{n}\rangle_{3}$ & $|Z_{n}\rangle_{3}$ &  $|XZ_{n}\rangle_{3}$ & $|X^{2}Z_{n}\rangle_{3}$ \\[3pt]
\hline
$n=0$ & $1153\pm 13$ & $1120\pm 15$ & $743\pm 10$ & $727\pm 11$ \\ [3pt]
\hline
$n=1$ & $1117\pm 12$ & $1138\pm 14$ & $737\pm 11$ & $731\pm 11$ \\ [3pt]
\hline
$n=2$ & $1120\pm 14$ & $1107\pm 14$ & $720\pm 10$ & $723\pm 11$ \\ [3pt]
\hline
\end{tabular}
\caption{
\textbf{Lifetime of the optimized GKP qutrit.} The solid grey lines are fits to an exponential decay. The dashed black line indicates a probability of $1/3$.
}\label{supp:fig:gkp3_lifetime}
\end{figure}

To evaluate the average channel fidelity of the optimal QEC protocol via Eq. \eqref{supp:eq:gkp3_avg_channel_fidelity}, we measure the probabilities $\langle P_{n}|\mathcal{E}_{\mathrm{opt}}^{N}(|P_{n}\rangle\langle P_{n}|)|P_{n}\rangle_{3}$ as a function of $N$ for $n=0,1,2$ and $P\in\mathcal{P}_{3} = \{X_{3},Z_{3},X_{3}Z_{3},X_{3}^{2}Z_{3}\}$, where $\mathcal{E}_{\mathrm{opt}}^{N}$ is the channel corresponding to $N$ rounds of the optimal QEC protocol. We measure these probabilities by preparing state $|P_{n}\rangle$, performing $N$ rounds of QEC with the optimal protocol (sweeping the value of $N$), and measuring the GKP qutrit in the basis of Pauli operator $P$. The results of these twelve experiments are presented in Fig. \cref{supp:fig:gkp3_lifetime}. We then fit each probability $\langle P_{n}|\mathcal{E}_{\mathrm{opt}}^{N}(|P_{n}\rangle\langle P_{n}|)|P_{n}\rangle_{3}$ to an exponential decay and find the decay rate $\gamma_{P_{n}}$, which enables us to calculate the effective short-time depolarization rate of the optimized logical GKP qutrit
\begin{equation}
\Gamma_{3}^{\mathrm{GKP}} = \left(886\pm 3 \: \mathrm{\mu s}\right)^{-1},
\end{equation}
using Eq. \eqref{supp:eq:gkp3_eff_depolarization_rate}. Comparing with $\Gamma_{3}^{\mathrm{Fock}} = (488\pm 7 \: \mathrm{\mu s})^{-1}$, we obtain a QEC gain of
\begin{equation}
G_{3} = \Gamma_{3}^{\mathrm{Fock}}/\Gamma_{3}^{\mathrm{GKP}} = 1.82\pm 0.03,
\end{equation}
well beyond the break-even point.

\begin{figure}[t!]
\includegraphics{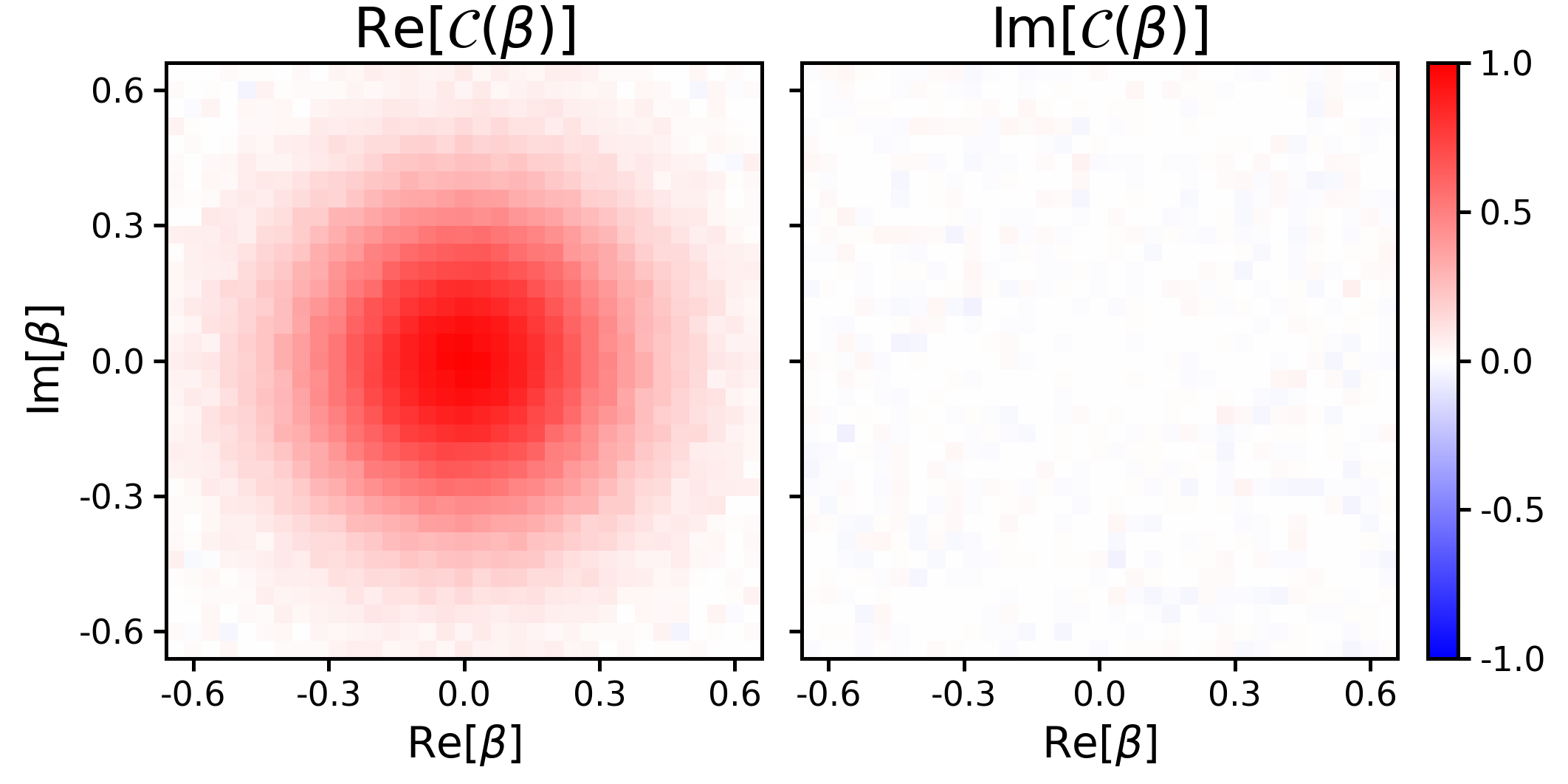}
\caption{
Steady-state characteristic function of the optimal GKP qutrit QEC protocol, sampled near the origin of reciprocal phase space.
}\label{supp:fig:gkp3_steady_state_cf}
\end{figure}

Lastly, we find the effective envelope size of our optimal GKP qutrit by measuring its steady-state characteristic function $\mathcal{C}(\beta)$ near the origin of reciprocal phase space. In this experiment we run the optimal protocol for $400$ rounds followed by characteristic function tomography, the result of which is shown in Fig. \cref{supp:fig:gkp3_steady_state_cf}. From this measurement we find an effective envelope size $\Delta_{\mathrm{eff}} = 0.261\pm 0.001$, corresponding to average photon number $\langle a^{\dagger}a \rangle = (1/\Delta_{\mathrm{eff}}^{2} - 1)/2 = 6.85\pm 0.06$. For more context on this analysis, see our discussion in Sec. \cref{supp:sec:gkp2_rl_optimization}.

\clearpage
\section{Logical GKP Ququart}
In this section we present the methods and results pertaining to our experimental realization of a logical GKP ququart ($d=4$) beyond break-even.

\subsection{State Preparation}
\begin{figure}[p]
\includegraphics{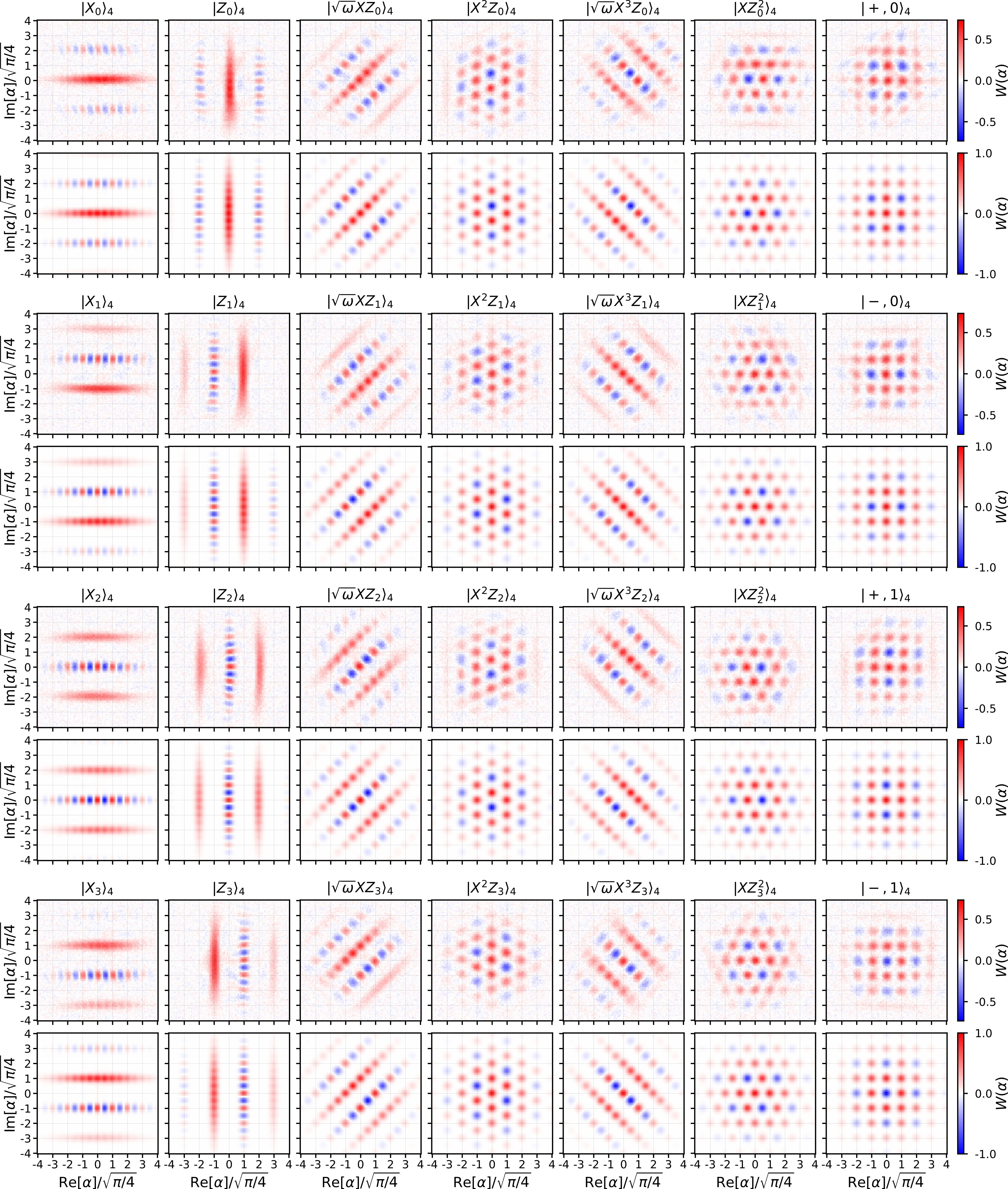}
\caption{
\textbf{GKP ququart state preparation.} For each state, the measured Wigner function of the prepared state is on top, and that of the target state is on the bottom. We use depth $8$ ECD circuits and a finite-energy envelope size $\Delta = 0.32$.
}\label{supp:fig:gkp4_ecd_prep}
\end{figure}

To prepare the Pauli eigenstates and parity states of the GKP ququart defined in Table \cref{supp:tab:qudit_states}, we again use interleaved sequences of ECD gates and ancilla qubit rotations \cite{eickbusch_fast_2022}, as described in Sec. \cref{supp:sec:state_preparation}. We optimize depth $8$ ECD circuits for each state with envelope size $\Delta = 0.32$, approximated numerically using the method described in Sec. \cref{supp:sec:gkp_qudits}. To verify that we are preparing the desired states, we run these ECD circuits experimentally and perform Wigner tomography of the prepared states, the results of which are presented in Fig. \cref{supp:fig:gkp4_ecd_prep}. We then reconstruct the density matrices $\rho_{\mathrm{prep}}$ of the prepared states from their measured Wigner functions and compute their fidelity with respect to the target states $|\psi_{\mathrm{target}}\rangle$ according to $\mathcal{F} = \langle \psi_{\mathrm{target}}|\rho_{\mathrm{prep}} |\psi_{\mathrm{target}}\rangle$, the results of which are presented in Table \cref{supp:tab:gkp4_state_prep_fidelities}.

\begin{center}
\begin{table}[b!]
\setlength\tabcolsep{5pt}
\renewcommand{\arraystretch}{2}
\begin{tabular}{ | c | c | c | c | c | c | c | c | c | c |} 
 \cline{2-7}
  \multicolumn{1}{c|}{$\mathcal{F}_{\mathrm{prep}}$ (\%)} & $|X_{n}\rangle_{4}$ & $|Z_{n}\rangle_{4}$ &  $|\sqrt{\omega}XZ_{n}\rangle_{4}$ & $|X^{2}Z_{n}\rangle_{4}$ & $|\sqrt{\omega}X^{3}Z_{n}\rangle_{4}$ & $|XZ^{2}_{n}\rangle_{4}$ & \multicolumn{1}{c}{} & \multicolumn{2}{c}{$\mathcal{F}_{\mathrm{prep}}$ (\%)} \\[3pt]
\cline{1-7}\cline{9-10}
$n=0$ & $92$ & $89$ & $87$ & $87$ & $87$ & $88$ && $|+,0\rangle_{4}$ & $83$ \\[3pt]
\cline{1-7}\cline{9-10}
$n=1$ & $94$ & $83$ & $89$ & $89$ & $93$ & $91$ && $|-,0\rangle_{4}$ & $90$ \\[3pt]
\cline{1-7}\cline{9-10}
$n=2$ & $93$ & $85$ & $88$ & $85$ & $91$ & $90$ && $|+,1\rangle_{4}$ & $74$ \\[3pt]
\cline{1-7}\cline{9-10}
$n=3$ & $86$ & $92$ & $89$ & $89$ & $86$ & $91$ && $|-,1\rangle_{4}$ & $84$ \\[3pt]
\cline{1-7}\cline{9-10}
\end{tabular}
\caption{Fidelity $\mathcal{F}_{\mathrm{prep}} = \langle \psi_{\mathrm{target}}|\rho_{\mathrm{prep}} |\psi_{\mathrm{target}}\rangle$ of prepared Pauli eigenstates and parity states of the GKP ququart, determined by reconstructing the prepared state from its measured Wigner function and comparing it to the target state.}
\label{supp:tab:gkp4_state_prep_fidelities}
\end{table}
\end{center}

\subsection{Pauli and Parity Measurements}

To measure our GKP ququart in the basis of Pauli operator $P_{4}$ we devised the two-part circuit depicted in Fig. 3b of the main text, which measures the state of our logical four-level system via two binary measurements of our ancilla. This is a simpler measurement than the case of the qutrit, since sequential binary measurements are more natural for a $2^{n}$ dimensional system. If the GKP ququart starts out in state $|\psi\rangle_{4} = c_{0}|P_{0}\rangle_{4} + c_{1}|P_{1}\rangle_{4} + c_{2}|P_{2}\rangle_{4} + c_{3}|P_{3}\rangle_{4}$ and the ancilla starts out in state $|+\rangle$, then the first part of the circuit entangles the two systems as
\begin{equation}
CP_{4}^{2}|+\rangle|\psi\rangle_{4} = |+\rangle\Bigl(c_{0}|P_{0}\rangle_{4} + c_{1}|P_{1}\rangle_{4}\Bigr) + |-\rangle\Bigl(c_{2}|P_{2}\rangle_{4} + c_{3}|P_{3}\rangle_{4}\Bigr),
\end{equation}
such that a subsequent measurement of the ancilla along its $x$ axis constitutes a projective measurement on the ququart distinguishing between the even subspace $\{|P_{0}\rangle_{4},|P_{2}\rangle_{4}\}$ and the odd subspace $\{|P_{1}\rangle_{4},|P_{3}\rangle_{4}\}$. Depending on the measurement result, our post-measurement state is given by one of the two states
\begin{equation}
\begin{split}
|\psi_{\mathrm{even}}\rangle_{4} =& \frac{1}{\sqrt{|c_{0}|^{2} + |c_{2}|^{2}}}\Bigl(c_{0}|P_{0}\rangle_{4} + c_{2}|P_{2}\rangle_{4}\Bigr) ,\\
|\psi_{\mathrm{odd}}\rangle_{4} =& \frac{1}{\sqrt{|c_{1}|^{2} + |c_{3}|^{2}}}\Bigl(c_{1}|P_{1}\rangle_{4} + c_{3}|P_{3}\rangle_{4}\Bigr) .
\end{split}
\end{equation}
Starting from these states, and the ancilla reinitialized in the $|+\rangle$ state, the second part of the measurement circuit entangles the two systems according to
\begin{equation}
\begin{split}
CP_{4}|+\rangle|\psi_{\mathrm{even}}\rangle_{4} =& \frac{1}{\sqrt{|c_{0}|^{2} + |c_{2}|^{2}}}\Bigl(c_{0}|+\rangle|P_{0}\rangle_{4} + c_{2}|-\rangle|P_{2}\rangle_{4}\Bigr),\\
CP_{4}|+\rangle|\psi_{\mathrm{odd}}\rangle_{4} =& \frac{1}{\sqrt{|c_{1}|^{2} + |c_{3}|^{2}}}\Bigl(c_{1}|+i\rangle|P_{1}\rangle_{4} + c_{3}|-i\rangle|P_{3}\rangle_{4}\Bigr),
\end{split}
\end{equation}
where $|\pm i\rangle = (|g\rangle \pm i|e\rangle)/\sqrt{2}$ are the $\sigma_{y}$ eigenstates of the ancilla. If we are in the even subspace, we perform a projective measurement distinguishing between the remaining two states by measuring the ancilla along its $x$-axis, whereas if we are in the odd subspace we do so by measuring the ancilla along its $y$-axis. All together, this measurement will project the ququart onto one of the eigenstates $|P_{n}\rangle_{4}$ with probability $|c_{n}|^{2}$.

For $P_{4}\in\{X_{4}, Z_{4}, \sqrt{\omega}X_{4}Z_{4}, X_{4}^{2}Z_{4}, \sqrt{\omega}X_{4}^{3}Z_{4}, X_{4}Z_{4}^{2}\}$, we compile the controlled Pauli gates $CP_{4}$ according to
\begin{equation}
\begin{split}
CX_{4} =& D(\sqrt{\pi/4}/2)~\ECD(\sqrt{\pi/4})~\sigma_{x} ,\\
CZ_{4} =& D(i\sqrt{\pi/4}/2)~\ECD(i\sqrt{\pi/4})~\sigma_{x} ,\\
C(\sqrt{\omega}X_{4}Z_{4}) =& D(e^{i\pi/4}\sqrt{\pi/2}/2)~\ECD(e^{i\pi/4}\sqrt{\pi/2})~\sigma_{x} ,\\
CX_{4}^{2}Z_{4} =& D((2+i)\sqrt{\pi/4}/2)~\sigma_{z}(-\pi/2)~\ECD((2+i)\sqrt{\pi/4})~\sigma_{x} ,\\
C(\sqrt{\omega}X_{4}^{3}Z_{4}) =& D(e^{3\pi i/4}\sqrt{\pi/2}/2)~\sigma_{z}(\pi/2)~\ECD(e^{3\pi i/4}\sqrt{\pi/2})~\sigma_{x} ,\\
CX_{4}Z_{4}^{2} =& D((1+2i)\sqrt{\pi/4}/2)~\sigma_{z}(-\pi/2)~\ECD((1+2i)\sqrt{\pi/4})~\sigma_{x} ,\\
\end{split}
\end{equation}
as described in the Methods section of the main text. In our experiments we do not implement the unconditional displacements $D(\beta_{nm}/2)$ associated with each $CP_{4}$, but here they all commute through the entire circuit so their combined effect can be considered at the end of the two-part circuit. This means that our Pauli measurement sequence will apply an overall displacement $D(-\beta_{nm}/2)$ that brings us out of the code space (ignoring the additional effect of $D(-\beta_{nm})$ from the first part of the circuit, which applies an overall phase and keeps us in the code space). However, the envelope of our finite-energy code remains centered since our conditional displacements are symmetric about the origin. 

To measure our GKP ququart in the parity basis $\{|s,n\rangle_{4} ; s=\pm 1, n=0,1\}$, defined to be the simultaneous eigenstates of $X_{4}^{2}$ and $Z_{4}^{2}$ according to $X_{4}^{2}|s,n\rangle_{4} = s|s,n\rangle_{4}$ and $Z_{4}^{2}|s,n\rangle_{4} = (-1)^{n}|s,n\rangle_{4}$, we devised the circuit depicted in Fig. 3d of the main text. This circuit sequentially measures the eigenvalues of $X_{4}^{2}$ and $Z_{4}^{2}$ via two binary measurements of our ancilla. If the GKP ququart starts out in state $|\psi\rangle_{4} = c_{+,0}|+,0\rangle_{4} + c_{-,0}|-,0\rangle_{4} + c_{+,1}|+,1\rangle_{4} + c_{-,1}|-,1\rangle_{4}$ and the ancilla starts out in state $|+\rangle$, then the first part of the circuit entangles the two systems as
\begin{equation}
CX_{4}^{2}|+\rangle|\psi\rangle_{4} = |+\rangle\Bigl(c_{+,0}|+,0\rangle_{4} + c_{+,1}|+,1\rangle_{4}\Bigr) + |-\rangle\Bigl(c_{-,0}|-,0\rangle_{4} +  + c_{-,1}|-,1\rangle_{4}\Bigr),
\end{equation}
such that a subsequent measurement of the ancilla along its $x$ axis constitutes a projective measurement on the ququart distinguishing between the $\pm 1$ eigenspaces of $X_{4}^{2}$. Similarly, the second part of the circuit entangles the two systems as
\begin{equation}
CZ_{4}^{2}|+\rangle|\psi\rangle_{4} = |+\rangle\Bigl(c_{+,0}|+,0\rangle_{4} + c_{-,0}|-,0\rangle_{4}\Bigr) + |-\rangle\Bigl(c_{+,1}|+,1\rangle_{4} +  + c_{-,1}|-,1\rangle_{4}\Bigr),
\end{equation}
such that a subsequent measurement of the ancilla along its $x$ axis constitutes a projective measurement on the ququart distinguishing between the $\pm 1$ eigenspaces of $Z_{4}^{2}$. Note that when we compile $CP_{4}^{2}$ gates in terms of $\ECD$ gates, omitting the unconditional displacement means that we apply an additional $P_{4}$ operation, which keeps us in the code space. However, unlike our previous examples of Pauli measurements, in this case the $X_{4}$ operation applied by the first part of the parity measurement circuit does not commute through the second part of the circuit. This must be considered when decoding measurement results and determining the measurement backaction, but the analysis for taking this into account is straightforward.

\begin{figure}[t!]
\includegraphics{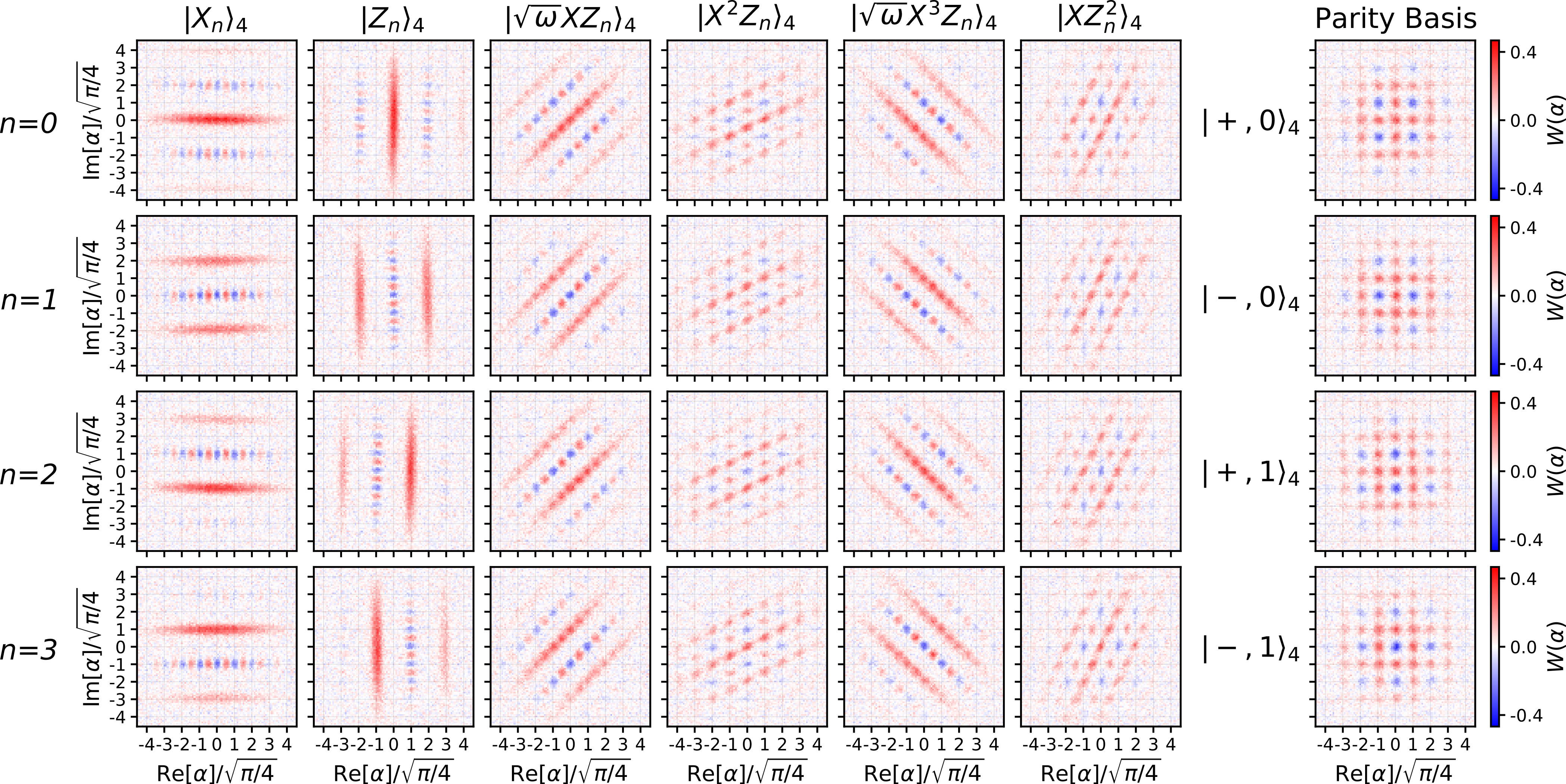}
\caption{
\textbf{GKP ququart Pauli and parity measurement backaction.} In each column, we implement a projective measurement with respect to the indicated basis and then perform Wigner tomography post-selected on the measurement outcomes.
}\label{supp:fig:gkp4_msmts}
\end{figure}

To verify that our logical ququart Pauli and parity measurements work as intended, we prepare the maximally mixed state of the GKP ququart $\rho_{4}^{\mathrm{mix}}$ by performing $200$ rounds of sBs starting from the cavity in vacuum, implement the two-part measurement circuit, and perform Wigner tomography of the cavity post-selected on the outcomes of the two binary measurements. The results of this experiment are shown in Fig. \cref{supp:fig:gkp4_msmts}, from which we can see that the measurements work as expected.

\begin{figure}[t!]
\includegraphics{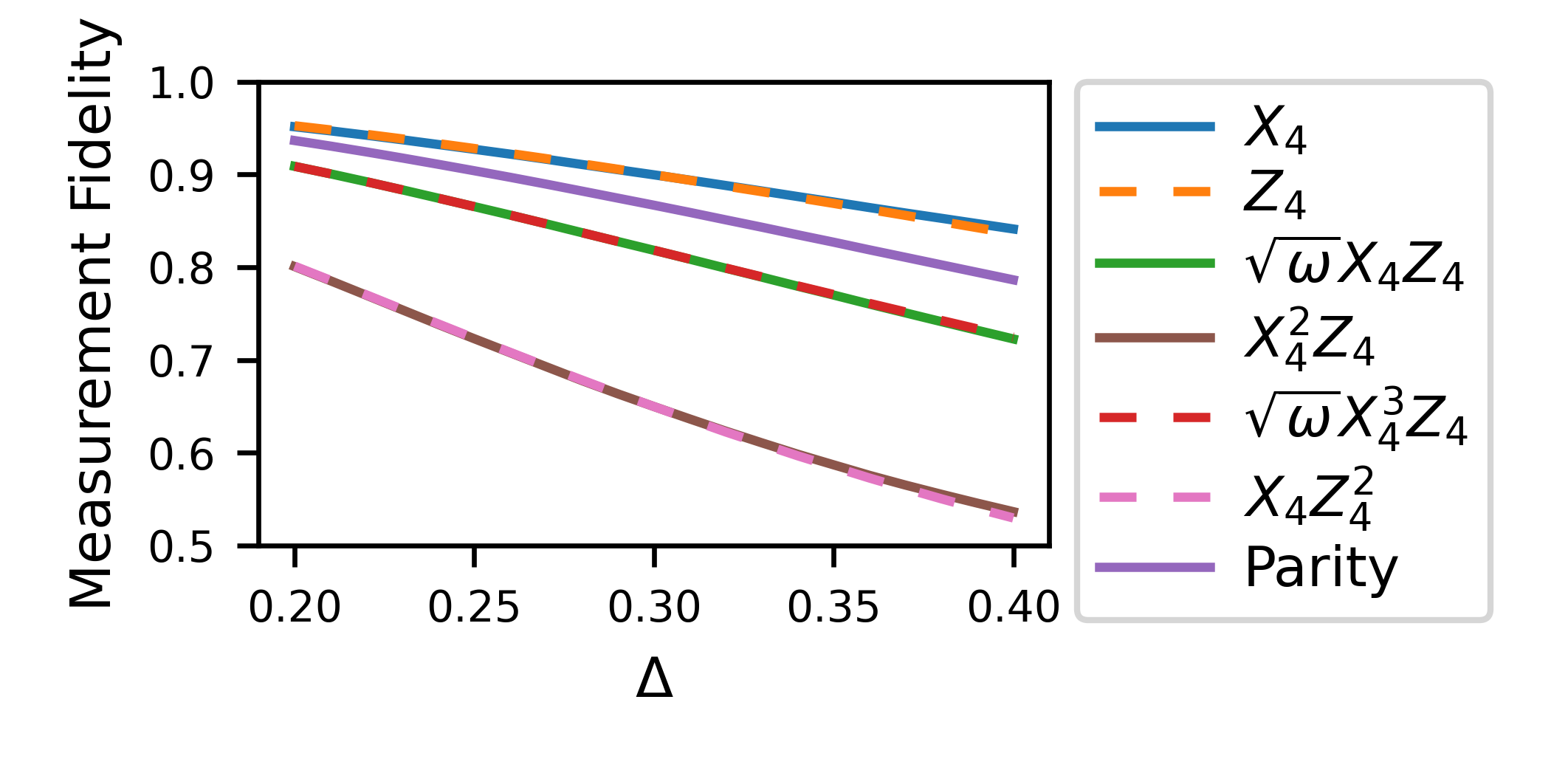}
\vspace{-0.25in}
\caption{
GKP ququart Pauli measurement fidelity (simulation)
}\label{supp:fig:gkp4_measurement_fidelity}
\end{figure}

We simulate the fidelity of these logical measurements on the ququart by preparing Pauli eigenstate $|P_{n}\rangle_{4}$ or parity state $|\pm,m\rangle$ with finite-energy envelope size $\Delta$, performing the corresponding logical measurement, and finding the probability $p(P_{n} | P_{n})$ or $p(\pm,m | \pm,m)$ that we measure the state correctly at the end. We quantify the average measurement fidelity by
\begin{equation}
\begin{split}
\mathcal{F}_{P_{4}} &= \frac{1}{4}\Bigl[p(P_{0} | P_{0}) + p(P_{1} | P_{1}) + p(P_{2} | P_{2}) + + p(P_{3} | P_{3})\Bigr], \\
\mathcal{F}_{\mathrm{Parity}} &= \frac{1}{4}\Bigl[p(+,0 | +,0) + p(-,0 | -,0)  + p(+,1 | +,1) + p(-,1 | -,1) \Bigr].
\end{split}
\end{equation}
The results of this simulation as a function of $\Delta$ are shown in Fig. \cref{supp:fig:gkp4_measurement_fidelity}. As noted in the main text, these measurements are designed for the ideal code, leading to limited fidelity when applied to finite-energy codewords. We expect it is possible to adapt these measurements to the finite-energy case \cite{royer_stabilization_2020, hastrup_measurement-free_2021, de_neeve_error_2022, rojkov_two-qubit_2024}, but we leave this for future work.

\subsection{Optimized Ququart Beyond Break-even}

\begin{figure}[t!]
\includegraphics{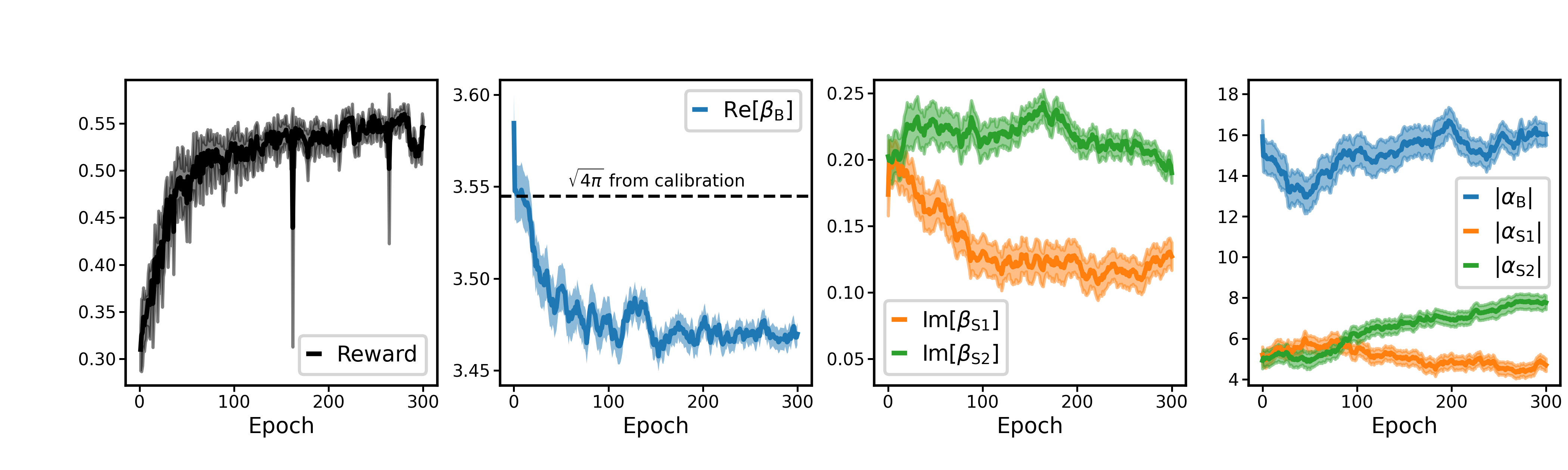}
\caption{
\textbf{GKP ququart RL training.} Lines indicate mean values while the shaded regions indicate the one-standard-deviation confidence interval. 
}\label{supp:fig:gkp4_rl_training}
\end{figure}

We optimize the QEC protocol of the GKP ququart using a reinforcement learning (RL) agent \cite{sivak_model-free_2022} following the method described in Ref. \cite{sivak_real-time_2023} (also, see the Methods section of the main text). In Fig. \cref{supp:fig:gkp4_rl_training} we present both the reward signal and key parameters of the sBs circuit as a function of training epoch. For our interpretation of these results, see our discussion in Sec. \cref{supp:sec:gkp2_rl_optimization}. The total duration of each sBs round is constrained to be $7$ $\mathrm{\mu}$s; the optimal protocol found by the RL training spends $3632$ ns measuring and resetting the ancilla qubit, $1648$ ns performing the sBs protocol, $1496$ ns idling, and $224$ ns executing FPGA sequencing instructions.

\begin{figure}[h!]
\includegraphics{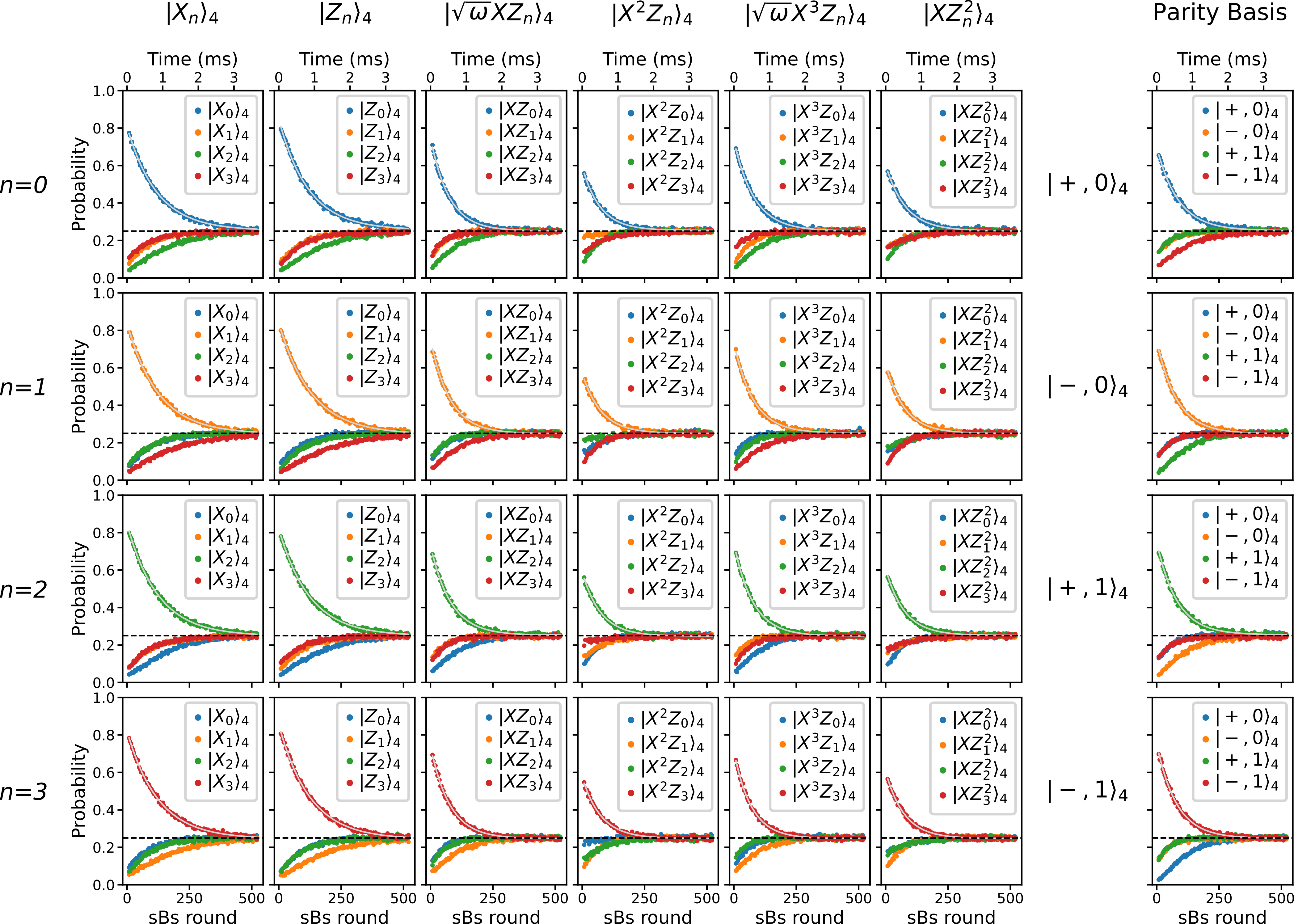}

\bigskip

\setlength\tabcolsep{6pt}
\renewcommand{\arraystretch}{2}
\begin{tabular}{ | c | c | c | c | c | c | c | c | c | c |} 
 \cline{2-7}
  \multicolumn{1}{c|}{$\gamma_{P_{n}}^{-1}$ ($\mathrm{\mu s}$)} & $|X_{n}\rangle_{4}$ & $|Z_{n}\rangle_{4}$ &  $|\sqrt{\omega}XZ_{n}\rangle_{4}$ & $|X^{2}Z_{n}\rangle_{4}$ & $|\sqrt{\omega}X^{3}Z_{n}\rangle_{4}$ & $|XZ^{2}_{n}\rangle_{4}$ & \multicolumn{1}{c}{} & \multicolumn{2}{c}{$\gamma_{\pm,m}^{-1}$ ($\mathrm{\mu s}$)} \\[3pt]
\cline{1-7}\cline{9-10}
$n=0$ & $840\pm 8$ & $836\pm 9$ & $519\pm 6$ & $507\pm 9$ & $571\pm 7$ & $562\pm 9$ && $|+,0\rangle_{4}$ & $607\pm 8$ \\[3pt]
\cline{1-7}\cline{9-10}
$n=1$ & $878\pm 10$ & $918\pm 9$ & $549\pm 7$ & $536\pm 11$ & $548\pm 7$ & $528\pm 9$ && $|-,0\rangle_{4}$ & $565\pm 6$ \\[3pt]
\cline{1-7}\cline{9-10}
$n=2$ & $867\pm 10$ & $867\pm 9$ & $570\pm 7$ & $520\pm 9$ & $531\pm 7$ & $525\pm 8$ && $|+,1\rangle_{4}$ & $568\pm 7$ \\[3pt]
\cline{1-7}\cline{9-10}
$n=3$ & $871\pm 9$ & $872\pm 9$ & $541\pm 6$ & $529\pm 10$ & $488\pm 6$ & $521\pm 9$ && $|-,1\rangle_{4}$ & $559\pm 6$ \\[3pt]
\cline{1-7}\cline{9-10}
\end{tabular}
\caption{
\textbf{Lifetime of the optimized GKP ququart.} The solid grey lines are fits to an exponential decay. The dashed black line indicates a probability of $1/4$.
}\label{supp:fig:gkp4_lifetime}
\end{figure}

To evaluate the average channel fidelity of the optimal QEC protocol via Eq. \eqref{supp:eq:gkp4_avg_channel_fidelity}, we measure the probabilities $\langle P_{n}|\mathcal{E}_{\mathrm{opt}}^{N}(|P_{n}\rangle\langle P_{n}|)|P_{n}\rangle_{4}$ as a function of $N$ for $n=0,1,2,3$ and $P\in\mathcal{P}_{4} = \{X_{4},Z_{4},\sqrt{\omega}X_{4}Z_{4},X_{4}^{2}Z_{4},\sqrt{\omega}X_{4}^{3}Z_{4},X_{4}Z_{4}^{2}\}$, and $\langle s,n|\mathcal{E}_{\mathrm{opt}}^{N}(|s,n\rangle\langle s,n|)|s,n\rangle_{4}$ as a function of $N$ for $s=\pm 1$ and $n=0,1$, where $\mathcal{E}_{\mathrm{opt}}^{N}$ is the channel corresponding to $N$ rounds of the optimal QEC protocol. We measure these probabilities by preparing Pauli eigenstate $|P_{n}\rangle$ (or parity state $|s,n\rangle_{4}$), performing $N$ rounds of QEC with the optimal protocol, and measuring the GKP ququart in the basis of Pauli operator $P$ (or the ququart parity basis). The results of these twenty-eight experiments are presented in Fig. \cref{supp:fig:gkp4_lifetime}. We then fit probabilities $\langle P_{n}|\mathcal{E}_{\mathrm{opt}}^{N}(|P_{n}\rangle\langle P_{n}|)|P_{n}\rangle_{4}$ and $\langle s,n|\mathcal{E}_{\mathrm{opt}}^{N}(|s,n\rangle\langle s,n|)|s,n\rangle_{4}$ to exponential decays and find the decay rates $\gamma_{P_{n}}$ and $\gamma_{s,n}$, which enable us to calculate the effective short-time depolarization rate of the optimized logical GKP ququart 
\begin{equation}
\Gamma_{4}^{\mathrm{GKP}} = \left(620\pm 2 \: \mathrm{\mu s}\right)^{-1},
\end{equation}
using Eq. \eqref{supp:eq:gkp4_eff_depolarization_rate}. Comparing with $\Gamma_{4}^{\mathrm{Fock}} = (332\pm 6 \: \mathrm{\mu s})^{-1}$, we obtain a QEC gain of
\begin{equation}
G_{4} = \Gamma_{4}^{\mathrm{Fock}}/\Gamma_{4}^{\mathrm{GKP}} = 1.87\pm 0.03,
\end{equation}
well beyond the break-even point.

\begin{figure}[t!]
\includegraphics{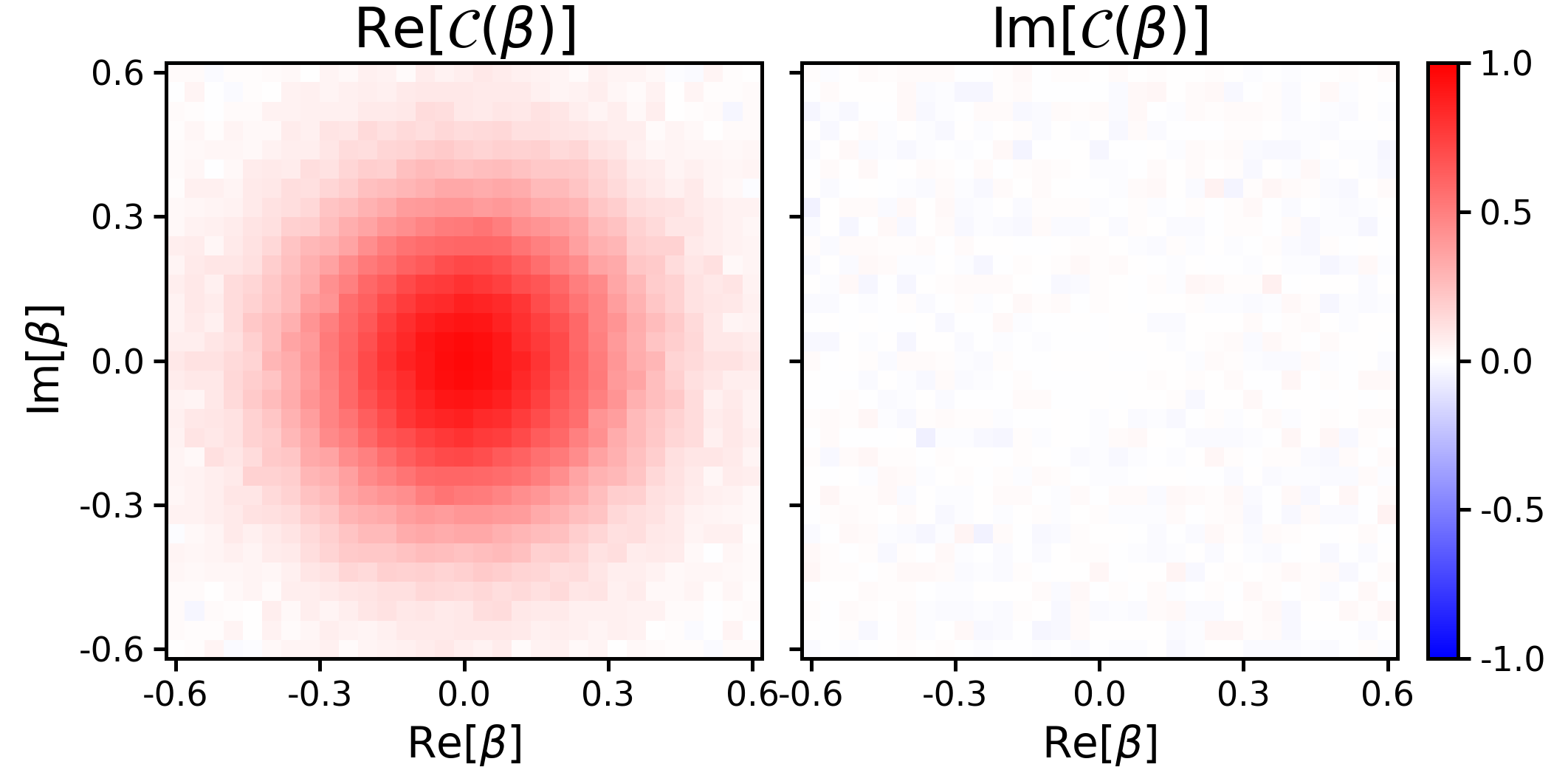}
\caption{
Steady-state characteristic function of the optimal GKP ququart QEC protocol, sampled near the origin of reciprocal phase space.
}\label{supp:fig:gkp4_steady_state_cf}
\end{figure}

Lastly, we find the effective envelope size of our optimal GKP ququart by measuring its steady-state characteristic function $\mathcal{C}(\beta)$ near the origin of reciprocal phase space. In this experiment we run the optimal protocol for $300$ rounds followed by characteristic function tomography, the result of which is shown in Fig. \cref{supp:fig:gkp4_steady_state_cf}. From this measurement we find an effective envelope size $\Delta_{\mathrm{eff}} = 0.246\pm 0.001$, corresponding to average photon number $\langle a^{\dagger}a \rangle = (1/\Delta_{\mathrm{eff}}^{2} - 1)/2 = 7.76\pm 0.07$. For more context on this analysis, see our discussion in Sec. \cref{supp:sec:gkp2_rl_optimization}.

\clearpage

\section{Sources of Logical Errors}

In this section we report the results of simulations aimed at estimating the relative contributions of different physical error sources to the effective decay rates $\Gamma_{d}^{\mathrm{GKP}}$ of our logical GKP qubit, qutrit, and ququart. We consider a simplified error model, consisting of transmon bit-flip errors during sBs at rate $\kappa_{1,q} = 1/T_{1,q}$ (i.e., combined heating and decay), transmon dephasing errors during sBs at rate $\kappa_{\phi,q} = 1/T_{2E,q} - 1/2T_{1,q}$, cavity photon loss during sBs at rate $\kappa_{1,c} = 1/T_{1,c}$, cavity photon loss during idling time at rate $\kappa_{1,c}$, and cavity dephasing during idling time at rate $\kappa_{\phi,c} = n_{\mathrm{th}}/T_{1,q}$. With this error model, we assume cavity dephasing primarily occurs during idling time, and can be fully attributed to thermal fluctuations of the transmon \cite{reagor_quantum_2016}.

We simulate a simplified version of our optimal QEC protocols, where each $\ECD(\beta=|\beta|e^{i\theta_{\beta}})$ gate in sBs is realized by evolving under the Hamiltonian
\begin{equation}
H_{CD} = \frac{i}{2}\chi \alpha\left(e^{i\theta_{\beta}} a^{\dagger} - e^{-i\theta_{\beta}} a\right)\sigma_{z},
\end{equation}
for time $t=|\beta|/\chi\alpha$ followed by an ancilla $\sigma_{x}$ gate, rather than performing a pulse-level simulation of the compiled ECD gate. For each ECD gate in sBs we use the $\alpha_{S1,B,S2}$ found by the reinforcement learning agent, and use $\beta_{S1,B,S2}$ scaled by a factor of $\sqrt{\pi d}/|\beta_{B}|$, since we assume the deviation we find between $\sqrt{\pi d}$ and $\beta_{B}$ is due to miscalibration. During idling time, we evolve under the identity.

To isolate the relative contributions of each type of physical error, we simulate our optimal QEC protocols in the presence of each individual source of error on its own. As a basis of comparison, we then simulate in the presence of all errors. Finally, to estimate the extent to which different errors compound with one another, we simulate in the presence of all errors occurring during sBs, and in the presence of all errors occurring during idling. We perform all of our simulations using a combination of Dynamiqs \cite{dynamiqs} and QuTiP \cite{qutip}. The results of our simulations for the optimal GKP qubit are presented in Table \cref{supp:tab:gkp2_error_budgeting}, those for the optimal GKP qutrit are presented in Table \cref{supp:tab:gkp3_error_budgeting}, and those for the optimal GKP ququart are presented in Table \cref{supp:tab:gkp4_error_budgeting}. We emphasize that with all of the simplifications and assumptions made in performing this simulation, these results are not meant to be directly compared with our experimental measurements. Rather, they should be viewed heuristically, providing a basis for analyzing the limiting factors of our QEC protocol and how they change with qudit dimension $d$.

We find that the most significant sources of logical errors are transmon bit-flips during sBs, cavity photon loss during idling, and cavity dephasing during idling. The rate of logical errors caused by transmon bit-flips remains fairly constant as we increase $d$ from $2$ to $4$, limiting the lifetime to about $6.5$ ms in $d=2$ and $d=3$, and to about $6$ ms in $d=4$. However, since the overall lifetime $\Gamma_{d}^{\mathrm{GKP}^{-1}}$ decreases as we increase $d$, the relative contribution of transmon bit-flips to the total logical error rate decreases. On the other hand, the rate of logical errors caused by cavity photon loss during idling increases as we go from $d=2$ to $4$, as does the relative contribution of these errors to the total logical error rate. As we argue in the main text, this is due to having a smaller $\Delta$ as we increase $d$, which means our codewords have more energy and lose photons more quickly as a result. We find that cavity dephasing during idling is the dominant source of error for $d=2,3$ and $4$, and its relative contribution increases as we increase $d$. For $d=4$, it is responsible for about half of the logical error rate. As discussed in the main text, we attribute this to the larger lattice spacing of GKP qudits, which means the codewords contain information further out in phase space, making them more susceptible to cavity dephasing errors. We find that all of these errors compound with one another, leading to larger logical error rates when combined than on their own. Interestingly, this is also true for all errors during sBs and all errors during idling; when these error rates are added, they only account for about $90\%$ of the total logical error rate in $d=2,3,$ and $4$ obtained when both are combined in the same simulation.

\begin{center}
\begin{table}[h!]
\setlength\tabcolsep{6pt}
\renewcommand{\arraystretch}{2}
\begin{tabular}{| c | c | c | c | c | c |} 
\cline{2-5} 
\multicolumn{1}{c|}{Error Type} & $\Gamma_{X}^{-1}\:(\mathrm{ms})$ & $\Gamma_{Z}^{-1}\:(\mathrm{ms})$ &  $\Gamma_{XZ}^{-1}\:(\mathrm{ms})$ & $\Gamma_{2}^{\mathrm{GKP}^{-1}}\:(\mathrm{ms})$ & \multicolumn{1}{c}{\%} \\[3pt]
\hline
All errors & $2.15$ & $2.15$ & $1.09$ & $1.62$ & $100$\\ [3pt]
\hline
Transmon $\kappa_{1,q}$ during sBs & $8.69$ & $8.69$ & $4.35$ & $6.52$ & $25$ \\ [3pt]
\hline
Transmon $\kappa_{\phi,q}$ during sBs & $3230$ & $3000$ & $1550$ & $2330$ & $0.07$ \\ [3pt]
\hline
Cavity $\kappa_{1,c}$ during sBs & $1210$ & $1890$ & $720$ & $1100$ & $0.15$ \\ [3pt]
\hline
All errors during sBs & $8.60$ & $8.60$ & $4.31$ & $6.46$ & $25$ \\ [3pt]
\hline
Cavity $\kappa_{1,c}$ during idling & $50.3$ & $50.3$ & $25.2$ & $37.8$ & $4.3$ \\ [3pt]
\hline
Cavity $\kappa_{\phi,c}$ during idling & $6.74$ & $6.74$ & $3.41$ & $5.08$ & $32$ \\ [3pt]
\hline
All errors during idling & $3.40$ & $3.40$ & $1.72$ & $2.56$ & $63$ \\ [3pt]
\hline
\end{tabular}
\caption{Error budgeting of the optimal GKP qubit. For each Pauli basis $P_{2}$, we only simulate the $|P_{0}\rangle_{2}$ state.}
\label{supp:tab:gkp2_error_budgeting}
\end{table}
\end{center}

\begin{center}
\begin{table}[h!]
\setlength\tabcolsep{6pt}
\renewcommand{\arraystretch}{2}
\begin{tabular}{| c | c | c | c | c | c | c |} 
\cline{2-6} 
\multicolumn{1}{c|}{Error Type} & $\Gamma_{X}^{-1}\:(\mathrm{ms})$ & $\Gamma_{Z}^{-1}\:(\mathrm{ms})$ &  $\Gamma_{XZ}^{-1}\:(\mathrm{ms})$ & $\Gamma_{X^{2}Z}^{-1}\:(\mathrm{ms})$ & $\Gamma_{3}^{\mathrm{GKP}^{-1}}\:(\mathrm{ms})$ & \multicolumn{1}{c}{\%} \\[3pt]
\hline
All errors & $1.08$ & $1.08$ & $0.55$ & $0.55$ & $0.73$ & $100$\\ [3pt]
\hline
Transmon $\kappa_{1,q}$ during sBs & $9.73$ & $9.67$ & $4.87$ & $4.86$ & $6.48$ & $11$ \\ [3pt]
\hline
Transmon $\kappa_{\phi,q}$ during sBs & $860$ & $1180$ & $660$ & $620$ & $780$ & $0.09$ \\ [3pt]
\hline
Cavity $\kappa_{1,c}$ during sBs & $370$ & $340$ & $180$ & $160$ & $230$ & $0.3$ \\ [3pt]
\hline
All errors during sBs & $9.42$ & $9.41$ & $4.73$ & $4.74$ & $6.3$ & $12$ \\ [3pt]
\hline
Cavity $\kappa_{1,c}$ during idling & $13.8$ & $13.8$ & $6.91$ & $6.91$ & $9.2$ & $7.9$ \\ [3pt]
\hline
Cavity $\kappa_{\phi,c}$ during idling & $2.42$ & $2.42$ & $1.22$ & $1.22$ & $1.62$ & $45$ \\ [3pt]
\hline
All errors during idling & $1.40$ & $1.40$ & $0.71$ & $0.71$ & $0.94$ & $78$ \\ [3pt]
\hline
\end{tabular}
\caption{Error budgeting of the optimal GKP qutrit. For each Pauli basis $P_{3}$, we only simulate the $|P_{0}\rangle_{3}$ state.}
\label{supp:tab:gkp3_error_budgeting}
\end{table}
\end{center}

\begin{center}
\begin{table}[h!]
\setlength\tabcolsep{6pt}
\renewcommand{\arraystretch}{2}
\begin{tabular}{| c | c | c | c | c | c | c | c | c | c |} 
\cline{2-9} 
\multicolumn{1}{c|}{Error Type} & \subcell{$\Gamma_{X}^{-1}$}{(ms)} & \subcell{$\Gamma_{Z}^{-1}$}{(ms)} &  \subcell{$\Gamma_{XZ}^{-1}$}{(ms)} & \subcell{$\Gamma_{X^{2}Z}^{-1}$}{(ms)} & \subcell{$\Gamma_{X^{3}Z}^{-1}$}{(ms)} & \subcell{$\Gamma_{XZ^{2}}^{-1}$}{(ms)} & \subcell{$\Gamma_{+,0}^{-1}$}{(ms)} & \subcell{$\Gamma_{4}^{\mathrm{GKP}^{-1}}$}{(ms)} & \multicolumn{1}{c}{\%} \\[3pt]
\hline
All errors & $0.73$ & $0.73$ & $0.39$ & $0.34$ & $0.39$ & $0.34$ & $0.39$ & $0.45$ & $100$\\ [3pt]
\hline
Transmon $\kappa_{1,q}$ during sBs & $9.63$ & $9.64$ & $4.85$ & $5.64$ & $4.85$ & $5.64$ & $7.38$ & $5.96$ & $7.6$ \\ [3pt]
\hline
Transmon $\kappa_{\phi,q}$ during sBs & $590$ & $580$ & $270$ & $250$ & $260$ & $260$ & $300$ & $320$ & $0.14$ \\ [3pt]
\hline
Cavity $\kappa_{1,c}$ during sBs & $190$ & $190$ & $98$ & $95$ & $97$ & $92$ & $100$ & $120$ & $0.38$ \\ [3pt]
\hline
All errors during sBs & $9.22$ & $9.21$ & $4.66$ & $5.37$ & $4.66$ & $5.37$ & $6.93$ & $5.72$ & $7.9$ \\ [3pt]
\hline
Cavity $\kappa_{1,c}$ during idling & $6.35$ & $6.35$ & $3.25$ & $3.05$ & $3.24$ & $3.06$ & $3.24$ & $3.91$ & $12$ \\ [3pt]
\hline
Cavity $\kappa_{\phi,c}$ during idling & $1.42$ & $1.41$ & $0.75$ & $0.67$ & $0.75$ & $0.67$ & $0.73$ & $0.88$ & $51$ \\ [3pt]
\hline
All errors during idling & $0.87$ & $0.87$ & $0.47$ & $0.40$ & $0.47$ & $0.40$ & $0.45$ & $0.54$ & $83$ \\ [3pt]
\hline
\end{tabular}
\caption{Error budgeting of the optimal GKP ququart. For each Pauli basis $P_{4}$, we only simulate the $|P_{0}\rangle_{4}$ state.}
\label{supp:tab:gkp4_error_budgeting}
\end{table}
\end{center}

\clearpage

%

\end{bibunit}
\makeatother

\end{document}